\documentclass[preprint,times,authoryear]{elsarticle}
\pdfoutput=1
\usepackage{amsmath}  
\usepackage{amssymb}
\usepackage{subfigure}
\usepackage{ecrc}
\usepackage{rotating}
\usepackage{xcolor}
\usepackage{latexsym}
\usepackage{bm}
\usepackage{floatrow}
\newfloatcommand{capbtabbox}{table}[][0.45\textwidth]

\volume{}
\jid{}
\firstpage{1}
\journalname{}
\jnltitlelogo{}
\runauth{}

\newcommand{\be}{\begin{equation}}
\newcommand{\ee}{\end{equation}}
\newcommand{\ba}{\begin{eqnarray}}
\newcommand{\ea}{\end{eqnarray}}

\newcommand{\al}{\alpha}
\newcommand{\bt}{\beta}
\newcommand{\gm}{\gamma}

\newcommand{\kp}{\kappa}

\newcommand{\red}{\color{black}}

\begin{document}

\begin{frontmatter}

 \title{Intergranular normal stress distributions in untextured polycrystalline aggregates
}

\author[JSI]{S.~El~Shawish\corref{cor1}}
\ead{samir.elshawish@ijs.si}
\author[CEA]{J.~Hure\corref{cor1}}
\ead{jeremy.hure@cea.fr}

\cortext[cor1]{{Corresponding author}}
\address[JSI]{{Jo$\check{z}$ef Stefan Institute, SI-1000, Ljubljana, Slovenia}}
\address[CEA]{{CEA, Universit\'{e} Paris-Saclay, DEN, Service d'\'Etudes des Mat\'{e}riaux Irradi\'{e}s, 91191 Gif-sur-Yvette cedex, France}}

\begin{abstract}
From a general point of view, InterGranular Stress-Corrosion Cracking
(IGSCC) 
%
%comes
results
from the interplay between mechanical loading and grain boundaries
opening. The former leads to intergranular stresses in polycrystalline
aggregates, the latter being either stress-accelerated or
stress-induced. This work aims at obtaining intergranular normal
stress distributions in uncracked polycrystalline aggregates, which is
considered as a key milestone towards IGSCC initiation predictive
modelling. Based on the finite element {method}, numerical simulations
have been performed on Voronoi polycrystalline aggregates considering
a wide variety of material constitutive equations: crystal elasticity
(cubic and hexagonal symmetries) with different anisotropy ratios and
crystal plasticity for different sets of slip systems \textcolor{black}{under the assumption of uniform critical resolved shear stress}: Face-Centered
Cubic (FCC), Body-Centered Cubic (BCC) and Hexagonal Close Packed
(HCP) {with or without hardening}, and for both uniaxial and
equibiaxial macroscopic loading conditions.
{In the elastic regime, a correlation between standard deviations of
intergranular normal stress distributions and a universal elastic
anisotropy index proposed recently is found and explained through a
simple model. For macroscopic strain larger than the yield strain, the
evolution of standard deviations with strain is rationalized by
accounting only for the macroscopic elastic strain and the standard
deviation of Taylor factor. These numerical results associated with
physically-based simple models allow to estimate easily intergranular
normal stress concentrations, constituting a tool for classifying
polycrystalline aggregates according to their potential susceptibility to
IGSCC.}

\end{abstract}

\end{frontmatter}

\section{Introduction}
\label{section1}
InterGranular Stress-Corrosion Cracking (IGSCC) is a material
degradation mode that has been observed experimentally for a wide
range of materials and applications, for example in austenitic
stainless steels \citep{nishioka2008,lemillier,stephenson2014,gupta}
and zirconium alloys \citep{cox,cox1990} used in nuclear power plant
environment, nickel-based alloys
\citep{rooyen1975,shen1990,panter2006,IASCC_IAEA} of steam generators,
high strength aluminium alloys used as structural materials in moist
air or aqueous environment \citep{speidel,burleigh}, {and}
ferritic steels used for pipelines in presence of carbon-dioxide
\citep{wang,arafin}. All these IGSCC examples correspond to the
initiation and propagation of cracks at grain boundaries
\citep{king08}. {Various} physical mechanisms have been proposed to
explain IGSCC \citep{burleigh} - \textit{e.g.} active path dissolution
\citep{parkins80}, slip-dissolution \citep{newman}, hydrogen
embrittlement \citep{beachem}, localized deformation
\citep{bieler,mdmcm} - that depend on the material, the corrosive
environment {and} the local stress state. As such, evaluation of
intergranular stresses is required in order to quantitatively predict
IGSCC. In the recent years, experimental assessment of grain
boundaries stresses has become available through High Resolution
Electron Back Scatter Diffraction (HR EBSD) \citep{gardner2010,mdmcm}
or Laue microdiffraction \citep{larson,guo} {which} remains up to
now limited to rather small number of grain boundaries and therefore
does not allow to get statistical information about intergranular
stress distributions.

On the modelling side using Crystal Plasticity constitutive equations,
efforts have been made to obtain local stresses in polycrystalline
aggregates using Finite Element Method (CPFEM)
\citep{barbe1,barbe2,sauzay} or FFT method \citep{lebensohn2012}.
CPFEM has been used in particular to get intergranular stresses
\citep{diard2002,diard2005,kanjarla,gonzalez2014,hure2016} providing
valuable information for IGSCC initiation, \textit{i.e.} for uncracked
material. Coupling CPFEM with cohesive zone modelling at grain
boundaries allowed to simulate both initiation and propagation of
intergranular cracks
\citep{diard2002,musienko2009,couvant2013,simono2014}. From an
engineering point of view, prediction of IGSCC initiation and
especially {its} dependence on applied macroscopic stress, is of
uttermost importance as it can provide safe operating range. Diard
\textit{et al.} \citep{diard2002,diard2005} provided the evaluation of
intergranular normal stresses as a function of the angle between the
grain boundary and loading direction, considering Hexagonal Close
Packed (HCP) crystal plasticity and uniaxial tension. Similar approach
was followed in \citep{gonzalez2014} for Face-Centered Cubic (FCC)
crystal plasticity in uniaxial tension, leading to the increase (or
decrease) of intergranular normal stresses - compared to isotropic
case - due to elastic and/or plastic anisotropy. In \citep{hure2016},
intergranular normal stress distributions were obtained for FCC
material for a realistic polycrystalline aggregate \citep{simono2011}
that give statistical information about intergranular
stresses. However, to the authors' knowledge, no systematic CPFEM
studies have been performed up to now to provide statistical
distributions of normal intergranular stresses as a function of
elastic and plastic anisotropy
%
%coming
arising
from the considered slip systems (HCP, FCC, BCC) and loading
conditions (uniaxial or biaxial) on polycrystalline aggregates. Owing
to the different materials on which IGSCC can be observed, such study
appears to be a key ingredient towards IGSCC initiation modelling and
is therefore the objective of this paper.

In the first part of the paper, finite element simulations are
described, including crystal-scale constitutive equations and
computations of intergranular stresses. In the second part,
probability density functions (\textit{pdf}) of normal intergranular
stresses for a wide range of situations - cubic and hexagonal
elasticity, (FCC, BCC, HCP) crystal plasticity, uniaxial and
equibiaxial loading - are presented. {Standard deviations are used to
quantify the broadening of the \textit{pdf}s due to elastic and/or
plastic mismatches. Simple formulas are proposed to rationalize the
main numerical observations.}
The results are finally discussed regarding {their} range of 
validity
and towards establishing a tool for a quick and reliable estimation of
material susceptibility to IGSCC initiation.

% Samir
%---------------------------------------
\section{Methods}
%---------------------------------------
\label{section2}

%---------------------------------------
\subsection{Assumptions}
%---------------------------------------
Grains in a polycrystalline aggregate are modeled as three-dimensional
(3D) homogeneous continua of convex shape. Voronoi partitioning of the
modelling space is used to define the topology of the grains. The
average grain is therefore assumed isotropic in shape, \textit{i.e.} 
not elongated in any direction. 3D Voronoi tessellation has been
extensively used over the last decade \citep{barbe1,barbe2,diard2005}
since it produces a random virtual microstructure that captures some
of the features of real microstructures {\citep{fan}}.

In single-phase polycrystals, the material properties are the same
for all the grains that are defined simply by their
crystallographic orientation. In this study, grain orientations are
assumed to be fully random, thus providing zero texture to the
aggregate.

Polycrystalline simulations are performed using finite element
discretization of the grains. To investigate the mechanical response
of the aggregate under uniaxial and equibiaxial loading conditions,
anisotropic elasticity and crystal plasticity material properties are
assigned to each grain. No constitutive modelling of the grain
boundaries is considered. Slip is furthermore assumed to be the only
mechanism for plastic deformation. For simplicity, all slip systems
(within FCC, BCC and HCP crystal systems) are assigned the same shear
flow law and hardening behavior.

%---------------------------------------
\subsection{Finite element simulations}
%---------------------------------------

%---------------------------------------
\subsubsection{Material constitutive equations}
%---------------------------------------
\label{sec_mat}

A single crystal is assumed to behave as an anisotropic continuum.
Constitutive relations in linear elasticity are governed by the
generalized Hooke's law
\be
  \sigma_{ij}=C_{ijkl} \epsilon_{kl}
  \label{eq_hooke}
\ee
where $\sigma_{ij}$ and $\epsilon_{ij}$ are, respectively, the
second-order stress and (elastic) strain tensors and $C_{ijkl}$ the
fourth-order stiffness tensor ($i,j,k,l=1\ldots 3$). Table \ref{tab1}
lists the elastic constants (in Voigt notation) used  most often
in this study.

\begin{table}[H]
\begin{tabular}{c|c|c|c|c|c}
 & $C_{11}$ & $C_{12}$ & $C_{13}$ & $C_{33}$ & $C_{44}$  \\
 & (GPa)   & (GPa)   & (GPa)   & (GPa)     &   (GPa)    \\
\hline
\hline
%Al           & 107.0   & 60.8    &  60.8  & 107.0  & 28.3  \\
Al           & 107.3   & 60.9    &  60.9  & 107.3  & 28.3  \\
Fe $\gamma$  & 197.5   & 125.0   &  125.0 & 197.5  & 122.0 \\
%Fe $\alpha$  & 230.0   & 135.0   &  135.0 & 230.0  & 117.0 \\
Fe $\alpha$  & 231.4   & 134.7   &  134.7 & 231.4  & 116.4 \\
Na           &  6.15   & 4.96    &  4.96  & 6.15   & 5.92  \\
\hline
Zn           & 161.0   & 34.2    & 50.1   & 61.0   &  38.3 \\
\hline
\end{tabular}
\caption{Elastic constants of cubic (Al, Fe $\gamma$, Fe $\alpha$,  Na) and
hexagonal (Zn) single crystals 
taken from \citep{sauzay,bower}. 
}
\label{tab1}
\end{table}

The plastic behavior of single crystal is described within crystal
plasticity theory where plastic deformation is governed by slip alone.
Crystal systems with associated slip systems used in this study are
listed in Table \ref{tab2}. Deformations by other mechanisms such as
diffusion, twinning or grain boundary sliding are not taken into
account.

\begin{table}[H]
\begin{tabular}{c|c|c|c|c}
Crystal system &  \multicolumn{3}{c}{Slip plane $\underline{n}$\ / direction $\underline{s}$}  &  Number of slip systems \\
\hline
\hline
FCC          &  $\{111\}$ $\langle 110 \rangle$    &  -  &  -  &  12 \\
\hline
BCC          &  $\{123\}$ $\langle 111 \rangle$    & \{112\} $\langle 111 \rangle$    &  \{110\} $\langle 111 \rangle$  &  48 \\
\hline
HCP1         &  $\{11\bar{2}2\}$ $\langle 11\bar{2}{\bar{3}} \rangle$    &  -  &  -  &   6  \\
HCP2         &  $\{0001\}$ $\langle 1\bar{2}10 \rangle$     & $\{11\bar{2}2\}$ $\langle 11\bar{2}{\bar{3}} \rangle$     &  $\{10\bar{1}0\}$ $\langle 1\bar{2}10 \rangle$    &  12   \\
\hline
\end{tabular}
\caption{Crystal systems and associated slip systems %used in this study 
(as, \textit{e.g.}, in \citep{mbiakop}).
}
\label{tab2}
\end{table}

The shear flow is modeled by visco-plastic behavior
\citep{Cailletaud1991,Hoc2001}
\be
  \dot{\gm}^\al=\left\langle\frac{|\tau^\al| - \tau_c^\al}{K_0}\right\rangle^n {\rm sign}(\tau^\al)
  \quad\hbox{with}\quad\langle x\rangle=\left\{ \begin{array}{ll}
	x &; x>0\\
        0 &; x\le 0
	\end{array}\right.
  \label{eq_1}
\ee
where $\gm^\al$ is the shear strain in slip system $\al$ and $\tau^\al$
and $\tau_c^\al$ are, respectively, the resolved shear stress and
critical resolved shear stress. Parameters $K_0$ and $n$ regulate the
viscosity of the shear flow. Here, $K_0=10$ MPa and $n=15$ have been
considered, leading to almost negligible strain rate dependency on
stresses for all simulations presented hereafter. This corresponds to
the rate-independent limit of visco-plastic behavior modeled by
Eq.~(\ref{eq_1}).

To account for hardening, the critical resolved shear stress is
additively decomposed into
\be
  \tau_c^\al=\tau_0 + H \Gamma \quad\hbox{with}\quad \Gamma=\sum_\al\int|\dot{\gm}^\al| dt
  \label{eq_2}
\ee
where $\tau_0$ denotes the initial critical resolved shear stress
assumed to be the same for all slip systems and $H\ge 0$ is the
amplitude of linear Taylor hardening.  An {arbitrary} value of
$\tau_0=100$ MPa is taken {and discussed }in Sec.
\ref{section3}. 
The purpose of using such a simple hardening behavior is to
demonstrate the effects of finite hardening ($H>0$) in comparison to
quasi-ideal plasticity ($H=0$). In Sec.~4 further
investigation towards more realistic hardening behavior
%
%will be
{is}
considered. \textcolor{black}{Similarly, the assumption of uniform critical resolved shear stress (CRSS) for HCP crystal systems listed in Table~\ref{tab2} is not intended to be realistic, but rather to investigate the effects of higher plastic anisotropy compared to FCC and BCC. A discussion about the effect of non-uniform CRSS is given in Sec.~4.}

The constitutive law was implemented into two numerical codes --
Abaqus \citep{abaqus} and Cast3M \citep{castem} -- to assess the
influences of different integration methods on the robustness of
results. In both implementations large deformation theory was
accounted for by the usual multiplicative decomposition of the
deformation gradient. The Abaqus implementation was accomplished
through a UMAT subroutine following the implementation of Huang
\citep{huang}, where slight modifications were needed to accommodate the
shear flow law of Eq. (\ref{eq_1}). In Cast3M the constitutive
equations were generated by the MFront code generator \citep{mfront}.
Both implementations have been shown to give equivalent
results. Details about numerical implementations can be found
elsewhere \citep{hure2016}.

%---------------------------------------
\subsubsection{Polycrystalline aggregates}
%---------------------------------------

\begin{figure}[H]
\subfigure[]{\includegraphics[height = 6cm]{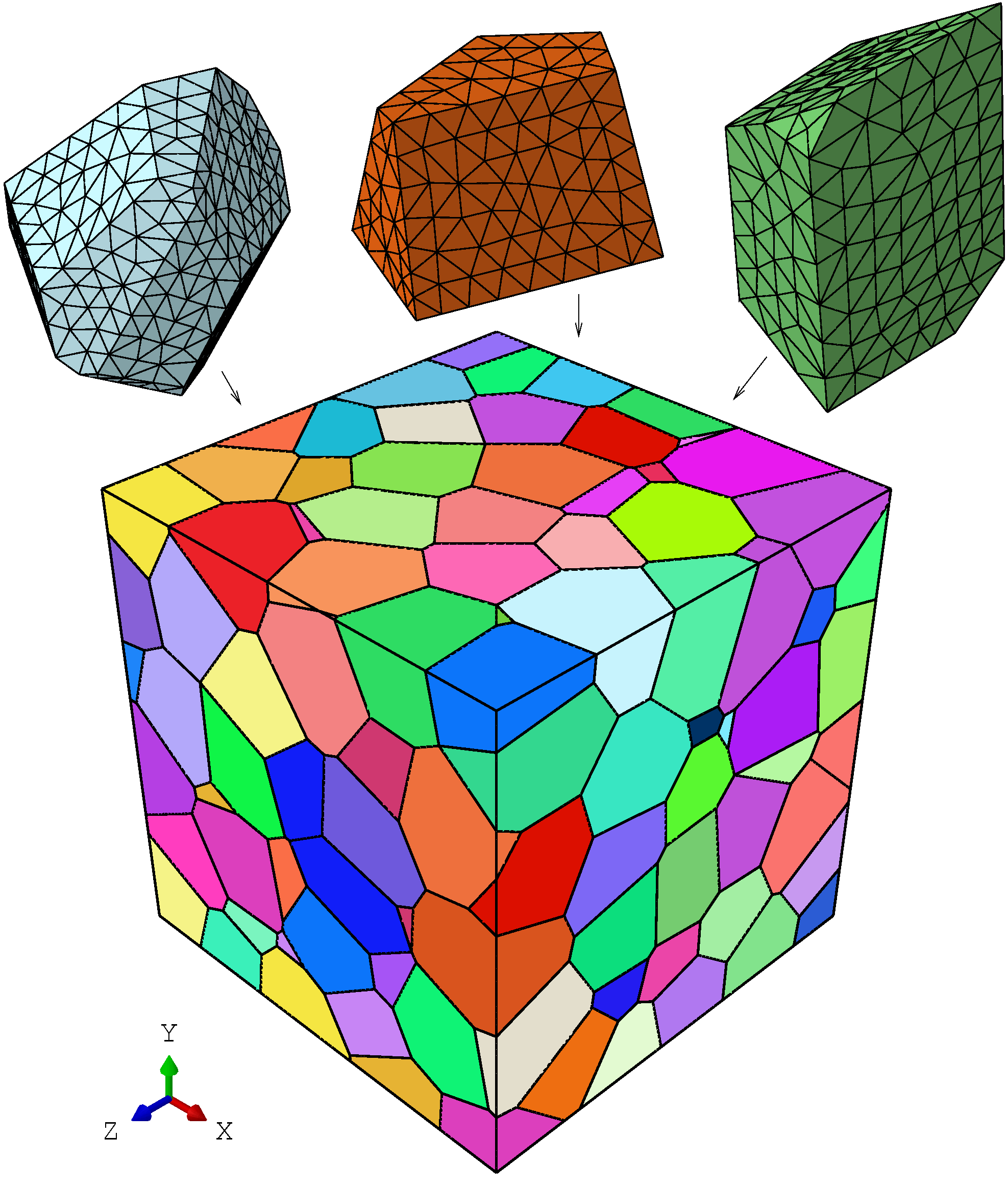}}
\hspace{1cm}
\subfigure[]{\includegraphics[height = 6cm]{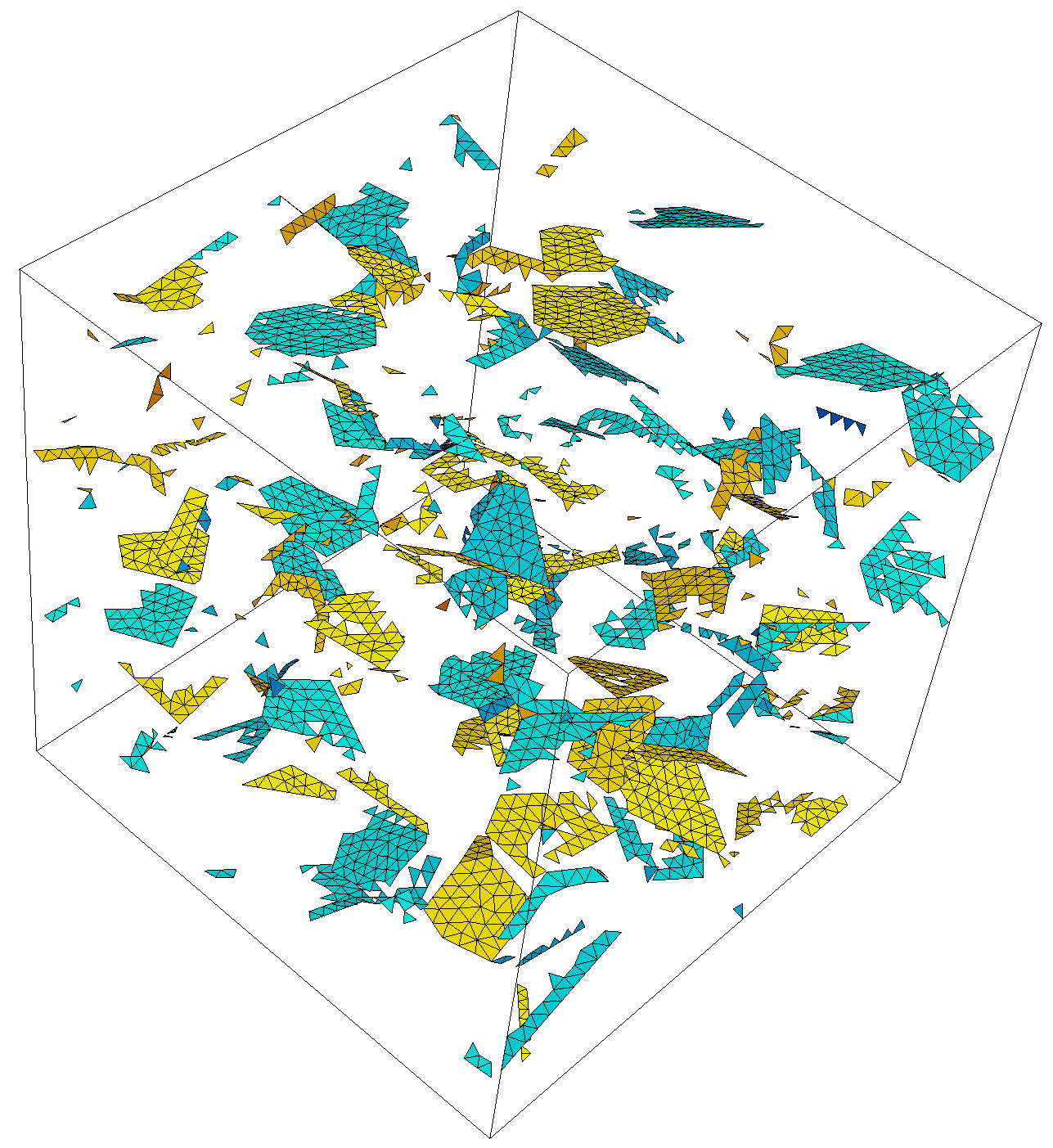}}
\caption{(a) Voronoi aggregate model with 200 grains with
corresponding mesh density indicated on the three grains.  (b) Set of
intergranular elements for which normal stresses are larger than
macroscopic stress $\Sigma$ (obtained considering cubic elasticity of
Fe $\gamma$ for uniaxial tensile loading in $y$ direction). Colors
correspond to different grain boundary orientations.}
\label{fig:1}
\end{figure}

Finite element models of polycrystalline aggregates were built from
Voronoi tessellations made of 
 100 to 1000
grains and composed of 
130k to 1400k
tetrahedral quadratic elements (C3D10 in Abaqus). Voronoi
tessellations were obtained using non-zero repulsion distance between
the seeds to avoid tiny edges which were shown in
\citep{gonzalez2014} to be the source of unphysically large
variability of intergranular stresses. Random initial crystallographic
orientations were assigned to the grains to introduce zero texture.
Figure \ref{fig:1}(a) shows {the Voronoi aggregate} with 200 grains
used in this study. {Except in Sec. \ref{size_effects} where the
influence of using different Voronoi aggregates is studied in more
detail, all results presented hereafter have been obtained on this
aggregate with one set of random crystallographic orientations. The
variability regarding intergranular stresses due to different sets of
random crystallographic orientations and/or aggregates through the
presence of small boundaries were evaluated in \citep{gonzalez2014}
for a given set of material properties. 
%
%At the opposite,
On the contrary,
the purpose of this study is to quantity separately the effects of
material properties, therefore requiring 
%
%considering 
%
the same aggregate/crystallographic orientations for all the
simulations.}

In the case of uniaxial loading, incremental tensile displacement is
applied along X axis to all the nodes on front surface, while keeping
the nodes on back surface constrained to have zero axial displacement.
In the case of equibiaxial loading, same displacements and constraints
are applied along the additional Y axis. In both cases nodes on the
lateral (free) surfaces are not constrained so as to study the free
surface effects, {relevant for IGSCC}. 
{Such boundary conditions, approximating periodic boundary conditions
by keeping the opposite surfaces (of normal X and Y) flat, have been
shown to lead to local disturbances of the stress field. However, this
artifact was checked to have a negligible effect on the stress
distributions shown hereafter (see \ref{app1}).} {Moreover, especially
in the cases where no grain-scale hardening is considered, local
necking is expected to appear leading to a strain gradient along the
loading direction that complicates the interpretation of the
results. Thus, small applied macroscopic strains are considered
in this study, which are relevant for IGSCC initiation phenomenon. In
addition, the deformed aggregates are checked at the end of each
simulation in order to assess the potential occurrence of local
necking.}
The applied nominal strain in the uniaxial and equibiaxial loading is
respectively $\Delta x/x_0=0.05$ and $\Delta x/x_0=\Delta y/y_0=0.05$,
while the associated strain rate is set to $10^{-3}$ s$^{-1}$.{The
macroscopic Cauchy stress tensor $\bm{\Sigma}$ {is computed by
dividing the resulting force on the boundaries where loading is
applied by the \textit{actual} section}, leading to $\bm{\Sigma}
\approx \Sigma e_X \otimes e_X$ for uniaxial loading (and $\bm{\Sigma}
\approx \Sigma (e_X \otimes e_X + e_Y \otimes e_Y)$ for equibiaxial
loading) as a result of the boundary conditions and the absence of
crystallographic texture. The scalar value $\Sigma$ is used in the
following as a measure of stress amplitude.}
All stress distributions shown in Sec.~\ref{section3}, once normalized
by the macroscopic stress $\Sigma$, were checked to be practically
independent of the number of grains and mesh density.

%---------------------------------------
\subsubsection{Intergranular stress computations}
%---------------------------------------

Simulations performed on polycrystalline aggregates allow to obtain
probability density functions (thereafter noted \textit{pdf}) of
normal stresses ($\sigma_{nn}$) between adjacent grains as a function
of elastic anisotropy, plastic anisotropy, applied macroscopic strain
and loading conditions (uniaxial or equibiaxial). As brittle cracking
of grain boundaries is dependent on local stress state, these
\textit{pdf}s are believed to be a key ingredient towards IGSCC
modelling.

For each pair of tetrahedral elements defining a boundary between two
grains%
\footnote{
Tetrahedral element can touch grain boundary either with its facet,
edge or node. In this study, only pairs of tetrahedra are considered
which share common grain boundary facet. In this case the associated
facet area and normal are well defined for further calculation of
\textit{pdf}.}%
, the Cauchy stress tensors $\bm{\sigma}$ at the closest Gauss points%
\footnote{
Tetrahedral element C3D10 has four Gauss points placed close to the
element corners. The three Gauss points belonging to the corners in
contact with the grain boundary were selected as closest Gauss
points.}
near the boundary are obtained. These are then converted to normal
stresses, $\sigma_{nn}=n.\bm{\sigma}.n$, knowing the normal $n$ of the
grain boundary facet. The projected stresses computed at various Gauss
points are then averaged (with equal weights) to yield one single
value for the normal stress per element pair. As elements of different
sizes are used in the mesh, the occurrence of $\sigma_{nn}$ in the
computation of \textit{pdf} is weighted by the surface of the grain
boundary facet on which it was obtained. One may note, however, that
such 
%
%rescaling 
weighting
has only small effect on the results.

The method introduced here to obtain intergranular stresses differs
slightly from those used in \citep{diard2002} and \citep{gonzalez2014}. In
the former, additional Gauss points were considered in elements close
to grain boundaries, while cohesive elements were added to the model
in the latter to obtain stresses exactly at grain boundaries. It is
also known \citep{elshawish2013} that, in addition to large scatter,
the accuracy of cohesive elements (in Abaqus) is questionable,
especially when the grains are plastically strained. The rather simple
method used in this study has, however, shown (see \citep{hure2016})
to be accurate enough to provide converged global shapes of the
distributions upon refining the mesh. {Moreover, the robustness of the
method used here is assessed in more details in \ref{app2} where it is
shown that using special tetrahedral elements (C3D10I in Abaqus) with
Gauss points located at the nodes (therefore exactly on the grain
boundary) leads to similar results regarding the moments (mean and
standard deviation) of \textit{pdf}s. Same conclusion is also obtained
by averaging intergranular normal stresses on grain facets composed of
numerous triangular elements (see Fig.~\ref{fig:1} b).}

%%%%%%%%%%%

It is important to note that in absence of material anisotropy the
\textit{pdf} of normal stress behaves as
$1/2\sqrt{\sigma_{nn}/\Sigma}$ for uniaxial and as
$1/2\sqrt{1-\sigma_{nn}/\Sigma}$ for equibiaxial loading conditions,
where $\sigma_{nn}$ is trivially bounded by $\Sigma$
 (see \ref{app2b} for derivation).
However, the upper bound can be easily exceeded when the mismatch
effects between adjacent grains become important. Figure
\ref{fig:1}(b) demonstrates such an example where only those
intergranular elements are shown for which normal stresses are larger
than macroscopic stress, $\sigma_{nn}>\Sigma$, assuming anisotropic
elasticity of the grains and uniaxial tensile loading
conditions. These elements are expected to either have small angle
between their grain boundary facet normal $n$ and loading direction
{and/or be at the boundary between two grains having strong
elastic/plastic mismatches} and/or be positioned close to geometric
discontinuities (edges or triple points) where stress concentrations
may be present.

%---------------------------------------
\section{Intergranular stress distributions}
%---------------------------------------
\label{section3}

In this section, intergranular normal stress distributions obtained
through finite element simulations on Voronoi aggregates are
described, considering crystal elasticity and/or plasticity (with
different slip systems), for both uniaxial and equibiaxial loading
conditions.

%---------------------------------------
\subsection{Crystal elasticity}
\label{crystalelasticity}
%---------------------------------------

Only crystal elasticity is considered in this part, which corresponds
to the case of polycrystalline aggregates loaded below their
macroscopic yield stress. 

To follow the same nomenclature as used later for crystal plasticity,
cubic symmetry is specified as FCC or BCC and hexagonal symmetry as
HCP.

\begin{figure}[H]
\centering
\subfigure[]{\includegraphics[height = 5cm]{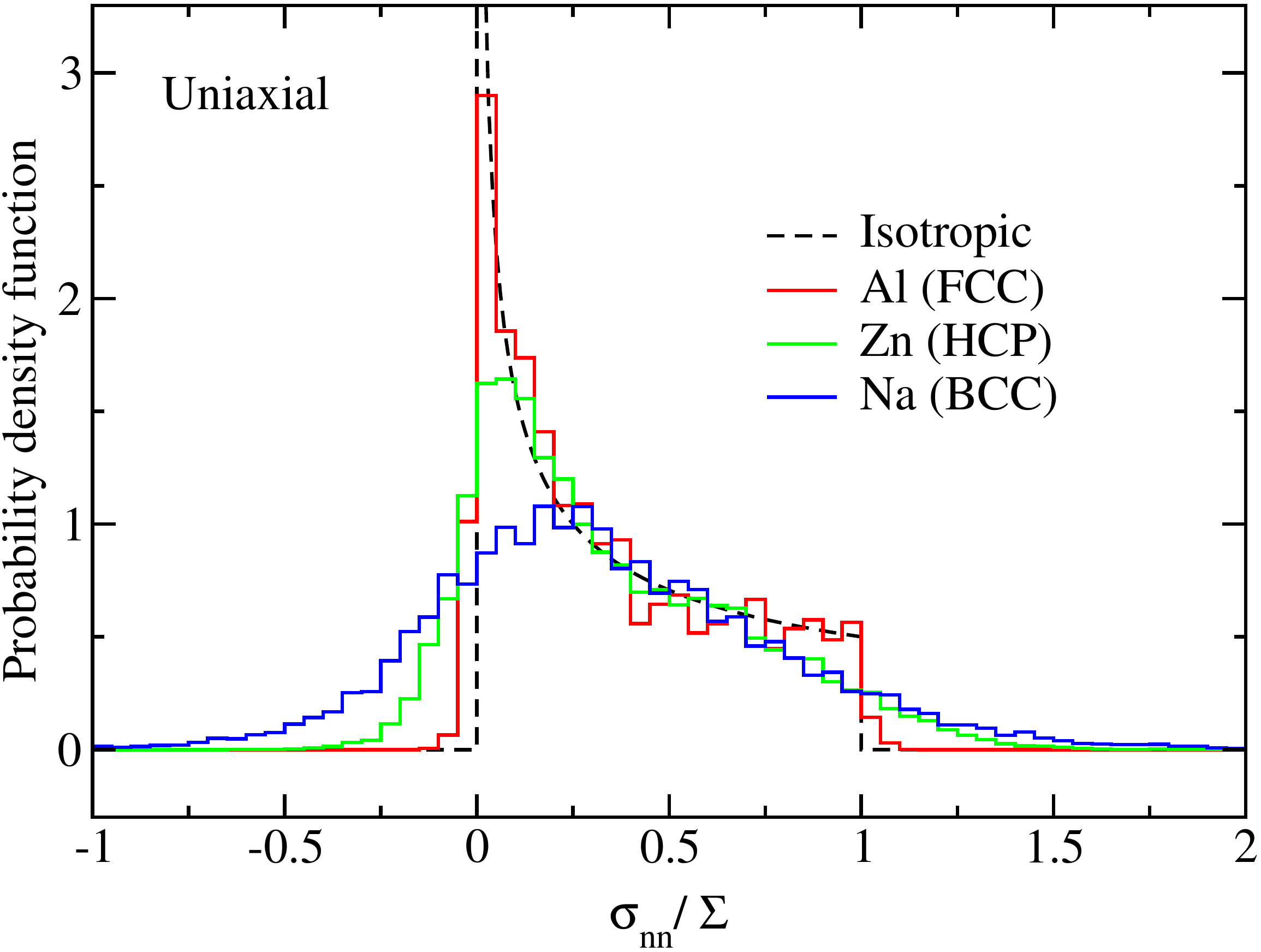}}
\hspace{1cm}
\subfigure[]{\includegraphics[height = 5cm]{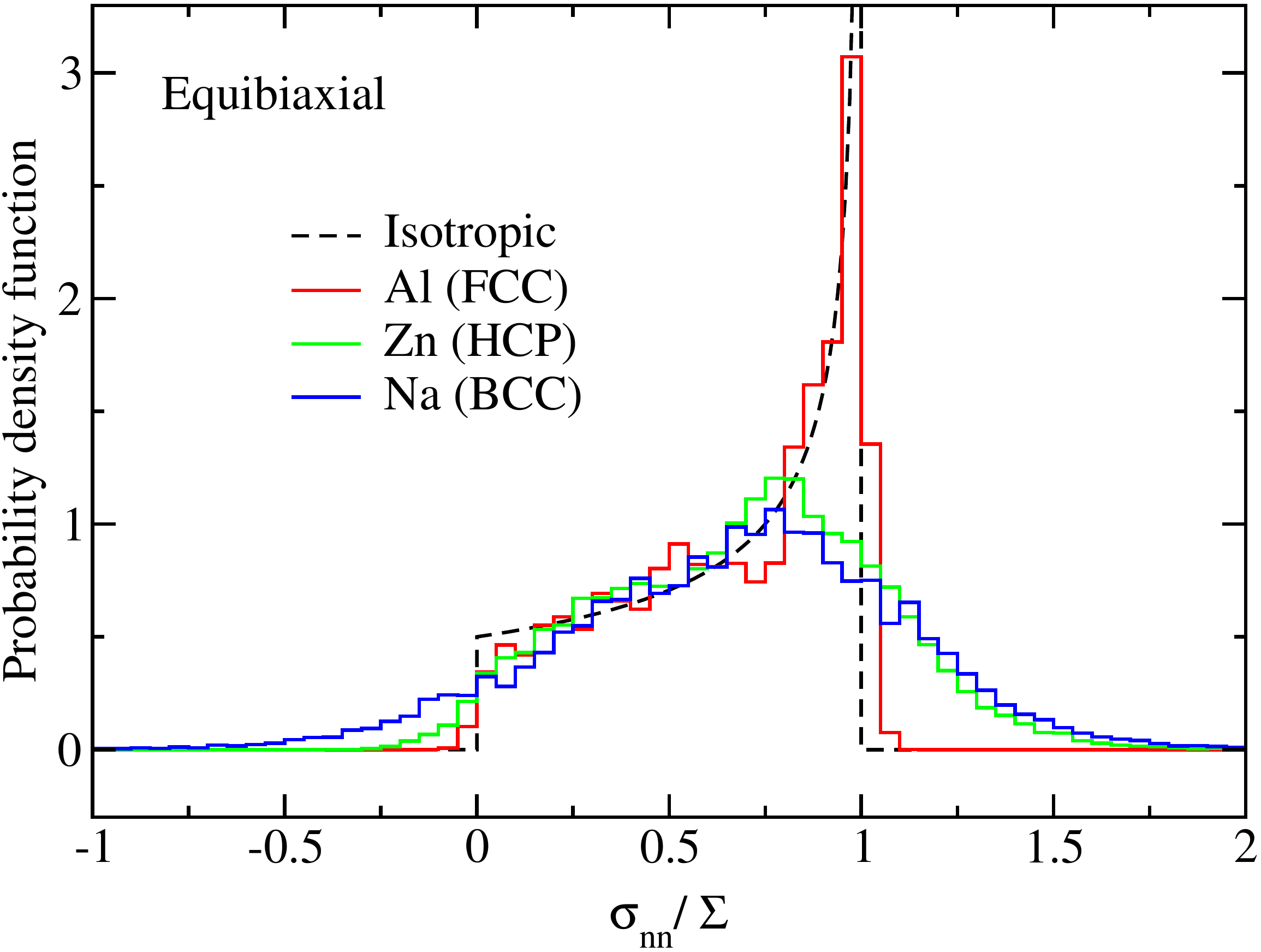}}
\caption{
Probability density functions of normalized intergranular normal
stress calculated with Voronoi finite element simulations assuming
crystal elasticity and (a) uniaxial and (b) equibiaxial loading
conditions.}
\label{fig:2}
\end{figure}

\begin{figure}[H]
\centering
\subfigure[]{\includegraphics[height = 5cm]{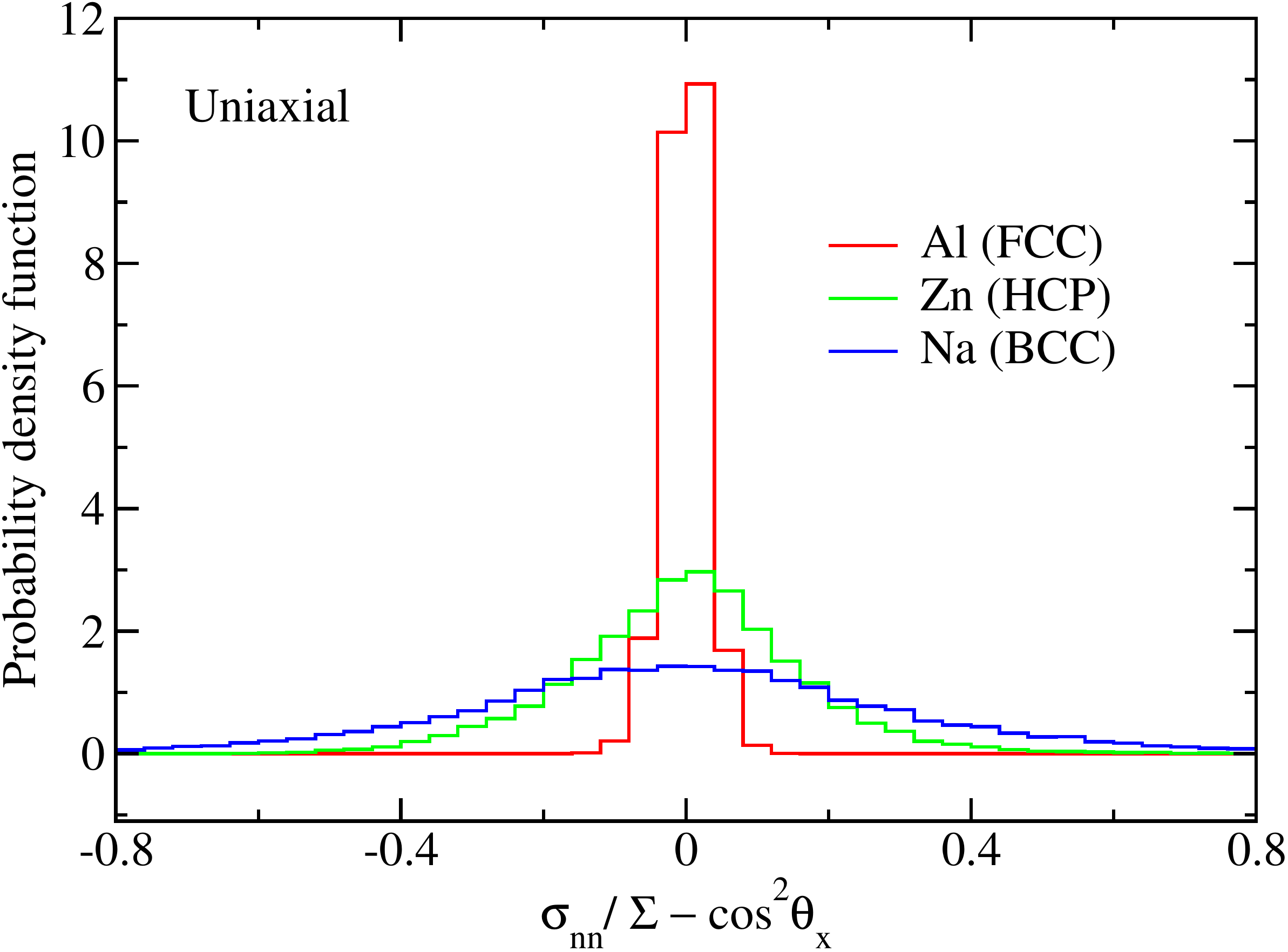}}
\hspace{1cm}
\subfigure[]{\includegraphics[height = 5cm]{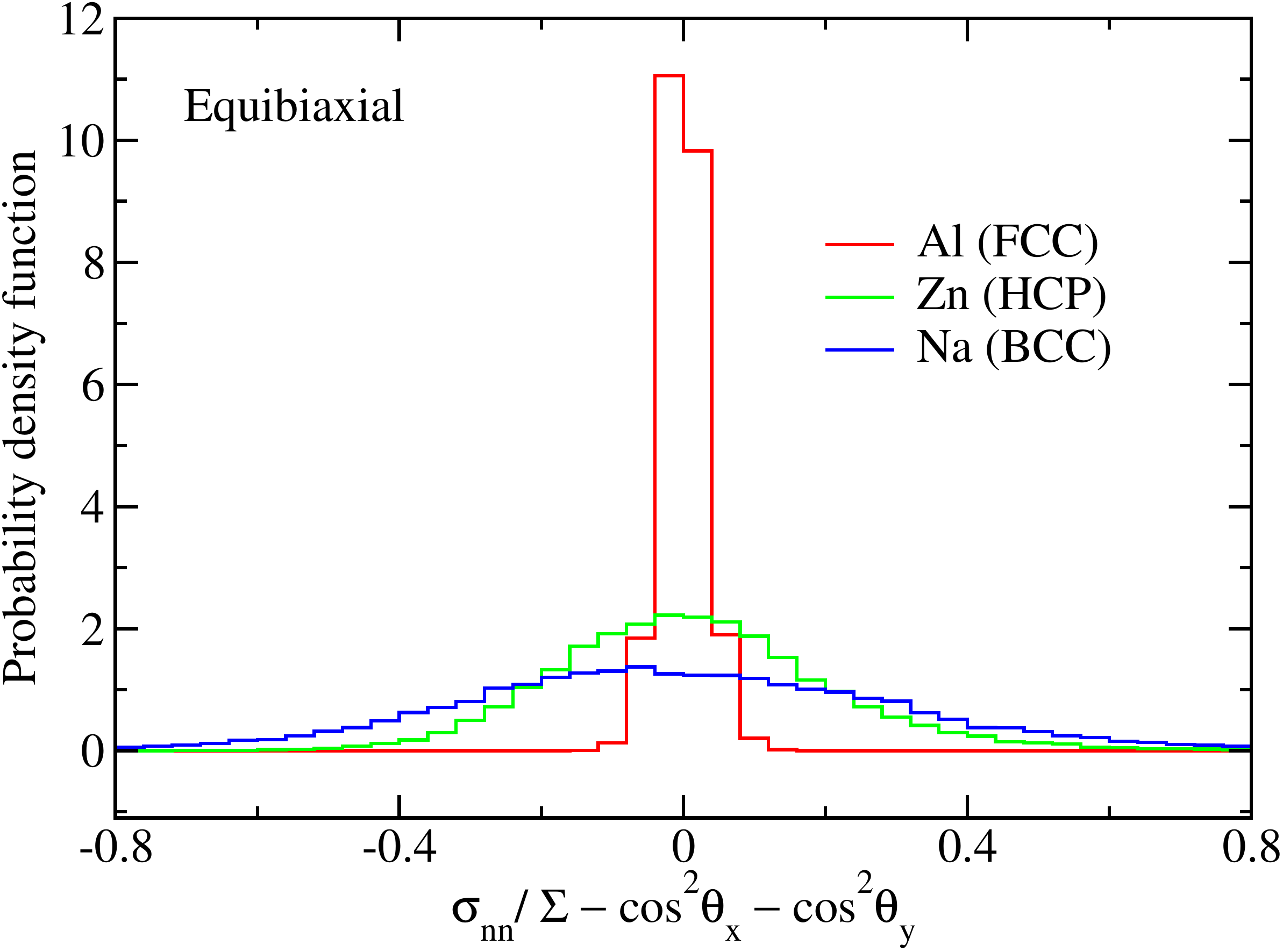}}
\caption{ 
Probability density functions of normalized intergranular normal
stress reduced by its isotropic contribution calculated with Voronoi
finite element simulations assuming crystal elasticity and (a)
uniaxial and (b) equibiaxial loading conditions. $\theta_x$ and
$\theta_y$ are the angles between grain boundary normal and loading
directions along X and Y axes, respectively.}
\label{fig:2b}
\end{figure}

As mentioned in the previous section, in the absence of material
anisotropy intergranular stresses $\sigma_{nn}$ are bounded by
$\Sigma$, where $\Sigma$ is the macroscopic stress. However, the upper
bound can be easily exceeded when the mismatch effects between the
adjacent grains become important, \textit{i.e.} when material
anisotropy is finite. Figures \ref{fig:2}(a) and \ref{fig:2}(b) show
the calculated intergranular normal stress distributions obtained
through finite element simulations on Voronoi aggregate for,
respectively, uniaxial and equibiaxial loading conditions, on three
different materials: Al with FCC, Na with BCC and Zn with HCP lattice
symmetry. 

The three materials were chosen to show the evolution of the
\textit{pdf} shape from isotropic-like (Al case) through bimodal-like
(Zn case) towards normal-like shape (Na case). 
{In the three cases, the mean values remain close to the theoretical
value of $1/3$ in the absence of material anisotropy.}

It seems reasonable to expect that broadening of the distributions
arises due to stronger mismatch effects between the grains which may
be characterized by elastic anisotropy of the grains. Being almost
isotropic at the crystal scale, polycrystalline Al has distributions
in close agreement with the assumption of uniform stress, as mismatch
effects between the grains are small (Figs. \ref{fig:2}(a) and
\ref{fig:2}(b)). On the contrary, one of the most anisotropic metallic
materials with BCC symmetry (Na) or HCP symmetry (Zn) show
considerably wider distributions where significant probabilities are
associated with intergranular normal stress larger than macroscopic
stress.

The influence of anisotropy is analysed further in Figs.
\ref{fig:2b}(a) and \ref{fig:2b}(b) where \textit{pdf}s of normalized
intergranular normal stress reduced by its isotropic counterpart,
$(\sigma_{nn}-\sigma_{nn}^{iso})/\Sigma$, are presented with
$\sigma_{nn}^{iso}=\Sigma \cos^2{\theta_x}$ for uniaxial tensile and
$\sigma_{nn}^{iso}=\Sigma (\cos^2{\theta_x} + \cos^2{\theta_y})$ for
equibiaxial loading conditions, respectively. Here, $\theta_x$ and
$\theta_y$ are the angles between grain boundary normal and loading
directions along X and Y axes, respectively. The obtained normal-like
distributions for all the cases show the corresponding distribution
widths growing with the increasing mismatch effects, in agreement with
the observations in Figs. \ref{fig:2}(a) and \ref{fig:2}(b). It thus
seems natural to
{use the first two statistical moments - mean and variance (or
standard deviation) - to describe the distributions of
$(\sigma_{nn}-\sigma_{nn}^{iso})/\Sigma$. The mean value is observed
to be close to zero (which is consistent with the constant mean value
of $\sigma_{nn}/\Sigma$), while the standard deviation (hereafter
denoted by $s$) depends on the elastic anisotropy of the grains.}

\begin{figure}[H]
\centering
\subfigure[]{\includegraphics[height = 5.cm]{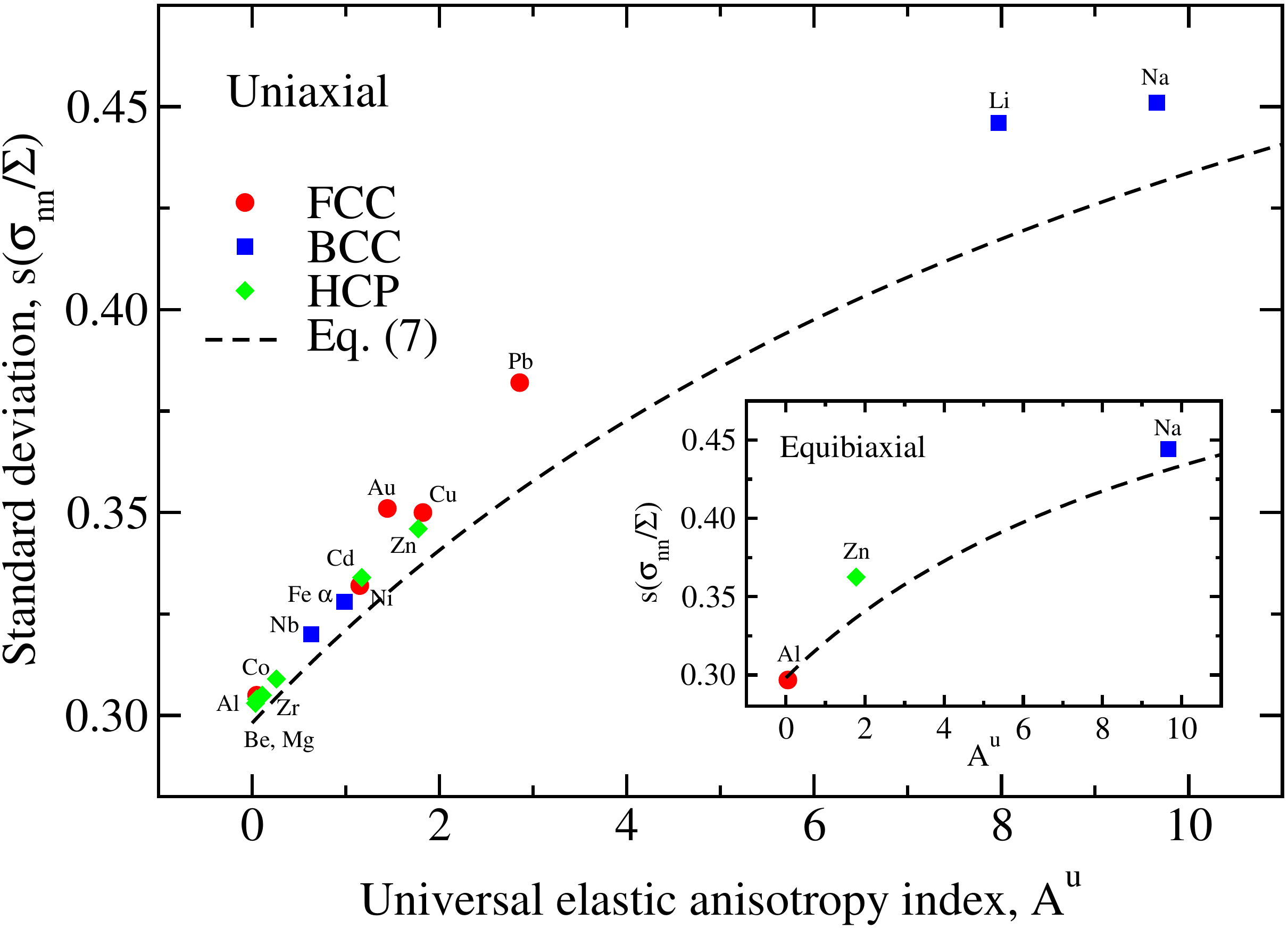}}
\hspace{0.5cm}
\subfigure[]{\includegraphics[height = 5.cm]{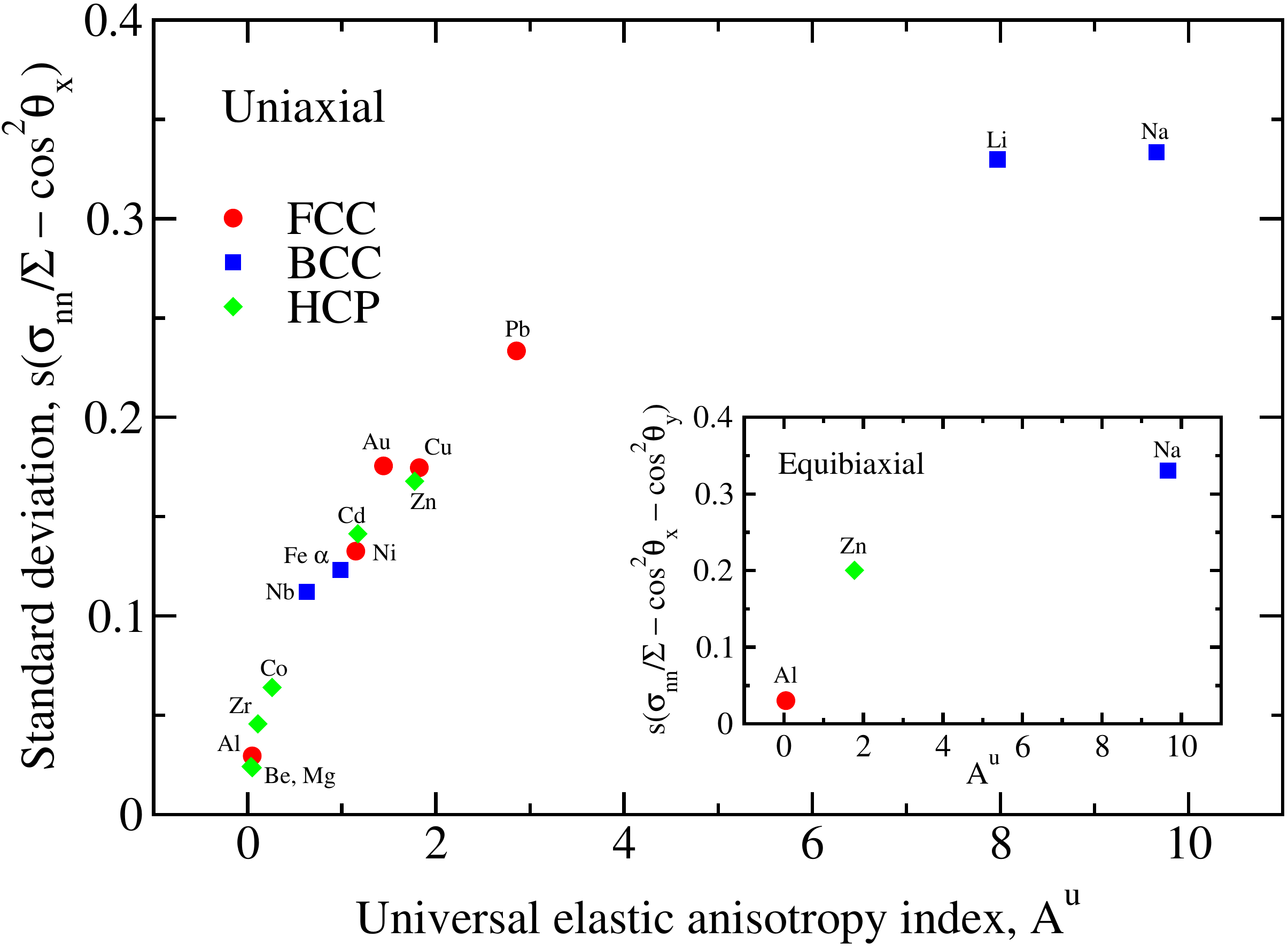}}
\caption{
Comparison of universal elastic anisotropy index $A^u$ with (a)
standard deviation of normalized intergranular normal stress for
uniaxial tensile loading (inset: equibiaxial loading) 
and (b) standard deviation of normalized intergranular normal stress
reduced by its isotropic contribution for uniaxial tensile loading
(inset: equibiaxial loading). Crystal elasticity was considered in
Voronoi finite element simulations to calculate standard
deviations. $A^u$ was calculated for infinite random polycrystal using
analytic formulae given in \ref{app3}. Dashed line in (a) represents a
relationship given in Eq.~(\ref{austd}). 
}
\label{fig:3}
\end{figure}

The elastic anisotropy can be characterized conveniently by the
universal elastic anisotropy index $A^u$ introduced in
\citep{ranganathan}  as
\be
  A^u=5\frac{G^V}{G^R}+\frac{K^V}{K^R}-6
\ee
where $G$ and $K$ denote shear and bulk moduli, and superscripts $V$
and $R$ represent the Voigt and Reuss estimates, respectively. Unlike
\textit{e.g.} Zener index \citep{zener} for cubic materials, $A^u$ can
be applied to any lattice symmetry to describe the elastic anisotropy
($A^u=0$ corresponds to elastic isotropy). {Roughly speaking, this
universal anisotropy index is proportional to the ratio of Voigt
\citep{voigt} to Reuss \citep{reuss} bounds of the elastic modulus for
an aggregate with no crystallographic texture.} 

In this study, index $A^u$ was calculated for several most common
cubic (FCC and BCC) and hexagonal (HCP) metallic materials assuming
infinitely large polycrystals with random grain orientations (see
\ref{app3} for more detail). For each material the calculated $A^u$ is
shown in Fig. \ref{fig:3} along with the corresponding standard
deviation $s$ of $\sigma_{nn}/\Sigma$ (in Fig. \ref{fig:3}(a)) and
$(\sigma_{nn}-\sigma_{nn}^{iso})/\Sigma$ (in Fig. \ref{fig:3}(b)),
both calculated on a finite Voronoi aggregate under uniaxial and
equibiaxial loading conditions. A clear monotonic relationship is
observed in Figs. \ref{fig:3}(a) and (b) for all considered materials
(FCC, BCC and HCP), which suggests that {the elastic anisotropy index
$A^u$ is able to faithfully describe standard deviation of normalized
intergranular normal stresses.}
%
%elastic anisotropy may be faithfully characterized by standard
%deviation of normalized intergranular normal stresses.
%
Since both variants, $s(\sigma_{nn}/\Sigma)$ and
$s((\sigma_{nn}-\sigma_{nn}^{iso})/\Sigma)$, seem to correlate equally
well with the universal elastic anisotropy index $A^u$, the former one
is selected for further investigation as \textit{pdf} of
$\sigma_{nn}/\Sigma$ allows to associate the tails of distributions
directly with larger intergranular stresses and with their probability
of appearance in the model. In this respect, for example, it is easy
to verify in Fig. \ref{fig:2} that equibiaxial loading condition, in
comparison to uniaxial tensile loading, produces higher \textit{pdf}
tails on a tensile (right) side, which may indicate, statistically,
more likely initiation of IGSCC. This observation, on contrary, could
not be deduced directly from the \textit{pdf}s of
$(\sigma_{nn}-\sigma_{nn}^{iso})/\Sigma$ in Fig. \ref{fig:2b}.

{The evolution of the standard deviation of normalized normal
intergranular stresses with the universal anisotropy index $A^u$ can
be understood as follows. Considering a polycrystalline aggregate in
uniaxial (or equibiaxial) tension with macroscopic strain $E$, the maximum
(intergranular normal) stress magnitude in the aggregate can be
estimated using Voigt's upper bound, $\sigma_{nn}^{max.} \approx
Y_{\mathrm{Voigt}}\,E$. The minimal value is assumed to be close to
zero $\sigma_{nn}^{min.} \approx 0$, which corresponds to grain
boundaries {parallel} to the loading direction. The macroscopic
stress of the aggregate is equal to $\Sigma = Y\,E$, where $Y$ is the
effective Young's modulus. For most aggregates, $Y$ can be fairly well
approximated as the arithmetic mean of Voigt and Reuss estimations of
Young's modulus, $Y \approx (Y_{\mathrm{Voigt}} +
Y_{\mathrm{Reuss}})/2$, as proposed by Hill. As the standard deviation
of $\sigma_{nn}/\Sigma$ scales with the measurement range,
\be
\begin{aligned}
  s\left(\frac{\sigma_{nn}}{\Sigma} \right) &\sim \frac{\sigma_{nn}^{max.} - \sigma_{nn}^{min.}}{\Sigma} \\
  &\approx \frac{4}{3\sqrt{5}} \frac{1}{ 1 + Y_{\mathrm{Reuss}}/Y_{\mathrm{Voigt}} }
\end{aligned}
\label{stdestim}
\ee
where the prefactor $4/3\sqrt{5}$ is used to recover the exact result
for isotropic crystals. Moreover, neglecting the differences between
Voigt and Reuss bounds of Poisson's ratio, the universal anisotropy
index can be approximated as
\be
  A^u \approx 6\left(Y_{\mathrm{Voigt}}/Y_{\mathrm{Reuss}} -1 \right).
  \label{auapprox}
\ee
Equation~(\ref{auapprox}) is shown in \ref{app3} to be a good
approximation (with error smaller than $\sim$15\%) of the universal
anisotropy index. Combining Eq.~(\ref{auapprox}) with
Eq.~(\ref{stdestim}) leads to the estimation of the standard deviation
as a function of the universal anisotropy index $A^u$
\be
  s\left(\frac{\sigma_{nn}}{\Sigma} \right)\approx \frac{4}{3\sqrt{5}}\frac{A^u + 6}{A^u + 12}.
  \label{austd}
\ee
Equation~(\ref{austd}) is compared to the numerical results in
Fig.~\ref{fig:3}a. Despite simplified assumptions used in the
derivation, Eq.~(\ref{austd}) is able to capture the evolution of the
standard deviation of the normalized intergranular normal stresses
with the universal anisotropy index $A^u$.}\\

The results of this section {concerning intergranular normal stresses
$\sigma_{nn}$ in an elastic untextured polycrystal} can be
summarized into the following few statements:
\begin{itemize}
\item 
Independently of the anisotropy of the single crystal, the mean value
of $\sigma_{nn}/\Sigma$ is equal to $1/3$ and $2/3$ for uniaxial and
equibiaxial loading conditions, respectively.

\item 
{Standard deviation of $\sigma_{nn}/\Sigma$ increases with the
anisotropy of the single crystal}: a clear monotonic relationship
between the standard deviation
and the universal elastic anisotropy index $A^u$ proposed by
\citep{ranganathan} is observed, {and a simple model (Eq.~(\ref{austd}))
is proposed to explain and describe this relationship.}

\item
{The universal anisotropy index $A^u$ can thus be used through
Eq.~(\ref{austd}) to assess the potential susceptibility of a
polycrystalline aggregate to IGSCC. }

\item
Equibiaxial loading condition is shown to be more damaging with
respect to the initiation of IGSCC than uniaxial tensile loading.
\end{itemize}

{The results presented in this section regarding the mean and standard
deviation of normalized normal intergranular stresses are valid for
untextured polycrystalline aggregates loaded below their macroscopic
yield stress.}

The influence of plastic effects on the evolution of
$s(\sigma_{nn}/\Sigma)$ at strains close or well above the yield
strain is going to be studied in the next section.

%---------------------------------------
\subsection{Crystal elasticity and plasticity}
\label{crystalplasticity}
%---------------------------------------

In this part, both crystal elasticity and plasticity are accounted for
in numerical simulations of polycrystalline aggregate, and probability
density functions of intergranular normal stress are shown as a
function of total strain up to 5\%, \textit{i.e.} well above the yield
strain. The focus is set on studying the evolution of
$s(\sigma_{nn}/\Sigma)$ with applied strain to identify main elastic
and plastic contributors to the broadening of the \textit{pdf}.

%-------------------------------------------------------
\subsubsection{Typical observations}

\begin{figure}[H]
\centering
\subfigure[]{\includegraphics[height = 5.5cm]{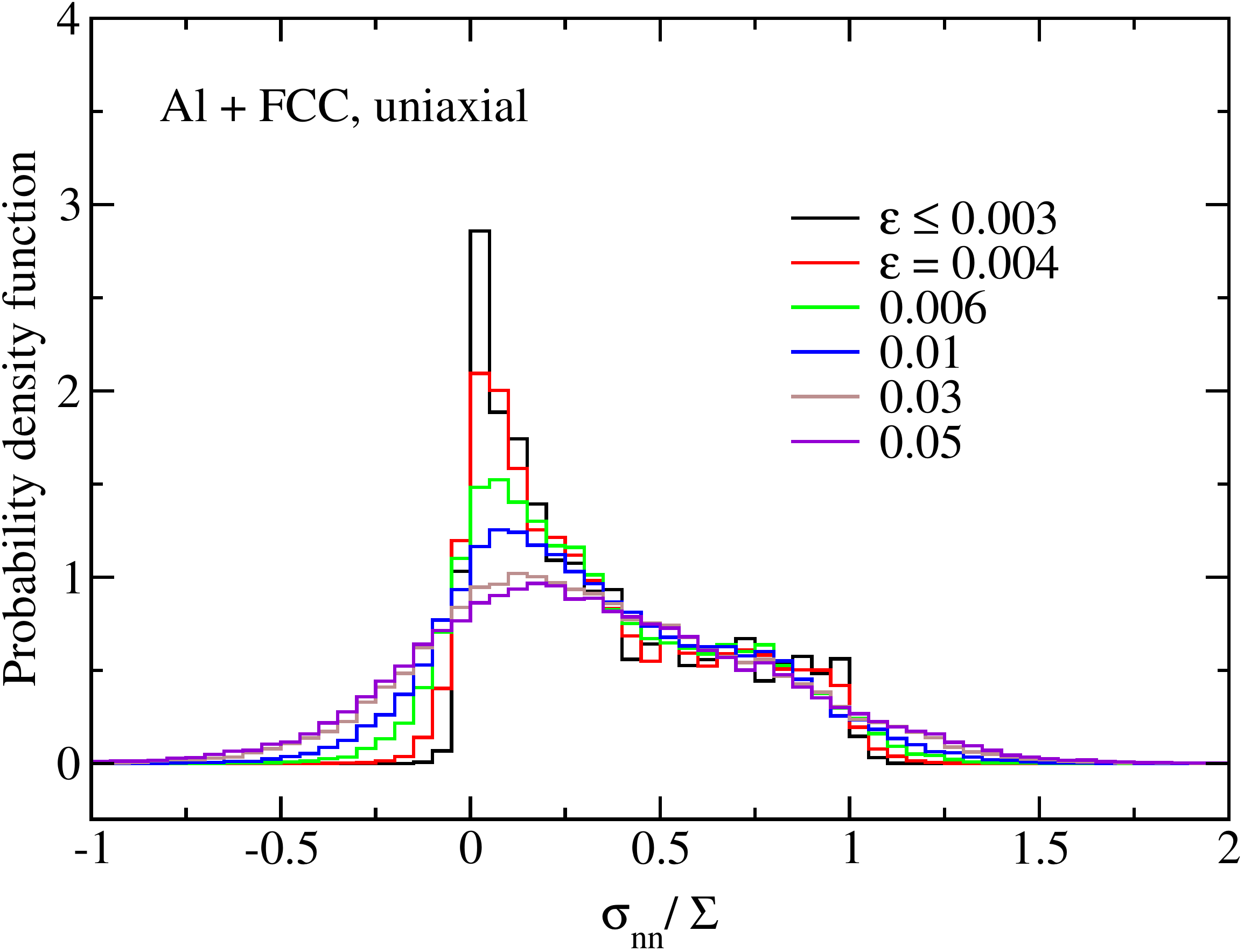}}
\hspace{0.5cm}
\subfigure[]{\includegraphics[height = 5.5cm]{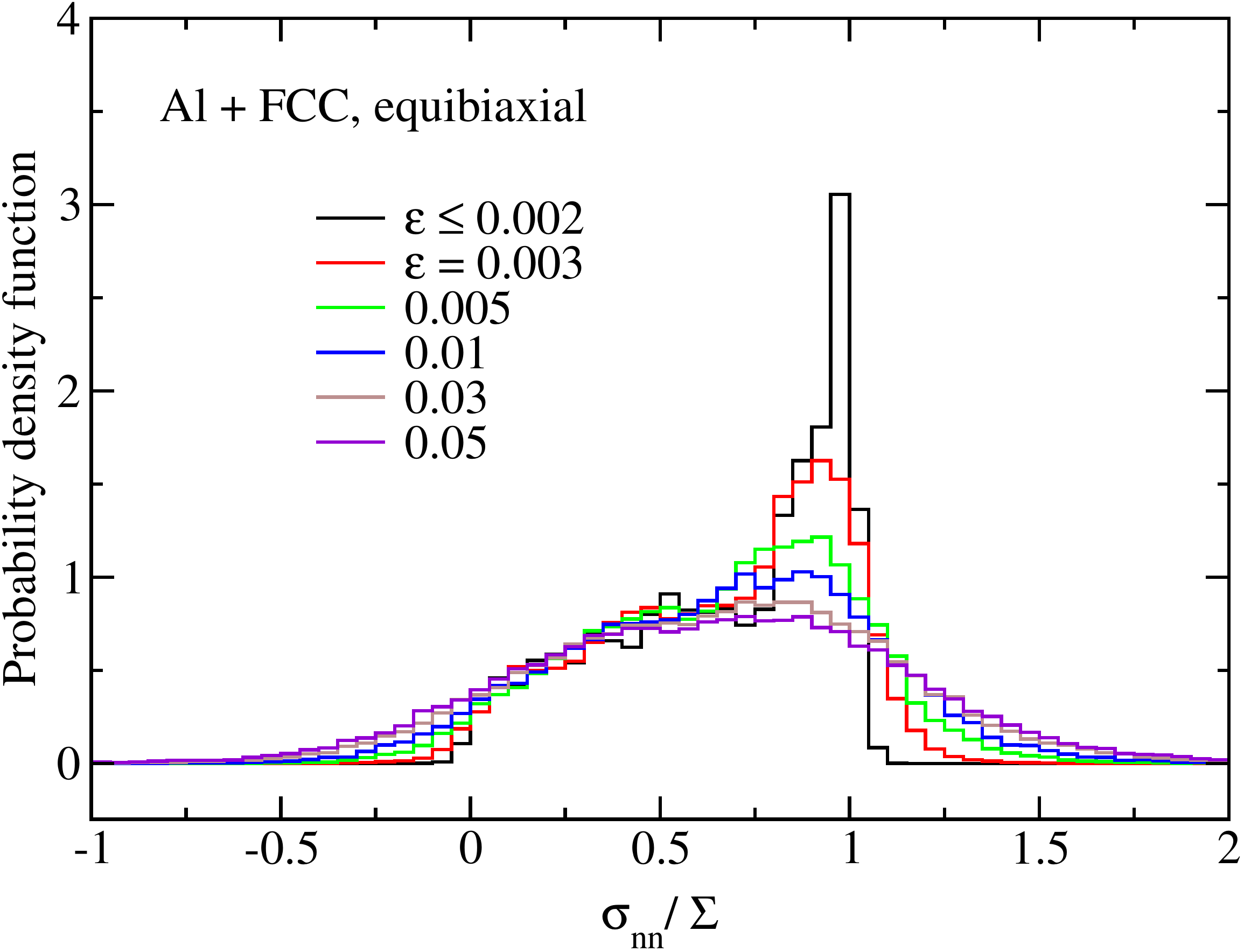}}
\caption{
Probability density functions of normalized intergranular normal
stress calculated with Voronoi finite element simulations at various
strains assuming crystal elasticity and ideal ($H=0$) plasticity for
Al and FCC under (a) uniaxial and (b) equibiaxial loading conditions.}
\label{fig:5}
\end{figure}

{Figure \ref{fig:5} shows the evolution of \textit{pdf} with strain
for two loadings (tensile and equibiaxial) considering Al elasticity
and FCC slip systems (and no hardening). For total strain below
yield strain, distributions are the same as the ones presented in
Sec.~\ref{crystalelasticity}. For larger strains, a significant
broadening of \textit{pdf} is observed, with a shape change from
iso-like (Al has small elastic anisotropy) through bi-modal (moderate
strain) towards normal-like shape (large strain). The evolution of pdf
shape with strain is very similar to the one observed in
Fig. \ref{fig:2} for different materials. Therefore, broadening of pdf
with strain, {also observed for other materials (with different
crystal elasticity parameters and slip systems) as shown in the
following sections,} is assumed to be the result of the increase of
(plastic) anisotropy of the grains. {Additional simulations (not shown
here) performed in the small strain limit
(to avoid any texture evolution)
or for different viscosity parameters used in the constitutive
equations lead to similar results, thus confirming that the broadening
is only due the local grain-grain interactions.} {A similar broadening
  of intergranular stress distributions with strain was reported
in previous studies, \textit{e.g.}, \citep{diard2002,gonzalez2014}.}
}

{For all simulations, independently of crystal elasticity parameters
and slip systems considered, the mean values of normalized
intergranular normal stresses remain approximately constant%
\footnote{
With larger strain the grains become elongated and may develop small
texture. Both effects may change the mean value of \textit{pdf}.}
at $\sim$1/3 and $\sim$2/3 for uniaxial and equibiaxial loading
conditions, respectively. As in Sec.~\ref{crystalelasticity}, the
broadening of distributions is quantified through the evolution of
standard deviations in the following sections.}\\

%-------------------------------------------------------
\subsubsection{Without hardening}
\label{nohardening}

\begin{figure}[H]
\centering
\subfigure[]{\includegraphics[height = 5.cm]{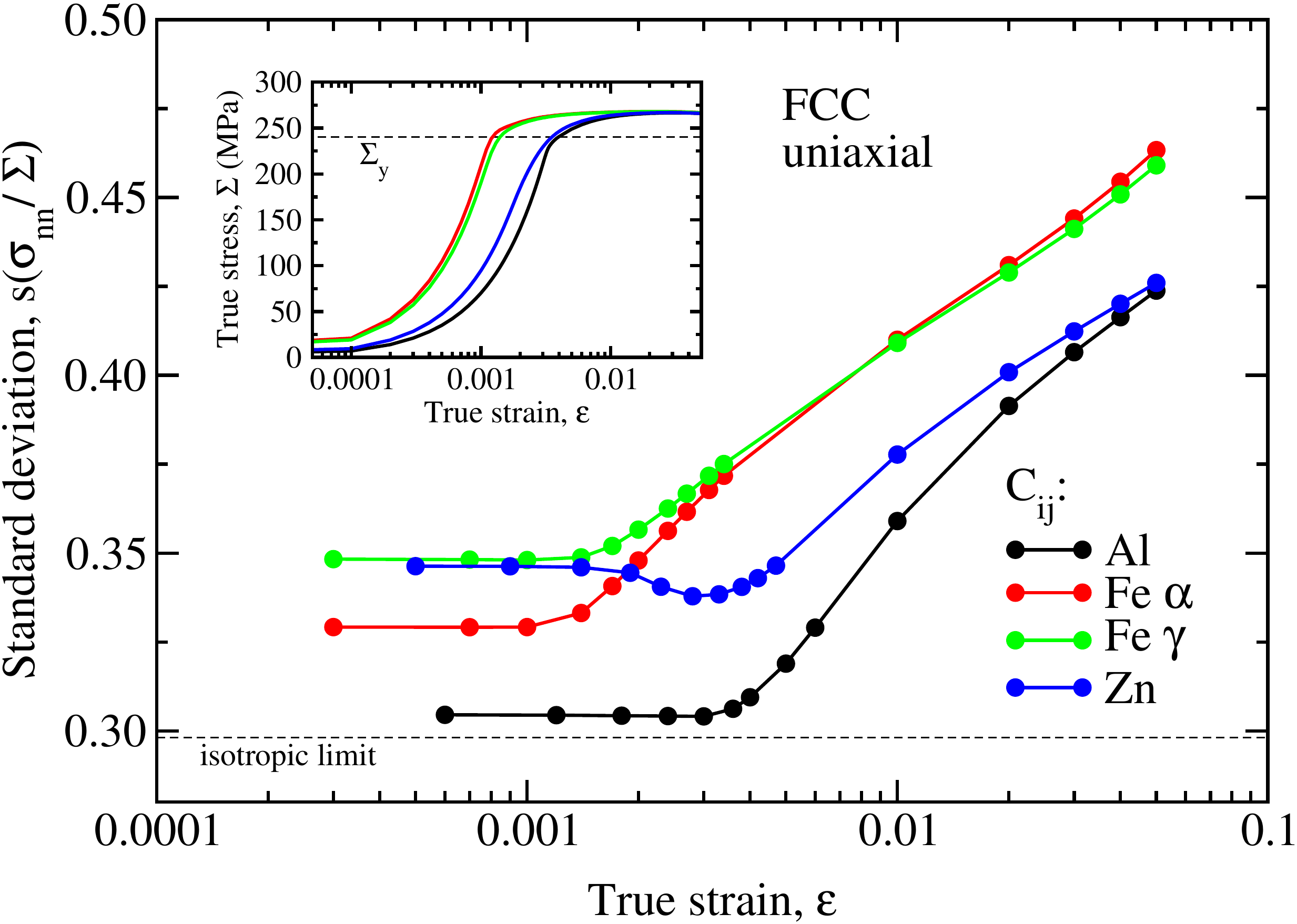}}
\hspace{0.5cm}
\subfigure[]{\includegraphics[height = 5.cm]{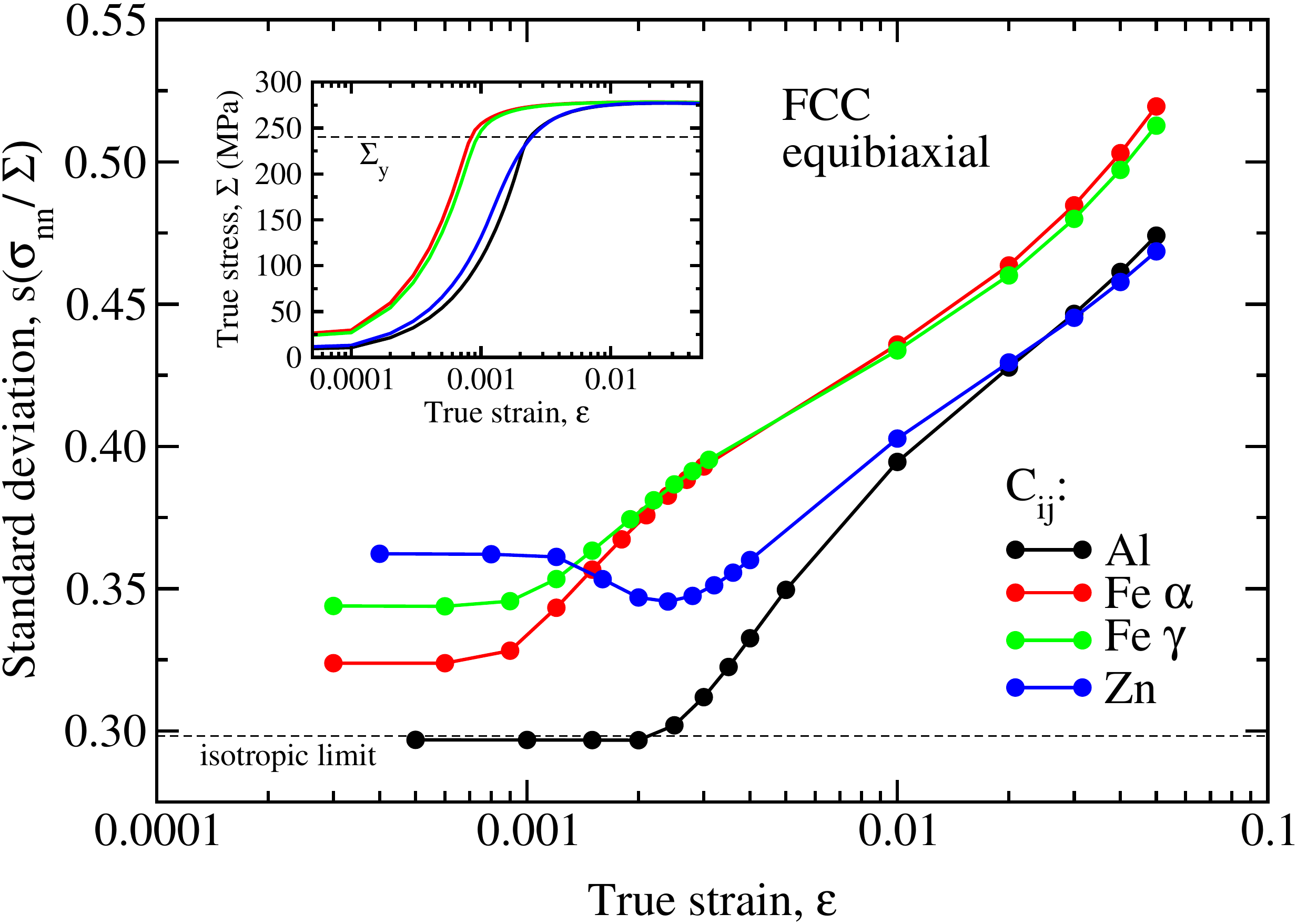}}
\caption{
Standard deviation of normalized intergranular normal stress (inset:
macroscopic tensile stress) as a function of applied strain calculated
with Voronoi finite element simulations assuming crystal elasticity
and ideal ($H=0$) plasticity for various materials and FCC under (a)
uniaxial and (b) equibiaxial loading conditions.}
\label{fig:6}
\end{figure}

The evolution of standard deviation of normalized intergranular stress
as a function of elastic properties is shown in Fig.~\ref{fig:6},
considering FCC slips systems with no hardening. For low applied total
strain, constant values are obtained in quantitative agreement with
purely elastic simulations presented in Sec.~\ref{crystalelasticity},
for both uniaxial and equibiaxial loading conditions. For large
applied strain, standard deviations are found to increase with
strain. For all cases considered, the value of strain at which
$s(\sigma_{nn}/\Sigma)$ starts to deviate from elastic value
corresponds to the onset of plasticity (macroscopic yield strain
$\epsilon_y$) observed in tensile curves (inset Fig.~\ref{fig:6}), in
agreement with what was reported in \citep{gonzalez2014}. This
defines an elastic regime where only elastic properties influence
intergranular normal stress distributions and an elasto-plastic regime
where both elastic properties and slip systems dictate the evolution
of $s(\sigma_{nn}/\Sigma)$.

\begin{figure}[H]
\centering
\subfigure[]{\includegraphics[height = 5.5cm]{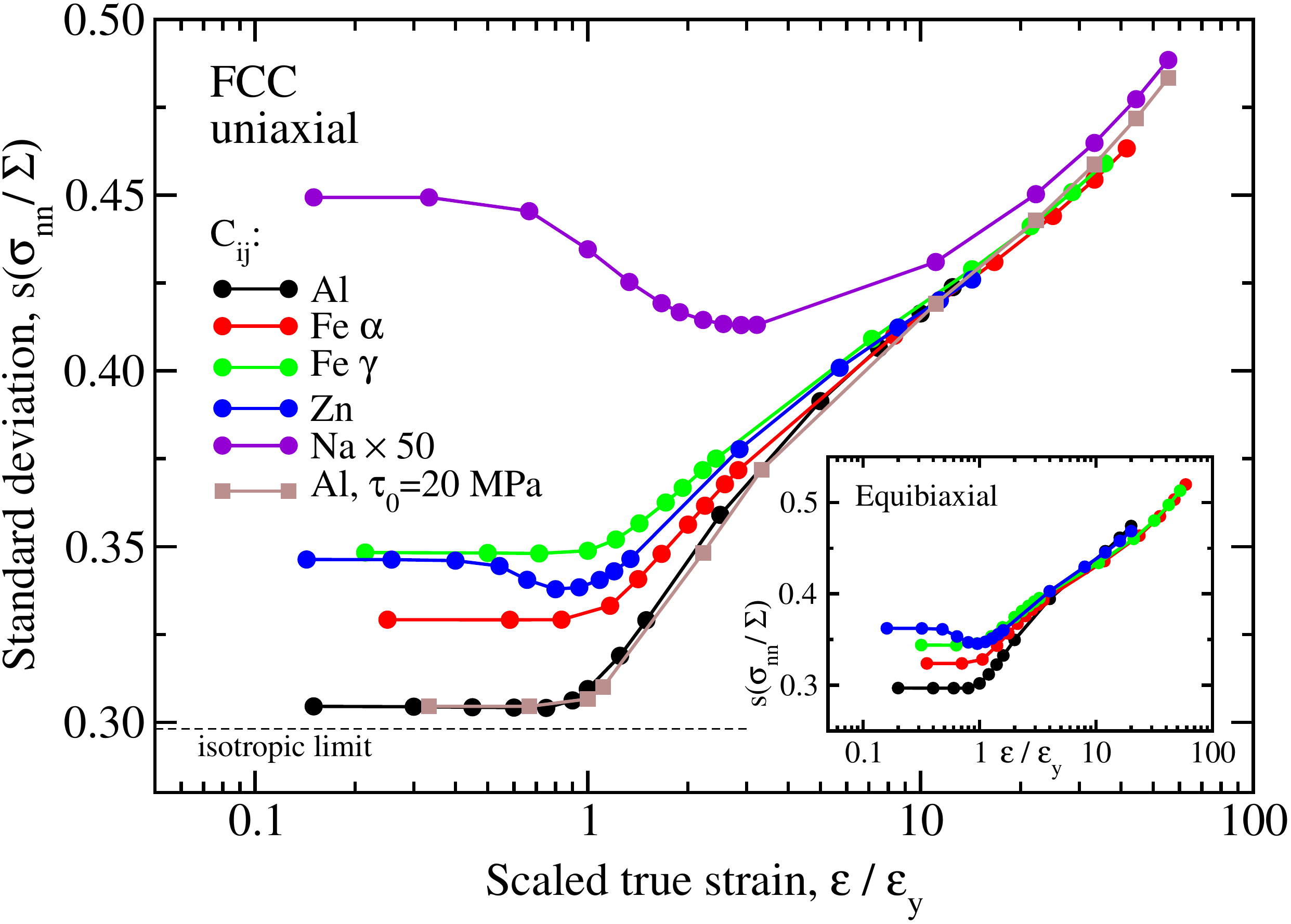}}
\caption{
Standard deviation of normalized intergranular normal stress as a
function of applied strain rescaled by yield strain calculated with
Voronoi finite element simulations assuming crystal elasticity and
ideal ($H=0$) plasticity for various materials and FCC under uniaxial
(inset: equibiaxial) loading conditions.
Note that elastic constants of Na were multiplied by 50 to reduce
$\epsilon_y$ (but keeping the same elastic anisotropy). All
simulations were performed with $\tau_0=100$ MPa except the last one
for Al and $\tau_0=20$ MPa.}
\label{fig:6c}
\end{figure}

As yield strain $\epsilon_y$ is a result of both elastic and plastic
properties, therefore being different for simulations shown in
Fig.~\ref{fig:6}, a natural rescaling is to consider
$\epsilon\to\epsilon/\epsilon_y$. With this rescaling, a universal
curve is observed in the plastic regime, as shown in
Fig.~\ref{fig:6c}, {where both elastic properties and initial critical
resolved shear stress have been changed}, for both uniaxial and
equibiaxial loading conditions. Thus, the evolution of
$s(\sigma_{nn}/\Sigma)$ with rescaled strain $\epsilon/\epsilon_y$ is
found to be {practically} independent of elastic properties in plastic
regime ($\epsilon \gg \epsilon_y$), {in accordance with the results
presented in \citep{gonzalez2014} for specific conditions.}

%-------------------------------------------------------

% Justification of e/ey rescaling

{The universal behavior observed in Fig.~\ref{fig:6c} for
$s(\sigma_{nn}/\Sigma)$ as a function of rescaled strain
$\epsilon/\epsilon_y$ can be understood using the following
reasoning. Let us consider two materials $I$ and $II$ such that
\be
  C_{ijkl}^{II} = k_C C_{ijkl}^{I},\quad \tau_0^{II} = k_{\tau} \tau_0^I 
  \label{materialsIII}
\ee
where $k_C$, $k_{\tau}$ are constant factors, and with both
materials having the same set of slip systems defined through Schmid
tensor $\mu_{kl}$. For arbitrary location ${\bm r}$ and time $t$, the
stress and strain components obey Eqs.~(\ref{eq_hooke}), (\ref{eq_1})
and (\ref{eq_2}) combined into (omitting ${\bm r}$ and $t$ hereafter
for clarity)
\be
  \dot{\sigma}_{ij} = C_{ijkl} \left( \dot{\epsilon}_{kl} - \sum_\al
  \left\langle\frac{|\sigma_{pq}\mu_{pq}^\al| - \tau_0}{K_0}\right\rangle^n {\rm  sign}(\sigma_{pq}\mu_{pq}^\al)\ \mu_{kl}^\al \right)
  \label{sigmadot0}
\ee
where same definitions for elastic ($C_{ijkl}$ for $i,j,k,l=1\ldots
3$) and plastic ($\tau_0$, $K_0$, $n$) material parameters are used as
in Sec. \ref{sec_mat}. At a given time $t$, assuming the following
rescaling for the stress and strain components and viscosity
parameters,
\be
  \sigma_{ij}^{II} = k_{\tau} \sigma_{ij}^{I},\quad 
  \epsilon_{ij}^{II} = \frac{k_{\tau}}{k_C} \epsilon_{ij}^{I},\quad
  K_{0}^{II} = k_{\tau} \left( \frac{k_{\tau}}{k_C}\right)^{-1/n} K_{0}^{I},\quad
  n^{II} = n^{I}, \label{rescale} 
\ee 
along with Eq.~(\ref{sigmadot0}) leads to $\dot{\sigma}_{ij}^{II} =
k_{\tau} \dot{\sigma}_{ij}^{I}$, indicating that Eqs.~(\ref{rescale})
hold in fact at any time. Equation~(\ref{rescale}$_2$) can be
rewritten as $\epsilon_{ij}^{II}/\epsilon_y^{II} =
\epsilon_{ij}^{I}/\epsilon_y^{I}$, suggesting to write the standard
deviation of normalized intergranular normal stress as a function of
applied strain rescaled by yield strain for materials satisfying
Eqs.~(\ref{materialsIII}). The condition set by
Eq.~(\ref{rescale}$_3$) is not met in the simulations where $K_0$ is
kept constant. However, this parameter has negligible influence on the
results in the rate-independent limit $K_0 \dot{\epsilon}_0^{1/n} \ll
\tau_0$. Moreover, despite the fact that the argument given here
suggests the rescaling applies only for a specific class of materials
(set by Eqs.~(\ref{materialsIII})), Fig.~\ref{fig:6c} shows that it can
be applied in a broader context that remains to be precisely defined
theoretically. \\

For small values of applied strain, $\epsilon \ll \epsilon_y$,
Eq.~(\ref{sigmadot0}) reduces to Hooke's law as
$|\sigma_{pq}\mu_{pq}^\al| \leq \tau_0$. In this regime, the
standard deviation of normalized intergranular stress is constant and
in quantitative agreement with the one described in
Sec.~\ref{crystalelasticity}. For large values of applied strain,
$\epsilon \gg \epsilon_y$, all grains are assumed to deform
plastically and Eq.~(\ref{sigmadot0}) reduces to
\be
   \dot{\epsilon}_{kl}{\approx} \sum_\al
  \left\langle\frac{|\sigma_{pq}\mu_{pq}^\al| - \tau_0}{K_0}\right\rangle^n { {\rm  sign}(\sigma_{pq}\mu_{pq}^\al)}\ \mu_{kl}^\al.
  \label{sigmadot1}
\ee
{Continuity requirement of normal stress and strain components at a
grain boundary associated with Eq.~(\ref{sigmadot1}) is believed to
lead to the increase of intergranular normal stress with strain. Note
that Eq.~(\ref{sigmadot1}) does not involve explicitly the elastic
properties which appears to be consistent with the numerical
observation that the evolution of $s(\sigma_{nn}/\Sigma)$ as a
function of $\epsilon/\epsilon_y$ does not depend on elastic
properties.

}

The transition between both
regimes is observed to depend on the elastic parameters, {as already
reported in
\cite{lebensohn2012},} and can be qualitatively understood considering
the strain-dependent anisotropy of grains. At the onset of
macroscopic plasticity, $\epsilon\sim \epsilon_y$, an increase 
% revision: Samir
%or decrease 
of $s(\sigma_{nn}/\Sigma)$ can be explained through the
gradual activation of slip systems (where regions with more
activated slip systems provide smaller plastic anisotropy), thus to
plastic slip that can only occur along few particular slip directions
(keeping the deformation in all other directions elastic). }
%
% revision: Samir
{\red On the contrary, a decrease of $s(\sigma_{nn}/\Sigma)$ observed
  in Fig.~\ref{fig:6c} for Na and Zn can be understood to appear due
  to specific alignment of stiffness and Schmid tensors which enables
  slip activation along the hardest grain direction, thus reducing
  maximum stress and therefore stress fluctuations.}

%-------------------------------------------------------
%\subsubsection{Slip systems effects}

\begin{figure}[H]
\centering
\includegraphics[height = 5.5cm]{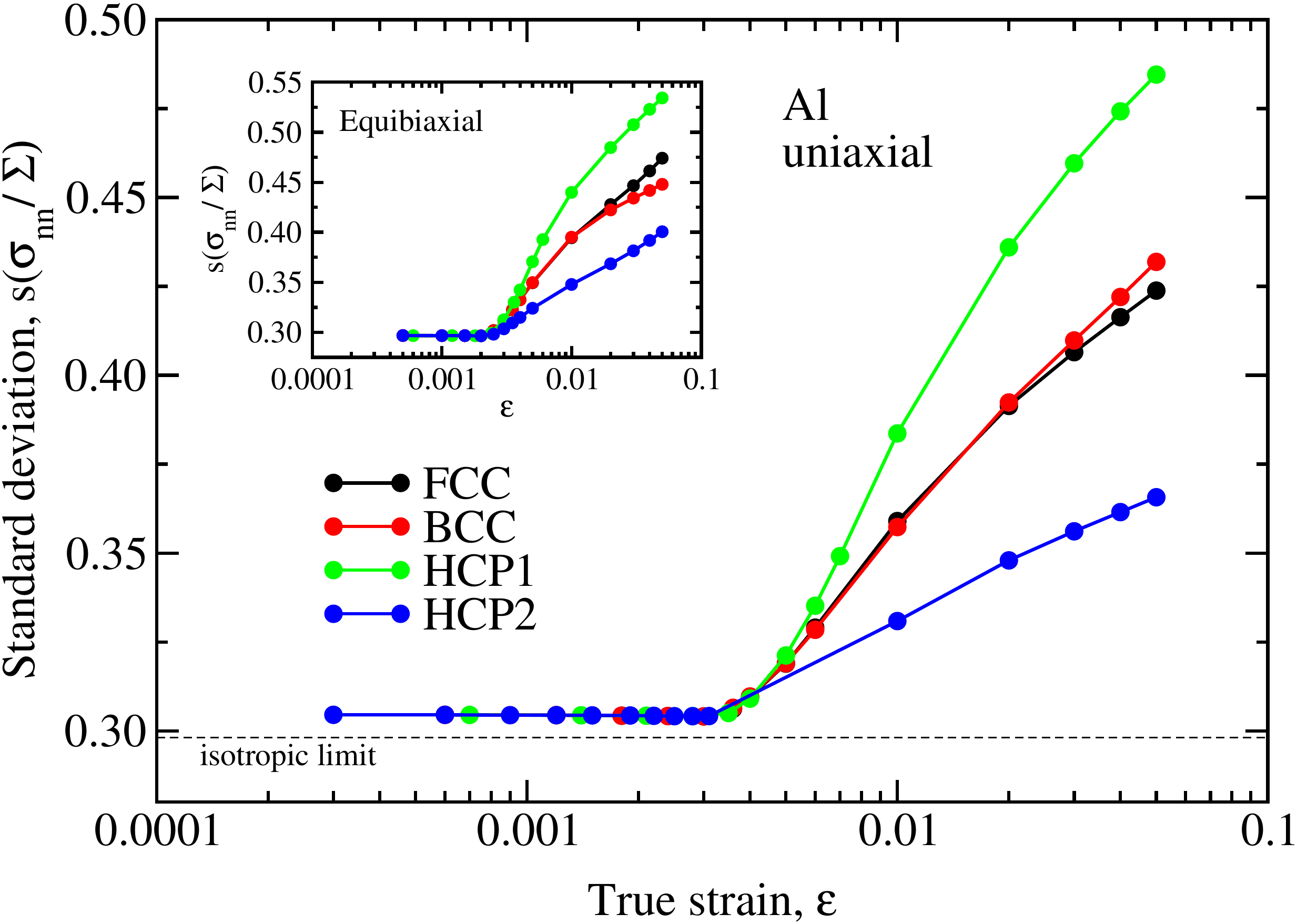}
\caption{
Standard deviation of normalized intergranular normal stress
calculated with Voronoi finite element simulations as a function of
applied strain assuming crystal elasticity and ideal ($H=0$)
plasticity for Al and various slip systems under uniaxial (inset:
equibiaxial) loading conditions.}
\label{fig:8}
\end{figure}

For a given set of elastic properties {(Al from
Tab. \ref{tab1})}, the evolution of intergranular normal stress
standard deviation with strain is assessed in Fig.~\ref{fig:8} for
different sets of slips systems. In the plastic regime, a log-like
behavior is observed for all simulations. However, the typical slope
{of the line (using lin-log type of plotting)} depends on the set of
slip systems considered: the highest slope is found for HCP1 and the
lowest for HCP2, while FCC and BCC lead to similar evolutions. Same
qualitative behavior is observed for uniaxial and equibiaxial loading
conditions.

\subsubsection{With hardening}

\begin{figure}[H]
\centering
\subfigure[]{\includegraphics[height = 5.cm]{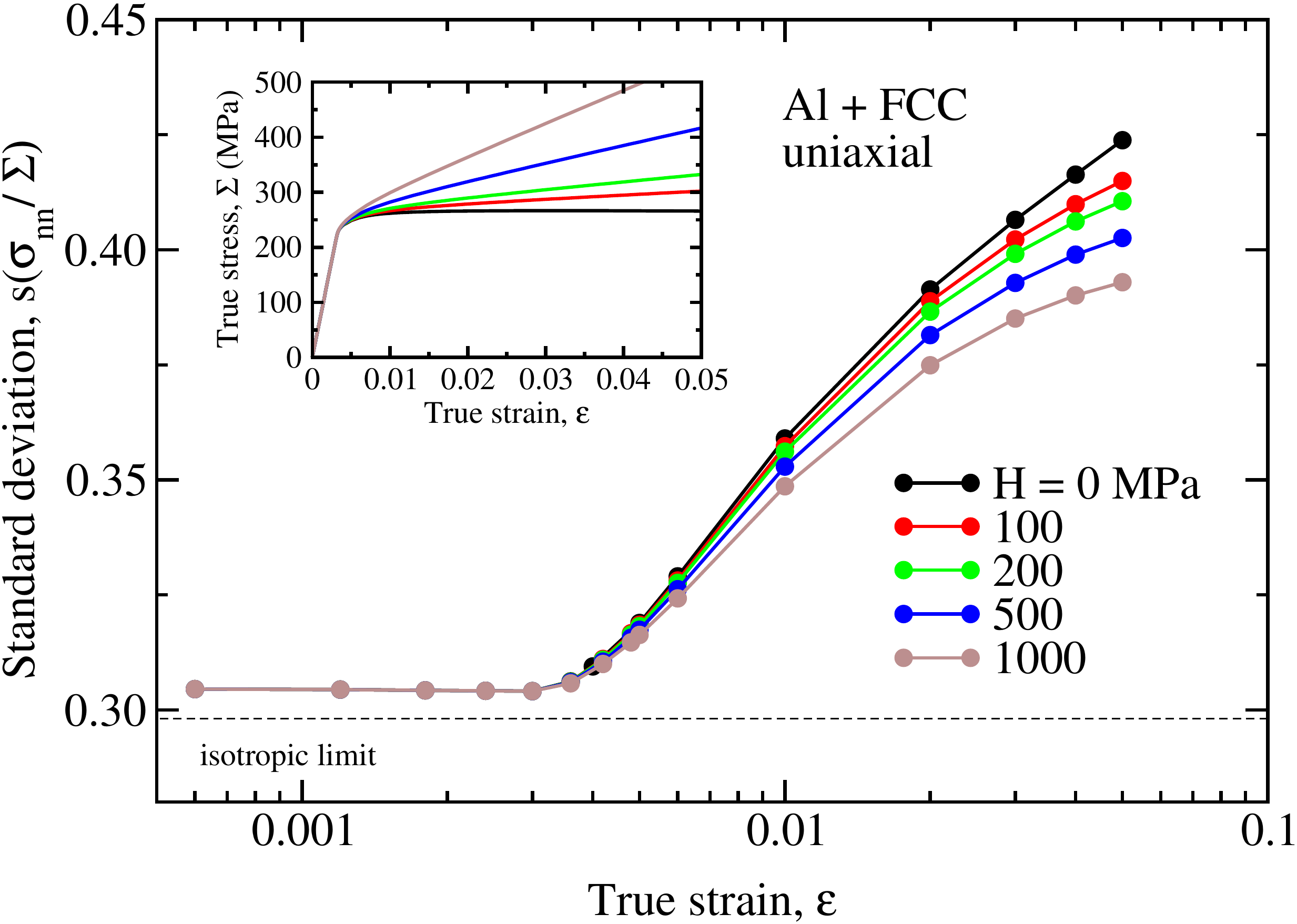}}
\hspace{0.5cm}
\subfigure[]{\includegraphics[height = 5.cm]{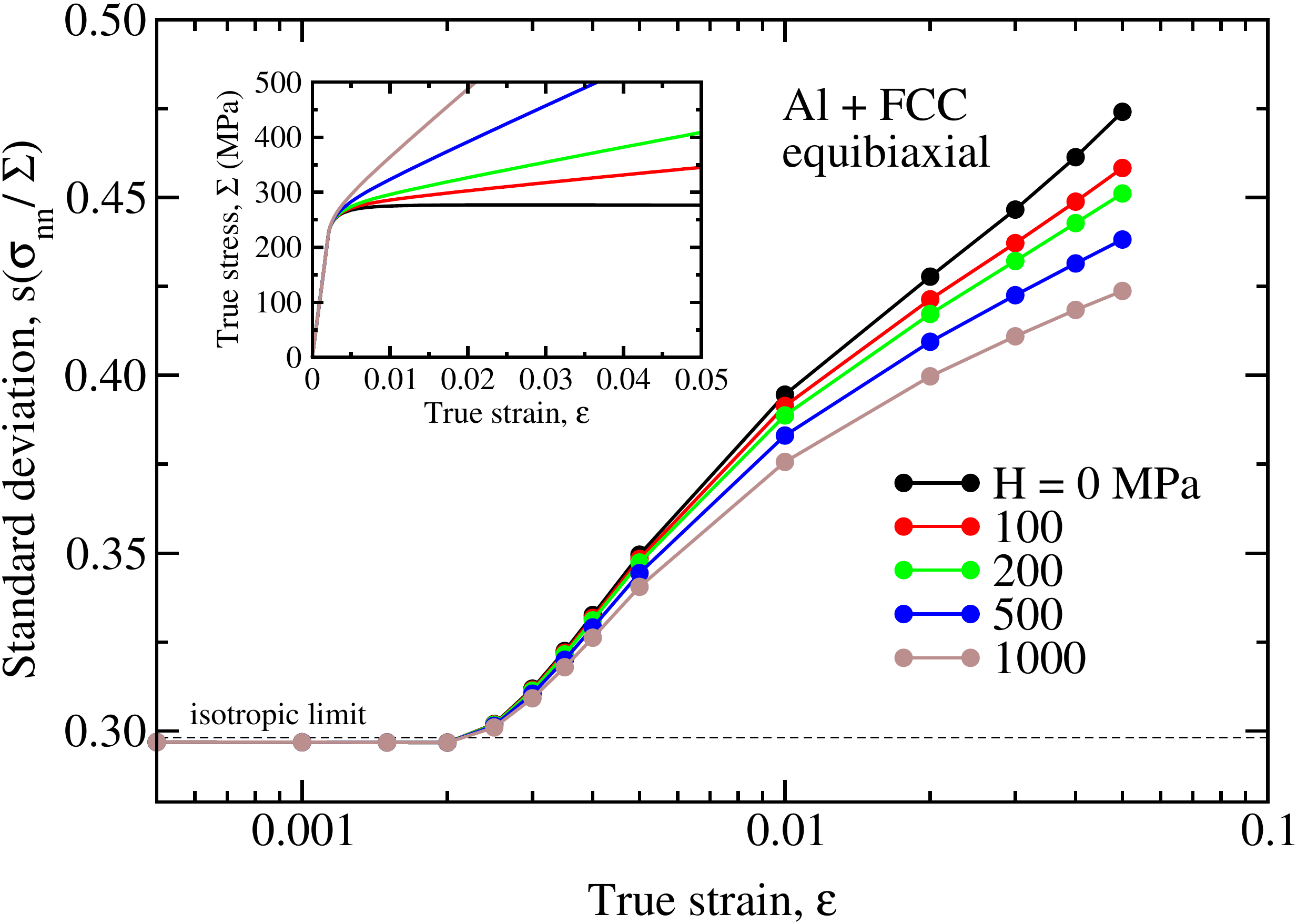}}
\caption{
Standard deviation of normalized intergranular normal stress (inset:
macroscopic tensile stress) as a function of applied strain calculated
with Voronoi finite element simulations assuming crystal elasticity
and plasticity for Al and FCC with various hardening strengths under
(a) uniaxial and (b) equibiaxial loading conditions.}
\label{fig:7}
\end{figure}

Results presented in Sec.~\ref{nohardening} regarding the effects of
elastic and plastic properties on the evolution of standard deviations
of intergranular normal stress distributions correspond to the case of
zero hardening ($H=0$ in Eq.~\ref{eq_2}). For a given set of elastic
property and slip system, the effect of (Taylor-)hardening on
$s\left(\sigma_{nn}/\Sigma\right)$ in plastic regime is evaluated in
Fig.~\ref{fig:7} for both uniaxial and equibiaxial loading
conditions. For a given strain value, the larger the hardening
modulus, the lower the standard deviation
$s\left(\sigma_{nn}/\Sigma\right)$. The same behavior is observed for
other combinations of elastic properties and slip systems {(not
shown)}.

\begin{figure}[H]
\centering
\subfigure[]{\includegraphics[height = 5.5cm]{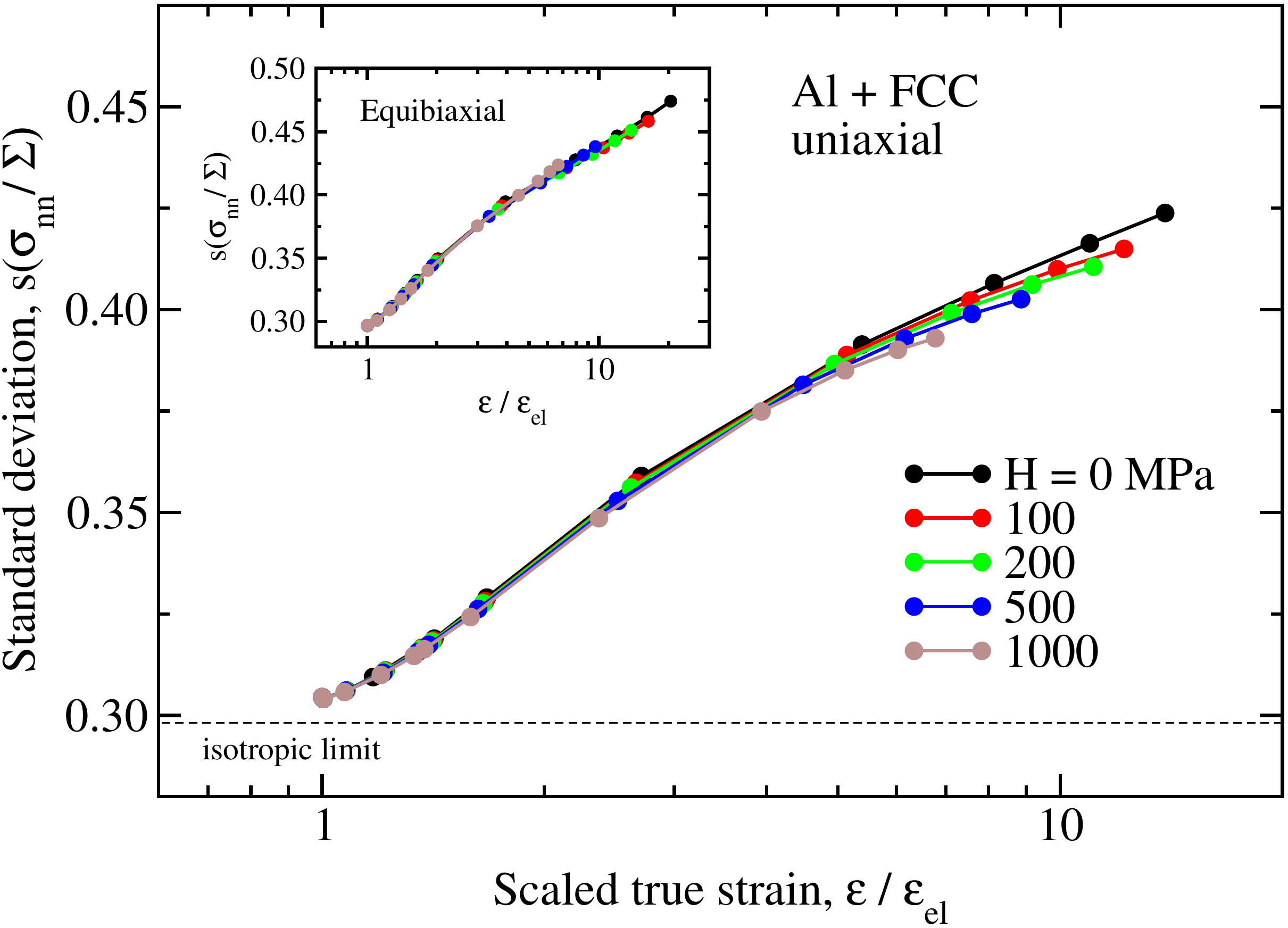}}
\caption{
Standard deviation of normalized intergranular normal stress as a
function of applied strain rescaled by elastic strain calculated with
Voronoi finite element simulations assuming crystal elasticity and
plasticity for Al and FCC with various hardening strengths under
uniaxial (inset: equibiaxial) loading conditions.}
\label{fig:7c}
\end{figure}

Hardening induces the evolution of critical resolved shear stress with
macroscopic strain, $\tau_c^{\alpha}(\epsilon) = \tau_0 +
H\Gamma(\epsilon)$, according to Eq.~(\ref{eq_2}), {which results} in
the evolution of macroscopic stress $\Sigma(\epsilon)$. For a given
macroscopic strain $\epsilon$, an identical value of macroscopic
stress can be {achieved} by considering a material with no hardening
($H=0$), but with {appropriately larger} constant critical resolved
shear stress.
{Since in this way both materials would share several macroscopic
quantities at a given strain (\textit{e.g.}, macroscopic stress
$\Sigma(\epsilon)$, macroscopic elastic strain
$\epsilon_{el}(\epsilon)$ and macroscopic plastic strain
$\epsilon_{pl}(\epsilon)$), we may assume that, at same strain and
restricting to monotonic loading, they would become equivalent also
with respect to standard deviation of intergranular normal stress
distribution $s\left(\sigma_{nn}/\Sigma\right)$.}
This assumption leads to a natural extension of the rescaling used in
Sec.~\ref{nohardening}, which is to consider
$\epsilon\to\epsilon/\epsilon_{el}$. %where $\epsilon_{el}$ corresponds
%to the macroscopic elastic strain. 
For $H=0$, $\epsilon\to\epsilon/\epsilon_{el}$ would give the same
result as shown in Fig. \ref{fig:6c} for other materials. Such a
rescaling allows to get a universal curve in the plastic regime as
shown in Fig.~\ref{fig:7c}.

As an intermediate conclusion, simulations performed on Voronoi
aggregates with different elastic properties, sets of slip systems and
(Taylor-)hardening have led to the following observations regarding
standard deviation of normalized intergranular normal stress
distributions $s\left(\sigma_{nn}/\Sigma\right)$:

\begin{itemize}
\item 
For strain values below macroscopic yield strain $\epsilon <
\epsilon_y$, $s\left(\sigma_{nn}/\Sigma\right)$ is constant and
depends solely on elastic properties (as described in
Sec.~\ref{crystalelasticity}).

\item 
For strain values {well} above macroscopic yield strain $\epsilon
\gg \epsilon_y$, $s\left(\sigma_{nn}/\Sigma\right)$ was found to follow a
{universal behavior} when plotted as a function of rescaled strain
$\epsilon/\epsilon_{el}$, where $\epsilon_{el}$ is the macroscopic
elastic strain.

\end{itemize}

The function $\mathcal{F}$ that relates the standard deviation of
intergranular normal stress to the normalized strain,
$s\left(\sigma_{nn}/\Sigma\right)=\mathcal{F}(\epsilon/\epsilon_{el})$,
depends on the set of slip systems considered, as shown in
Fig.~\ref{fig:8}. The aim of the next section is to provide a simple
model for predicting such behavior.

%-------------------------------------------------------
\subsubsection{Combined effects -- universal behavior}

\begin{figure}[H]
\centering
\subfigure[]{\includegraphics[height = 5.cm]{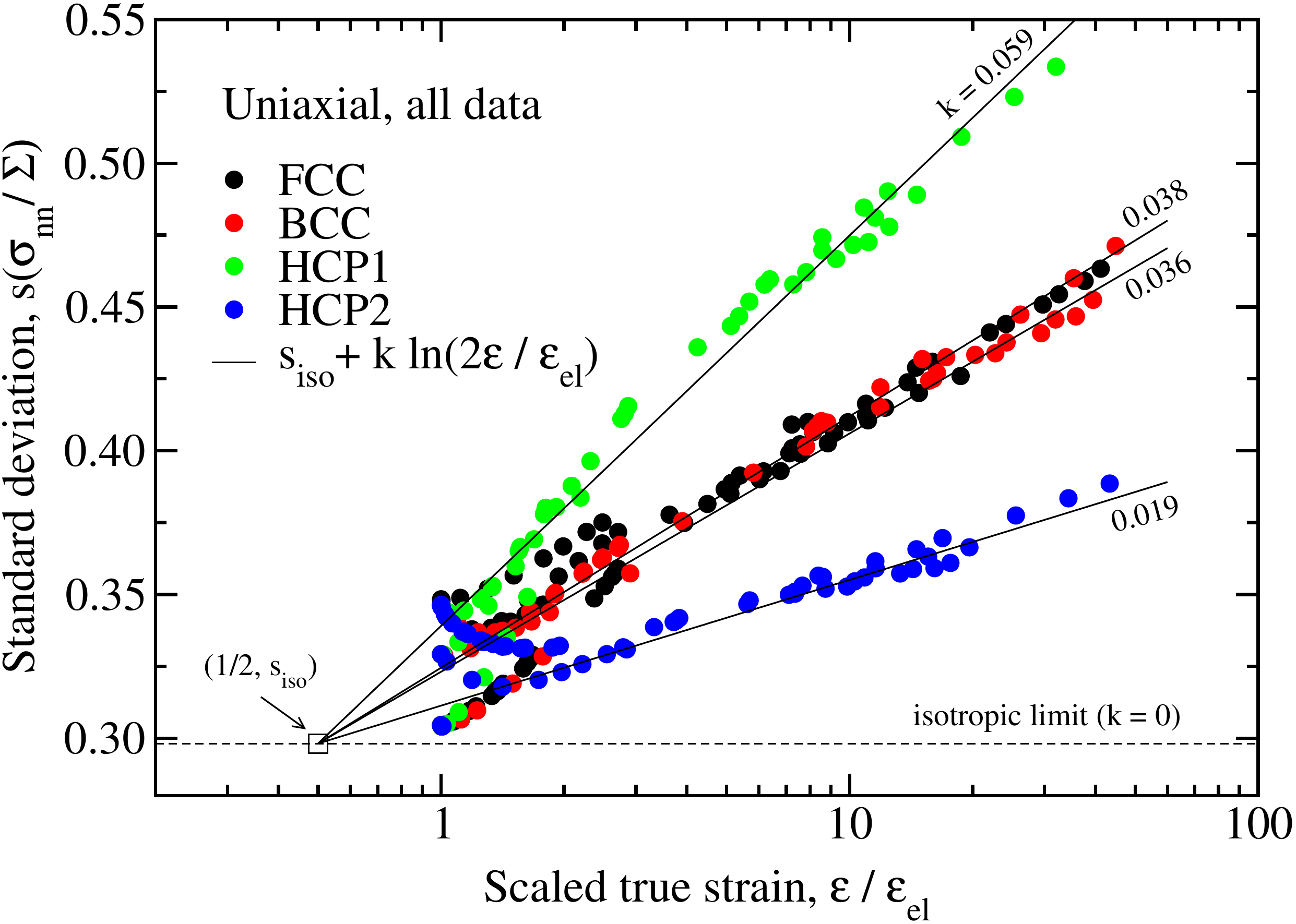}}
\hspace{0.5cm}
\subfigure[]{\includegraphics[height = 5.cm]{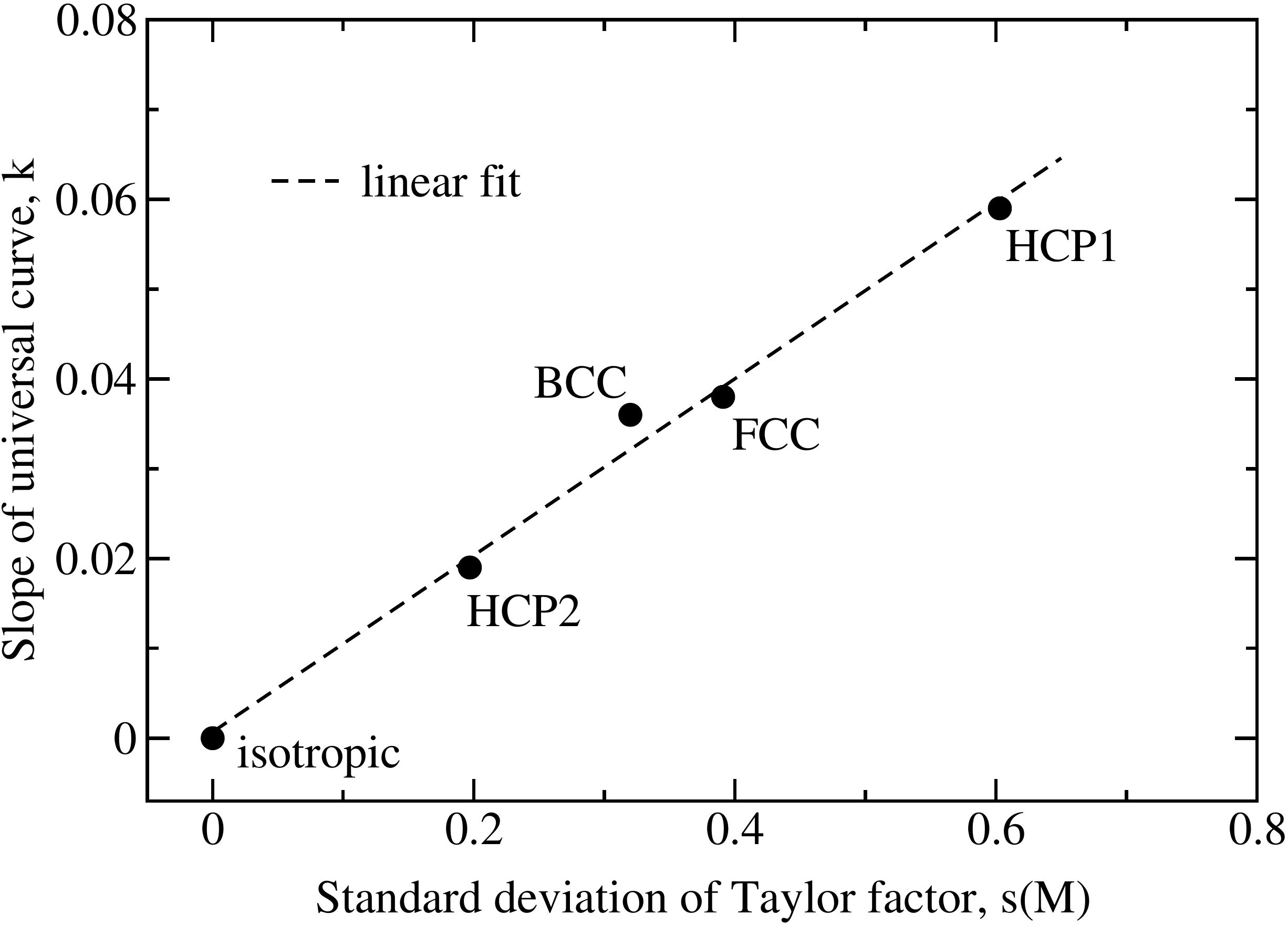}}
\caption{
(a) Standard deviation of normalized intergranular normal stress
calculated with Voronoi finite element simulations as a function of
applied strain rescaled by elastic strain assuming crystal elasticity
and plasticity for various materials {(from Tab. \ref{tab1})} and
slip systems under uniaxial loadingconditions. Solid lines represent
linear fits (in terms of $\ln(\epsilon/\epsilon_{el})$) going through
a common point denoted by a square.
(b) Estimated slope of the log-like behavior, $k$ in Eq.~\ref{eqk}, versus
standard deviation of Taylor factor, $s(M)$, under uniaxial loading
conditions. $k$ is taken from the fits in (a) and $s(M)$ is calculated
for a 1000-grain Lin model (\textit{i.e.}, polycrystal with assumed
uniform total strain) assuming crystal elasticity and ideal
plasticity.}
\label{fig:8c}
\end{figure}

Figure~\ref{fig:8c}a summarizes the evolution of
$s\left(\sigma_{nn}/\Sigma\right)$ in plastic regime with rescaled
strain $\epsilon/\epsilon_{el}$ in uniaxial loading conditions, for
different elastic parameters, sets of slip systems and hardening
modulus. As discussed previously, this evolution appears to depend
practically only on the chosen slip system (independently of elastic
properties and hardening strength) \textcolor{black}{under the assumption of uniform critical resolved shear stress}. In the plastic regime
(\textit{i.e.}, {for $3 \lesssim \epsilon/\epsilon_{el} \lesssim 10$
}), the evolution of $s\left(\sigma_{nn}/\Sigma\right)$ can be {well
approximated} by the following log-like equation
\begin{equation}
   s(\sigma_{nn}/\Sigma) \approx s_{iso} +  k \ln (2\epsilon/\epsilon_{el})
\label{eqk}
\end{equation}
{with $k$ parameter defining the {strain-independent part} of
plastic anisotropy}, and where $s_{iso}=2/3\sqrt{5}$ is the standard
deviation of $s(\sigma_{nn}/\Sigma)$ in an {aggregate with isotropic
grains} ($k=0$), {corresponding to the case of an infinite number of
slip systems (\textcolor{black}{Tresca plasticity}) or von Mises plasticity. In the latter case, it has
been checked that $s(\sigma_{nn}/\Sigma) = s_{iso}$.}

Moreover, Fig. \ref{fig:8c}(b) shows a very good linear correlation
between $k$ and standard deviation of Taylor factor, $s(M)$,
{(calculated in \ref{app4})}
\be
   s(\sigma_{nn}/\Sigma) \approx s_{iso} + 0.10 s(M) \ln (2\epsilon/\epsilon_{el})
\label{eq:uni}
\ee
for uniaxial loading conditions.
{A very similar behavior has been observed for equibiaxial loading
conditions (not shown), however with a prefactor $0.12 s(M)$
identified in Eq.~(\ref{eq:uni}), where $s(M)$ is the same as for
uniaxial loading conditions.}
The proportionality of standard deviation of
intergranular normal stress distribution can be understood as follows:
using Taylor's homogeneous strain assumption in
the aggregate (see \ref{app4}), the local {(von Mises)} stress
can be written as $\sigma \sim M \tau$ where $M$ is Taylor factor and
$\tau$ the critical resolved shear stress. Similar relation holds also
at the macroscopic scale: $\Sigma \sim \langle M\rangle \tau$ where
$\langle M\rangle$ denotes average Taylor factor. Standard deviation
of $\sigma_{nn}/\Sigma$ therefore scales as standard deviation of
Taylor factor, which is in accordance with results of the simulations.

Combining the results form Secs.~\ref{crystalelasticity}
and~\ref{crystalplasticity} leads to the following estimation of the
standard deviation of intergranular normal stress distribution in
untextured aggregates {under uniaxial loading conditions}
\be
  s(\sigma_{nn}/\Sigma) \approx \left\{
  \begin{array}{ll}
     \displaystyle
     \frac{4}{3\sqrt{5}}\frac{A^u + 6}{A^u + 12} &;\epsilon \leq \epsilon_y\\
     \displaystyle
     \frac{2}{3\sqrt{5}} + 0.10 s(M) \ln \left(\frac{2\epsilon}{\epsilon_{el}}\right) &;\mathrm{otherwise}
  \end{array}\right.
  \label{eqfinal}
\ee
where the first equation corresponds to the elastic anisotropy described
through the universal anisotropy index $A^u$ proposed by
\citep{ranganathan} and the second equation relates to the plastic
anisotropy described through Taylor factor $M$ {and macroscopic ratio
$\epsilon/\epsilon_{el}$}. {As shown in Figs.~\ref{fig:8c}a and
b},
Eq.~(\ref{eqfinal}) allows to (accurately) estimate the width of the
intergranular normal stress distribution under tensile loading from
well-known macroscopic values: macroscopic stress $\Sigma$,
macroscopic total strain $\epsilon$, macroscopic elastic strain
$\epsilon_{el}$ and standard deviation of Taylor factor $s(M)$ (see
Tab. \ref{tabA1} in \ref{app4}). As $s(\sigma_{nn}/\Sigma)$ is
directly related to the shape of pdf (exactly for normal distribution,
approximately for other kind of distributions), the estimated
$s(\sigma_{nn}/\Sigma)$ can be used to calculate, \textit{e.g.}, the
ratio of grain boundaries in a model with normal stress larger than
some critical stress, {as detailed in the conclusion section.} This
could lead to a powerful tool for quick {and reliable} estimation
of material potential susceptibility to IGSCC initiation without
actually performing finite element simulations.

%---------------------------------------
\section{Discussion}
%---------------------------------------
\label{discussion}

It is the purpose of this section to elaborate further on the observed
universality and to perform sensitivity analyses to establish the
limits of its validation. In most cases presented hereafter only
intergranular normal stresses under uniaxial loading conditions will
be discussed, but it has been checked that similar conclusions can be
drawn also for equibiaxial loading conditions.

%---------------------------------------
\subsection{Sensitivity to the aggregate size}
\label{size_effects}
%---------------------------------------

\begin{figure}[H]
\centering
\includegraphics[height = 5.5cm]{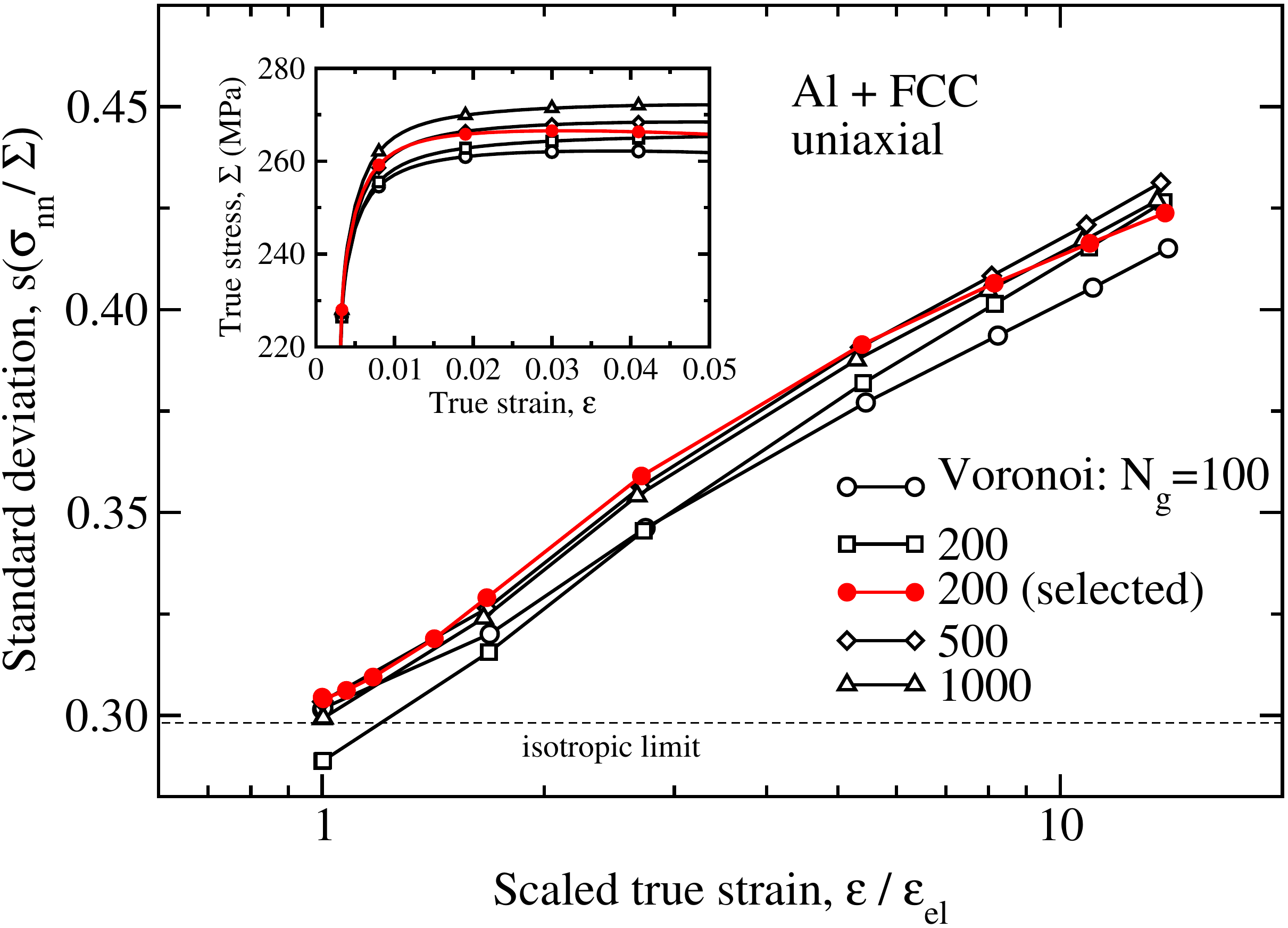}
\caption{
Standard deviation of normalized intergranular normal stress as a
function of applied strain rescaled by elastic strain calculated for
different Voronoi models ($N_g$ denotes the number of grains),
assuming crystal elasticity and ideal plasticity for Al and FCC under
uniaxial loading conditions. Two aggregates with $N_g=200$ share the
same number of grains but different grain topologies and orientations:
the model shown in red is selected to be the representative model in
this study. Inset: tensile response.}
\label{fig:10}
\end{figure}

In Fig. \ref{fig:10} finite size analysis is performed with respect to
different Voronoi aggregate models used in the calculation of standard
deviation $s(\sigma_{nn}/\Sigma)$. Four different aggregate sizes,
measured in terms of the number of grains $N_g$, are studied and
compared using random orientations for the grains and approximately
same mesh density of finite element discretization (\textit{i.e.},
$\sim$1300 elements per grain), see Tab.  \ref{tab5}. For one
aggregate size ($N_g=200$), however, two different grain topologies
and mesh densities are compared in addition.
\begin{table}[H]
\begin{tabular}{c|c|c}
$N_g$ &  $N_e$ &  $N_e/N_g$\\
\hline
\hline
100    &  133k & 1330 \\
200    &  271k & 1355 \\
200    &  396k & 1980 \\
500    &  685k & 1370 \\
1000   &  1355k & 1355\\
\hline
\end{tabular}
\caption{
A list of Voronoi aggregate models used in the finite size analysis.
Aggregates of cubic shape are composed of $N_g$ grains of random
orientations and meshed by $N_e$ tetrahedral quadratic elements (C3D10
in Abaqus). The model with $N_g=200$ and $N_e=396$k (\textit{i.e.},
$\sim$2000 elements per grain) is used as the representative model
throughout this study.}
\label{tab5}
\end{table}

In Fig. \ref{fig:10} a maximum scatter of $\sim$0.02 (or $\sim$6\%) is
observed among the five resulting $s(\sigma_{nn}/\Sigma)$
evolutions. The results obtained for $N_g=500$ and 1000 grains seem to
be already converged on a given scale, while moderate fluctuations can
still be observed for $N_g=100$ and 200 models. However, {the model
with $N_g=200$ grains and $N_e/N_g=1980$ elements per grain} (shown in
red in Fig. \ref{fig:10}) provides a very similar response as the
larger two models. This model has therefore been selected to be the
representative model in this study.

%---------------------------------------
\subsection{Free surface effects: Application to stress-corrosion cracking}
%---------------------------------------

The probability density functions presented so far were obtained by
post-processing entire aggregates, \textit{i.e.}, independently of the
location of the grain boundaries with respect to the distance from the
free surface. As stress corrosion cracking initiates at free surface,
the \textit{pdf}s of intergranular stresses close to free surface are
calculated and presented in this section. A comparison is made in a
fully plastic regime at $\epsilon=0.05$%
\footnote{Note that macroscopic strain of 0.05 is sufficient to
describe SCC initiation in most of the experimental cases.} %
using the entire aggregate model and outer sections of the model where
grain boundaries are located close to a free surface within distance
$r$. The case with Al elasticity and FCC ideal plasticity is
considered in the calculation of \textit{pdf}s.

\begin{figure}[H]
\centering
\subfigure[]{\includegraphics[height = 5.cm]{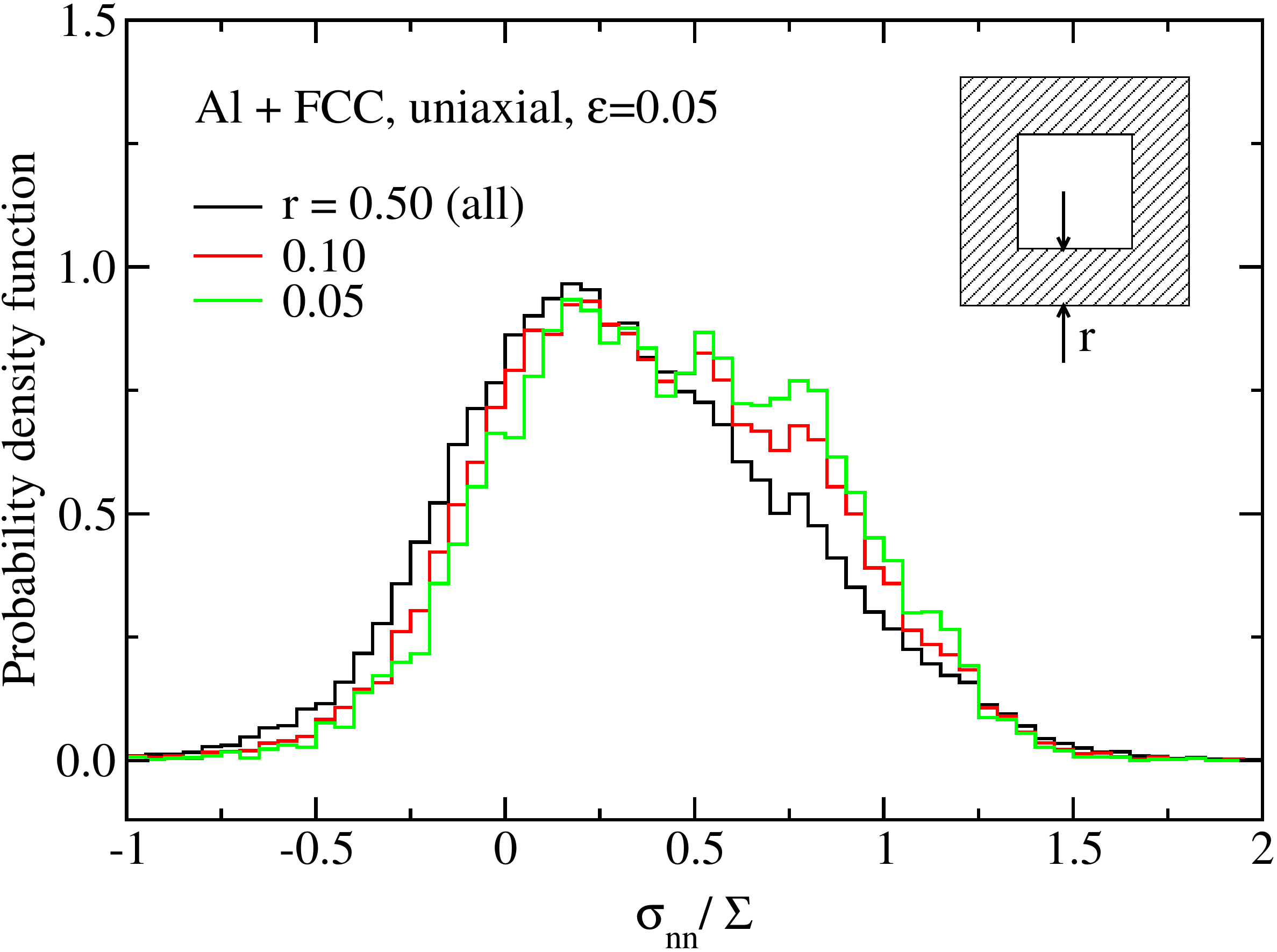}}
\hspace{0.5cm}
\subfigure[]{\includegraphics[height = 5.cm]{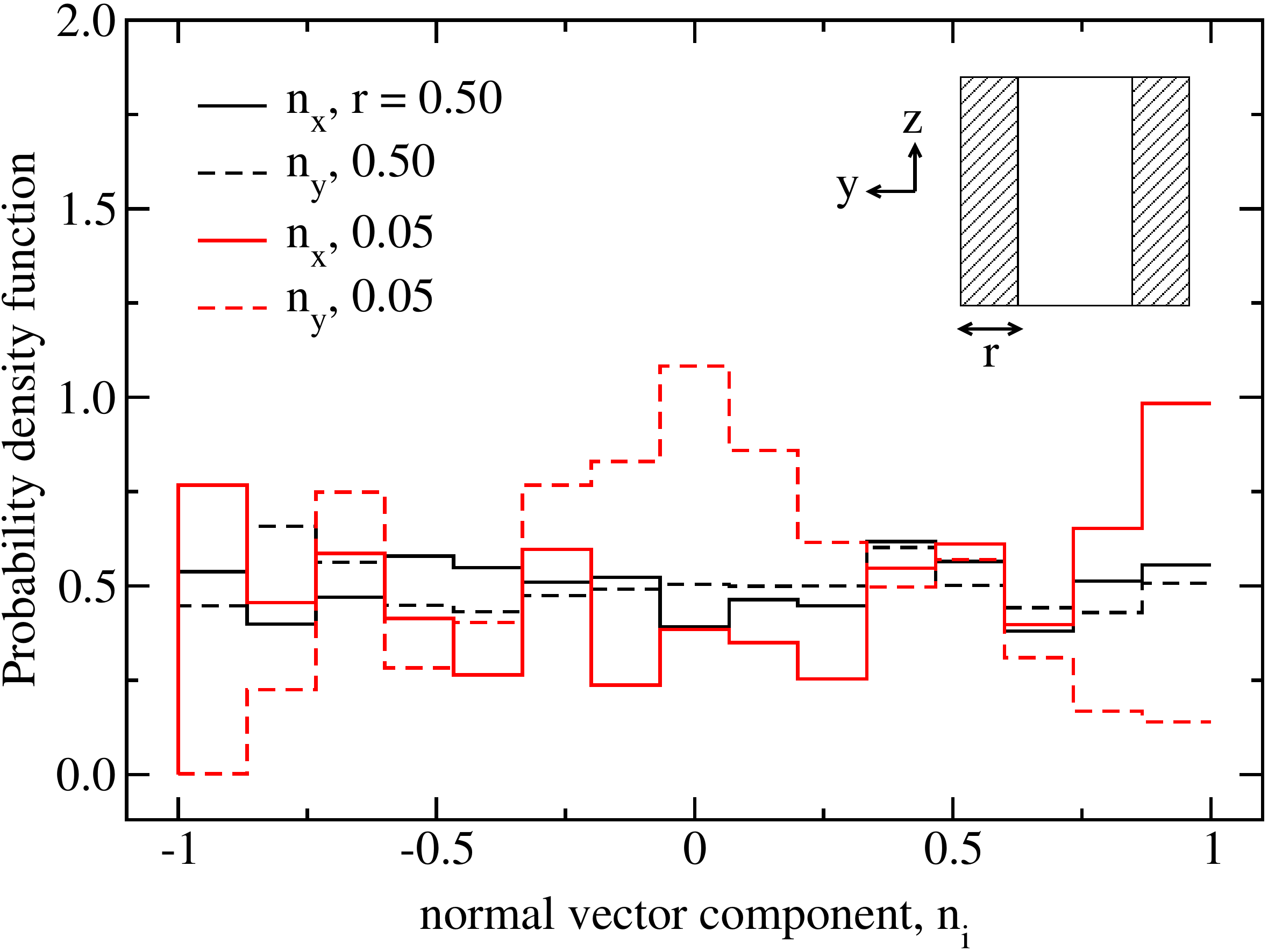}}
\caption{
(a) Probability density functions of normalized intergranular normal
stress calculated on grain boundaries with a maximum distance $r$ from
a free surface ($r=0.5$ denotes entire aggregate). Voronoi finite
element simulations were performed for $\epsilon=0.05$ assuming
crystal elasticity and ideal plasticity for Al and FCC under uniaxial
loading conditions.
(b) Probability density functions of grain boundary normal components
$n_x$ and $n_y$ calculated within the entire aggregate ($r=0.5$) and
within a thin outer layer ($r=0.05$) facing the free surface in Y
direction (note that tensile loading is assumed in X direction).}
\label{fig:12}
\end{figure}

In Fig. \ref{fig:12}(a) the free surface effects are manifested mostly
in bulk stress region by the appearance of additional peaks at
$\sigma_{nn}/\Sigma\sim 0.5$ and $0.8$. However, the upper tail of
\textit{pdf}s, relevant for IGSCC initiation, is not considerably
affected. {The corresponding differences in mean and standard
deviation of intergranular normal stress are listed in
Tab. \ref{tab6}}

\begin{table}[H]
\begin{tabular}{c|c|c}
$r$ &  $\langle\sigma_{nn}/\Sigma\rangle$ &  $s(\sigma_{nn}/\Sigma)$\\
\hline
\hline
0.50    &  0.335 & 0.445 \\
0.10    &  0.401 & 0.421 \\
0.05    &  0.440 & 0.409 \\
\hline
\end{tabular}
\caption{
Mean and standard deviation of intergranular normal stress
distributions from Fig. \ref{fig:12}(a).}
\label{tab6}
\end{table}

The observed changes in Fig. \ref{fig:12}(a) {and Tab. \ref{tab6}} may
be attributed to two different causes. The first one can be related to
the transition from bulk (3D) to plane (2D) stress state realized at a
free surface where softer material response is expected due to fewer
constraints applied there. However, this effect has been estimated to
be relatively small and should affect the \textit{pdf} only in the
$\sigma_{nn}\sim 0$ region.

The second cause can be connected to topology changes of the grain
boundaries when approaching free surfaces from the bulk. The fact that
grain boundary topology evolves with $r$ is confirmed in Fig.
\ref{fig:12}(b) by showing extracted distributions of grain boundary
normal components $n_x$ and $n_y$ from the (undeformed) Voronoi model.
An expected uniform distribution is observed for both $n_x$ and $n_y$
when accounting for the entire aggregate ($r=0.5$), while a
non-uniform distribution is clearly developed when considering grain
boundaries located within a thin outer layer ($r=0.05$). This
evolution is a consequence of using Voronoi model generator which, by
default, avoids making sharp corners between the grain boundaries and
free surfaces. In this way, small and highly distorted finite elements
are avoided in the discretization of the model. As a result, hence,
practically no grain boundaries parallel to the free surfaces are
generated on the outer most region of the aggregate.

In realistic aggregates, however, grain (boundary) topology of the
free-surface region should not differ considerably from the one
existing in the bulk (neglecting the effects of sample surface
preparation). Therefore, negligible free-surface effects are expected
in the \textit{pdf}s of realistic polycrystals. This suggests that
results obtained for the bulk aggregates (Sec.~\ref{section3}) can as
well be used to describe the stress distributions close to the free
surfaces, and therefore be used to model IGSCC initiation.

Strong influence of the grain topology (as well as overall texture) on
the evolution of the \textit{pdf}s was highlighted also in
\citep{hure2016}.

%---------------------------------------
\subsection{Deviations from universal plastic behavior}
%---------------------------------------

{The universal behavior for the evolution of standard deviation of
normalized intergranular normal stress with strain described in
Section~\ref{crystalplasticity} and modeled through
Eq.~(\ref{eqfinal}) has been proposed to {arise} solely from
grain-grain interactions {resulting from grain anisotropy}. {At higher
ratios of $\epsilon/\epsilon_y$ the observed universal behavior may no
longer be valid. This limit is investigated and} demonstrated in
Fig.~\ref{fig:sqrt} where additional simulations have been performed
for sets of material properties leading to very small values of yield
strain {(corresponding to the limit of rigid-plastic materials)},
therefore in the limit $\epsilon/\epsilon_y \gg 1$, while keeping the
macroscopic strain small {($\epsilon\le 0.05$) and grain orientations
fixed at zero texture}. {In this way, the effects of geometry and
texture evolution have been avoided.}
The universality of the evolution of $s(\sigma_{nn}/\Sigma)$ with
strain for different standard deviations of Taylor factor,
\begin{equation}
   s(\sigma_{nn}/\Sigma) = s_{iso} + \alpha\, s(M)\, \mathcal{F}\left(\frac{\epsilon}{\epsilon_{y}} \right)
   \label{eqgeneric}
\end{equation}
{is retained} up to $\epsilon/\epsilon_y \sim 1000$, but with
function $\mathcal{F}$ well approximated by square-root rather than
log-like behavior observed for lower values of $\epsilon/\epsilon_y$
(Eq.~(\ref{eqfinal})). {While Eq.~(\ref{eqgeneric}) has been
established for the conditions used in Fig.~\ref{fig:sqrt}}, such high
values of $\epsilon/\epsilon_y$ are not likely to occur in practical
situations, and deviations from Eq.~(\ref{eqfinal}) are likely to
develop at large strains. The purpose of the following two subsections
is thus to describe {the conditions under which the deviations}
from Eq.~(\ref{eqfinal}) are observed.

\begin{figure}[H]
\centering
\includegraphics[height = 5.5cm]{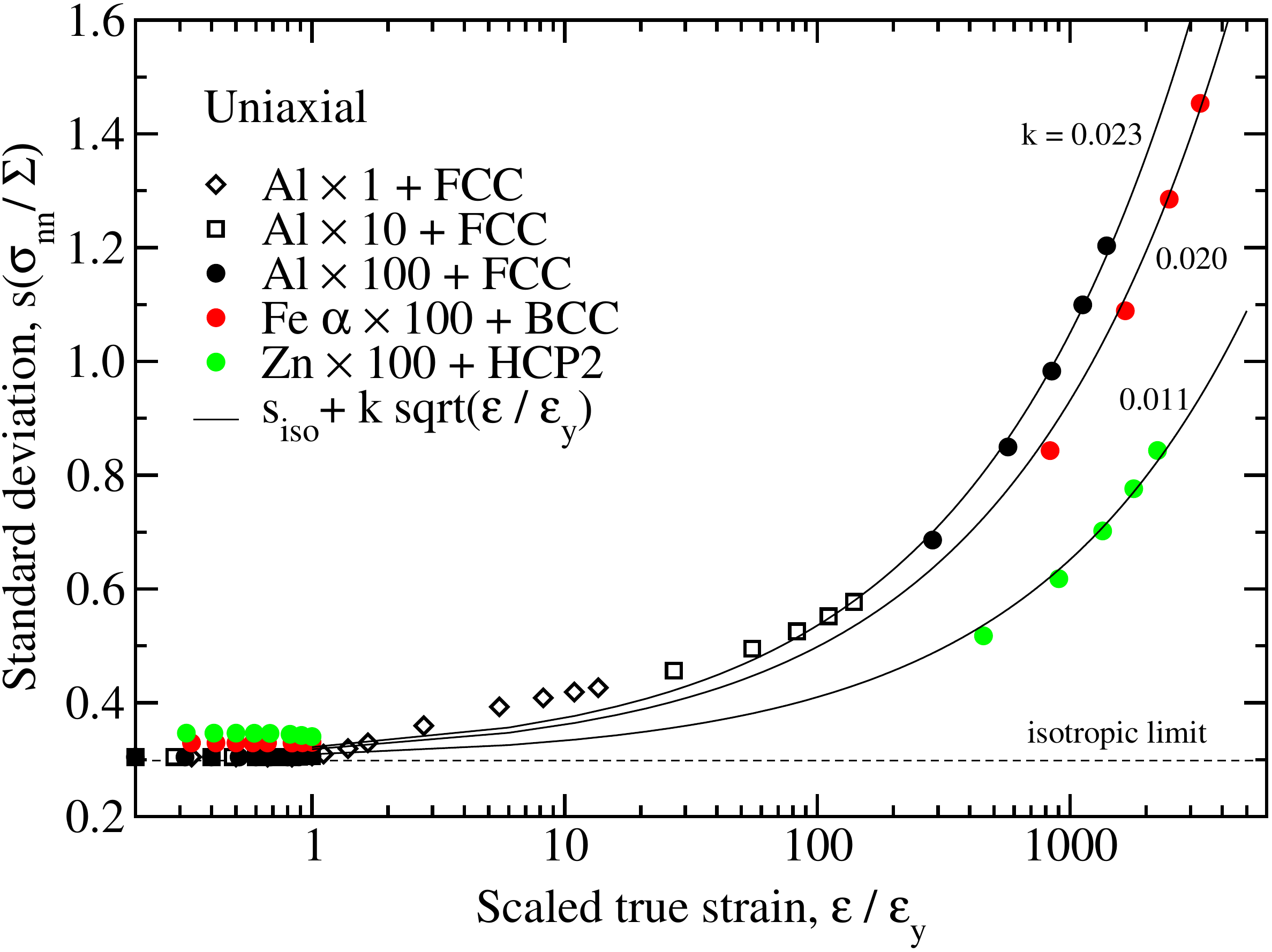}
\caption{
Standard deviation of normalized intergranular normal stress
calculated with Voronoi finite element simulations as a function of
applied strain rescaled by yield strain assuming crystal elasticity
and ideal ($H=0$) plasticity for various materials (with rescaled
elastic properties - by factor 10 or 100) and slip systems under
uniaxial loading conditions. Solid lines represent square-root fits of
Eq. (\ref{eqgeneric}) {on data points with
$\epsilon/\epsilon_y>100$}. The simulations were performed using small
deformation theory (\textit{i.e.}, fixed orientations) with
corresponding maximum strain of 0.05.}
\label{fig:sqrt}
\end{figure}

}

\subsubsection{Deviations due to geometrical changes}

At large strains the assumptions of zero texture and random grain
shapes become no longer valid. The geometrical changes (elongation of
the grains) as well as the generation of small non-zero texture may
affect the universal \textit{pdf} behavior of the intergranular
stresses in the plastic regime described in Sec.~\ref{section3}.
Large strain simulations have thus been performed to assess the strain
above which universal log-like behavior starts to fail. A case with Al
elasticity and FCC %ideal 
plasticity {with finite hardening ($H=500$ MPa)} is considered
for uniaxial % and equibiaxial 
loading conditions with strains up to $\epsilon\sim 0.4$.

\begin{figure}[H]
\centering
\includegraphics[height = 5.5cm]{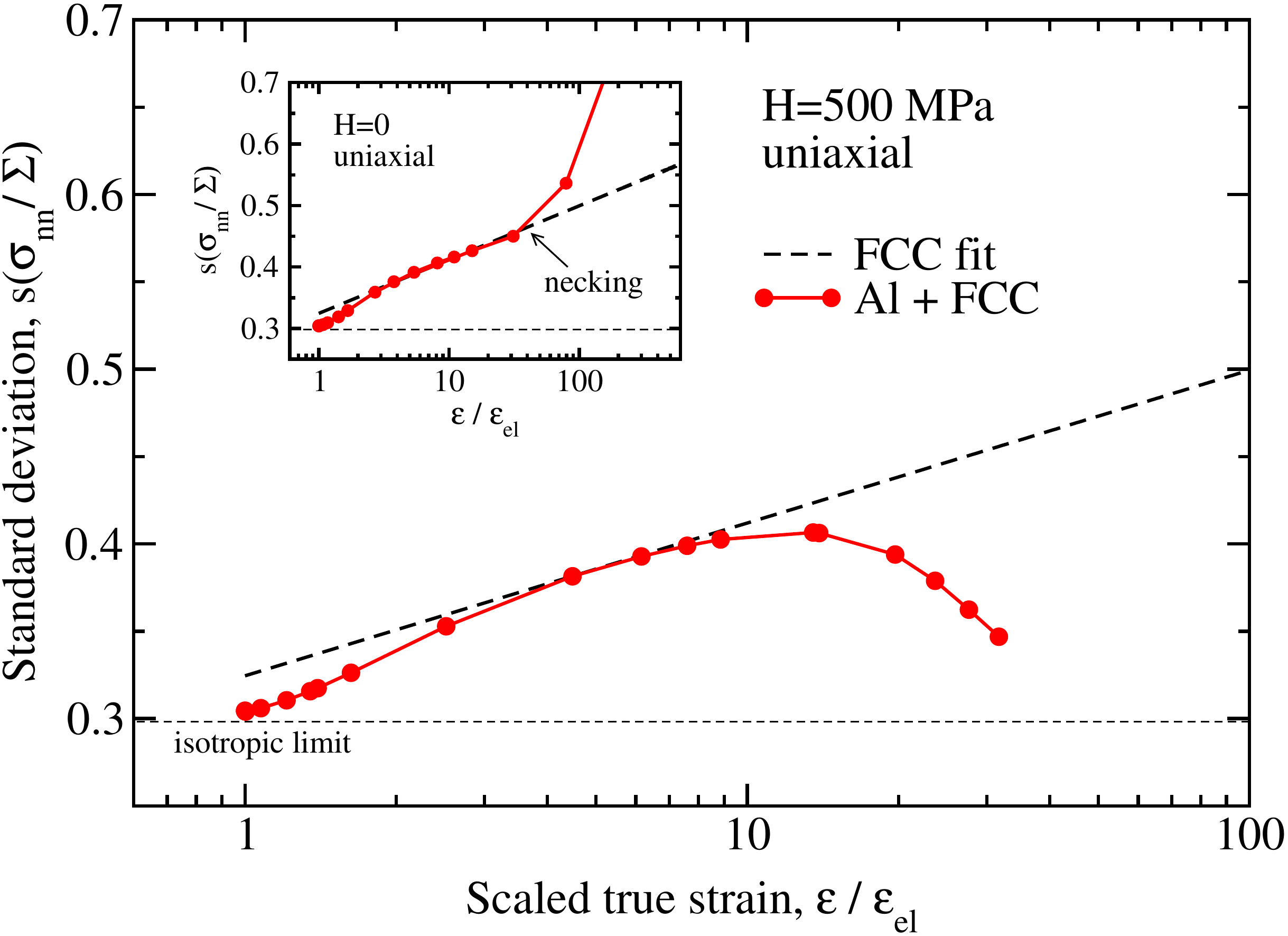}
\caption{
Standard deviation of normalized intergranular normal stress
calculated with Voronoi finite element simulation as a function of
applied strain (up to $\epsilon\sim 0.4$) rescaled by elastic strain
assuming crystal elasticity {and plasticity for Al and FCC with finite
hardening strength H=500 MPa (inset: H=0) under uniaxial loading
conditions.} The fitted logarithmic curve for FCC from
Fig. \ref{fig:8c}(a) is shown for comparison.}
\label{fig:11}
\end{figure}

{The results shown in Fig. \ref{fig:11} indicate that logarithmic
behavior of $s(\sigma_{nn}/\Sigma)$ is retained up to
$\epsilon/\epsilon_{el}\sim 10$, which corresponds to $\epsilon\sim
0.06$. Beyond this strain a down-turn of $s(\sigma_{nn}/\Sigma)$ is
observed which can be attributed to the accumulated changes of grain
shapes (grain elongations) as well as crystallographic texture
evolution.

With the evolution of texture the misorientations between neighboring
grains become smaller on average%
\footnote{
In the extreme limit of fully developed texture the misorientations
between the grains become zero as all the grains share the same
orientation.},
which reduces the mismatch effects (and thus the stresses) between
adjacent grains. It is important to note that finite hardening modulus
($H=500$ MPa) is used in the simulation to avoid the initiation of
macroscopic necking. In this way the stress state is kept uniaxial in
the entire model domain for all $\epsilon\lesssim 0.4$.

In the inset of Fig. \ref{fig:11}, however, a similar analysis but
with zero hardening ($H=0$) is performed to study the large-strain
response of the ideally plastic material. The observed up-turn of
$s(\sigma_{nn}/\Sigma)$ at $\epsilon/\epsilon_{el}\sim 30$ (or
$\epsilon\sim 0.1$) is attributed here mostly to the initiation of
macroscopic necking. At the onset of necking the macroscopic stress
$\Sigma$ starts to decrease with further straining due to strong
reorientations of the grains in the necking region. The effect of
$\Sigma$ decline with strain is strong enough to result in the up-turn
of $s(\sigma_{nn}/\Sigma)$.

Note that both uniaxial and equibiaxial loading conditions provide the
same qualitative behavior of $s(\sigma_{nn}/\Sigma)$ with strain. }

%---------------------------------------
\subsubsection{Deviations from Taylor hardening}
%---------------------------------------

{
  The influence of different hardening models on the observed
  universal behavior of $s(\sigma_{nn}/\Sigma)$ with
  $\epsilon/\epsilon_{el}$ (see, \textit{e.g.}, Fig. \ref{fig:7c}) is
  investigated in Fig. \ref{fig:9}(a) for Al and FCC {\red and in Fig.
    \ref{fig:9}(b) for Zn and HCP2} under uniaxial loading conditions.
  In addition to linear Taylor hardening, Eq.~(\ref{eq_2}), two more
  realistic hardening models are introduced: an empirical hardening
  model of Peirce, Asaro and Needleman (PAN) \citep{pierce83}, used to
  describe Stage I hardening in FCC metals, where
\be
  \tau_c^\al=\tau_0 + \int\!\sum_\bt h_{\al\bt}\ |\dot{\gm}^\bt| dt
  \quad\hbox{with}\quad
  h_{\al\al}=H\ \textrm{sech}^2\left(\frac{H}{\tau_s-\tau_0}\Gamma\right)
  \quad\hbox{and}\quad
  h_{\al\bt}=q\ h_{\al\al}\quad (\al\ne\bt),
  \label{eq_pan}
\ee
and a recently introduced physics-based hardening model of Han
\citep{phdHan}, used to describe the behavior of (non-irradiated)
austenitic stainless steel in \citep{phdHan,hure2016}, where
\be
  \tau_c^\al = \tau_0+\mu\sqrt{\sum_\bt a^{\al\bt}r_D^\bt}
  \quad\hbox{with}\quad
  \dot{r}_D^\al = \left(\frac{1}{\kp}\sqrt{\sum_\bt b^{\al\bt}r_D^\bt}-G_c r_D^\al\right)|\dot{\gm}^\al|.
  \label{eq_han}
\ee
With respect to Taylor hardening, two additional parameters are
introduced in PAN model, Eq.~(\ref{eq_pan}): $\tau_s$ denotes
saturation shear stress and $q$ a ratio of latent to self hardening.
In the limit of $q=1$ and $H \Gamma/(\tau_s-\tau_0)\ll 1$, PAN model
simplifies to Taylor hardening model.

The hardening model of Han, Eq.~(\ref{eq_han}), accounts for
dislocation density evolution (with corresponding normalized variable
$r_D^\al$).  Parameter $\mu$ is macroscopic shear modulus,
$a^{\al\bt}$ and $b^{\al\bt}$ are dislocation interaction matrices
composed of six independent parameters ($a_i$ and $b_i$, for
$i=1,\ldots,6$), $\kp$ is a value proportional to the number of
obstacles crossed by a dislocation before being immobilized and $G_c$
is a proportional factor that depends on the annihilation mechanism of
dislocation dipoles \citep{phdHan}.

The hardening constitutive equations were implemented numerically into
finite element solvers Abaqus \citep{abaqus} and Cast3M
\citep{castem}. Both implementations have been shown to give
equivalent results. The parameters of the three hardening models used
in Fig. \ref{fig:9}(a) are listed in Tab. \ref{tab_pan}.

\begin{table}[H]
\begin{tabular}{c|c|c|c|c|c|c|c|c|c|c|c|c|c|c}
Model  & $\tau_0$ & $\tau_s$ & $H$ & $q$ & $\mu$ & $\kp$ & $G_c$ & $a_1$ & $a_2$ & $a_3$ & $a_4$ & $a_5$ & $a_6$ & $b_i$ \\
\hline
\hline
Taylor &    100  &       & 200 &       &     &       &      & & & & & & & \\
\hline
PAN    &    100  &  200  & 200 &  0    &     &       &      & & & & & & & \\
PAN    &    100  &  200  & 200 &  1    &     &       &      & & & & & & & \\
PAN    &    100  &  200  & 200 &  2    &     &       &      & & & & & & & \\
\hline
Han    &    88   &       &     &       & 65.6& 42.8  & 10.4 & 0.124 & 0.124 & 0.070 & 0.625 & 0.137 & 0.122 & $1-\delta_{i1}$ \\
\hline
\end{tabular}
\caption{
  Hardening parameters used in Fig. \ref{fig:9}(a). Parameters $\tau_0$,
  $\tau_s$ and $H$ are shown in units of MPa, parameter $\mu$ in units
  of GPa. Parameters of Han model correspond to 304L stainless steel
  \citep{hure2016}.}
\label{tab_pan}
\end{table}

\begin{figure}[H]
\centering
\subfigure[]{\includegraphics[height = 5.cm]{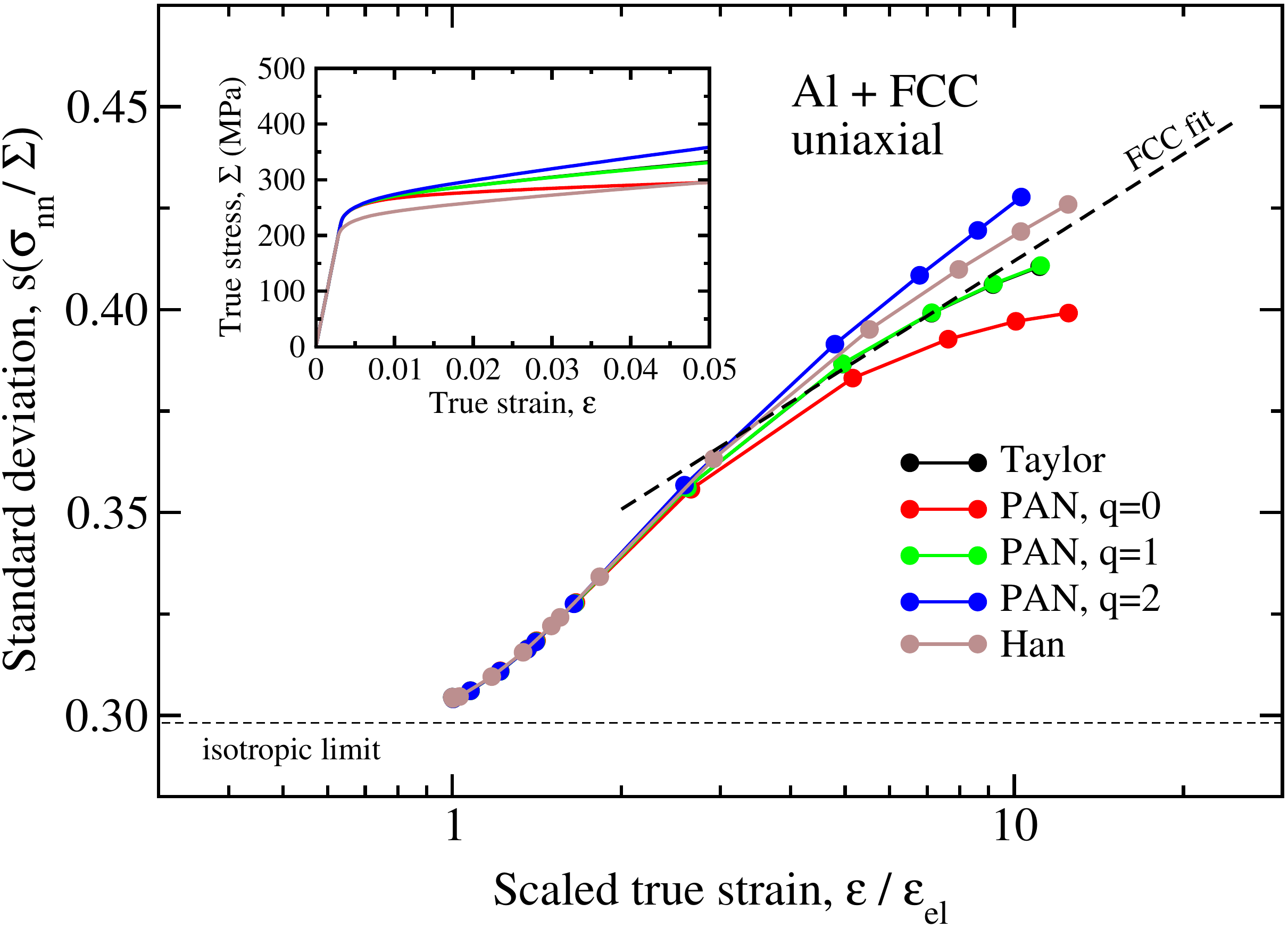}}
\hspace{0.5cm}
\subfigure[]{\includegraphics[height = 5.cm]{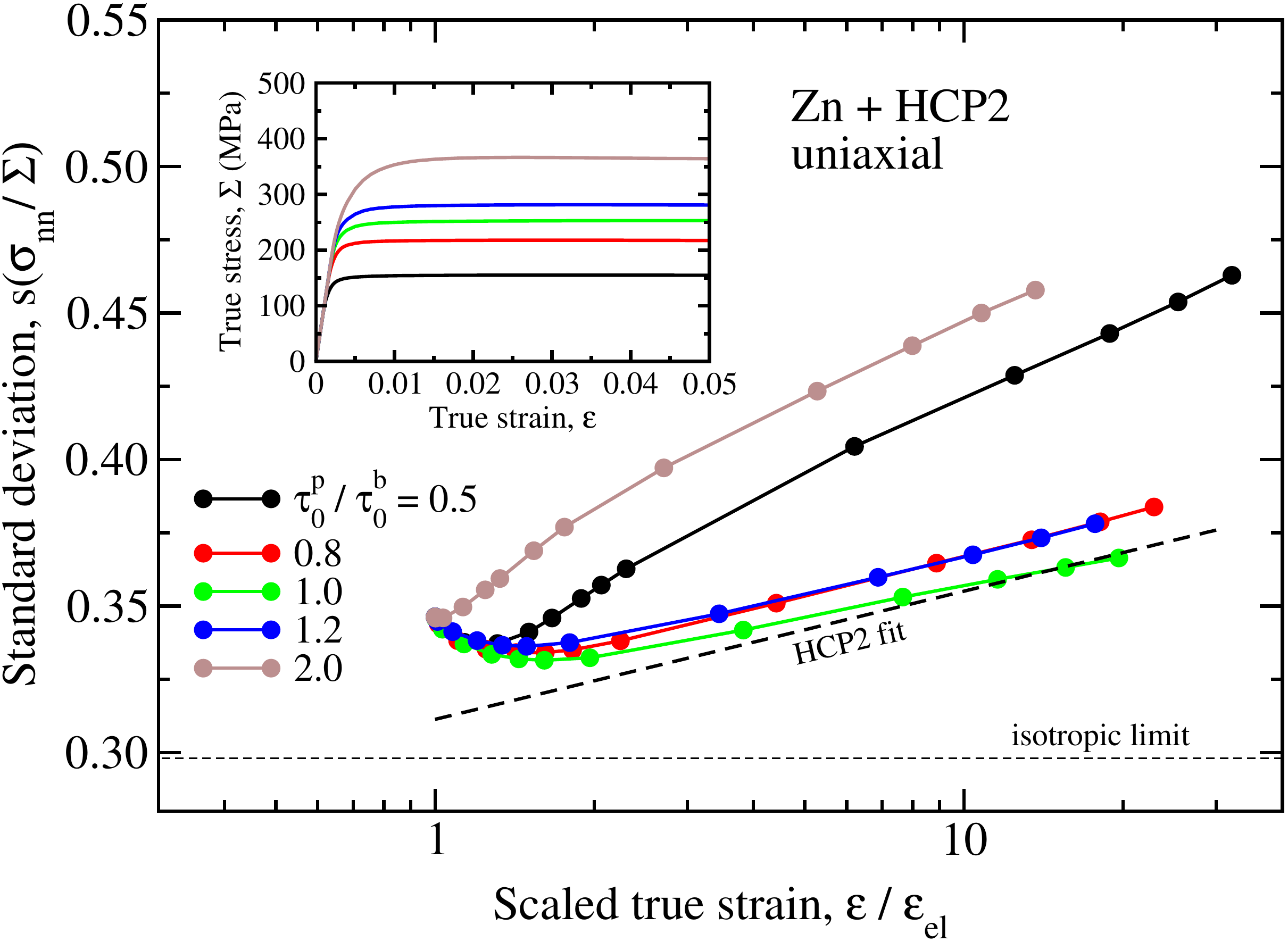}}
\caption{ Standard deviation of normalized intergranular normal stress
  (inset: macroscopic tensile stress) as a function of applied strain
  rescaled by elastic strain calculated with Voronoi finite element
  simulations assuming crystal elasticity and plasticity for {\red
    (a)} Al and FCC with various hardening models under uniaxial
  loading conditions (see the text for models definitions and material
  parameters used in simulations) {\red and (b) Zn and HCP2 with ideal
    ($H=0$) plasticity but with different ratios of initial critical
    resolved shear stresses $\tau_0^{pyr}/\tau_0^{bas}$ (assuming
    $\tau_0^{pri}=\tau_0^{pyr}$ and $\tau_0^{bas}=100$ MPa).} The
  fitted logarithmic curves for FCC and HCP2 from Fig. \ref{fig:8c}(a)
  are shown for comparison.}
\label{fig:9}
\end{figure}

A comparison of the three hardening models presented in
Fig. \ref{fig:9}(a) shows that universality of $s(\sigma_{nn}/\Sigma)$ is
maintained for all models up to $\epsilon/\epsilon_{el}\sim 5$ where
larger discrepancies from log-like behavior start to develop. In this
respect, the $q$ value is identified as the most influencing parameter
of PAN model%
\footnote{
Note that same qualitative results are obtained also with Bassani and
Wu hardening model \citep{bassani91} which accounts for Stage I and II
hardening.}%
. For $q=1$ a Taylor-like hardening is restored which results in
almost identical $s(\sigma_{nn}/\Sigma)$ curves. In this limit all
slip systems experience the same strength, while the balance among
slip systems is lost when $q\ll 1$ or $q\gg 1$. This is manifested as
down-turn or up-turn of the $s(\sigma_{nn}/\Sigma)$ curve at
$\epsilon/\epsilon_{el}\gtrsim 5$.} {{Such deviations from
Eq.~(\ref{eqfinal}) are in fact expected as the standard deviation of
Taylor factor, which is the key parameter of Eq.~(\ref{eqfinal}), was
computed assuming 
%
%constant
{same ($\alpha$-independent)}
critical resolved shear stress for all slip systems. For $q\ll 1$ or
$q\gg 1$ such assumption breaks down.
%
%{(for example with minimum work principle as described in Appendix~E).}
}} {However, it should be noted that typical values for common cubic
materials, like austenitic stainless steel, have been reported to be
in the range of $1\le q<2$
\citep{elshawish2017}, implying that a reasonably good agreement with
observed universal log-like behaviour can be expected in this
material.

The same reasoning may be used to justify the similar responses of Han
and Taylor hardening models observed in Fig. \ref{fig:9}(a). Since all
the terms but one of the dislocation interaction matrix $a^{\al\bt}$
in Han model are close to one single value ($\sim$0.1, see
Tab. \ref{tab_pan}), the critical resolved shear stress may be
approximated as $\al$-independent, $\tau_c^\al\sim\tau_0 +
\mu\sqrt{0.1\sum_\bt r_D^\bt}$. In this way, slip systems in Han model
should experience similar strengths, like in $q\sim 1$ case of PAN
model. 

% revision: Samir

{\red To account for even larger plastic anisotropy distinctive of
  realistic hcp materials
  , the
  influence of non-uniform initial critical resolved shear stress,
  $\tau_0=\mathcal{F}(\alpha)$, is studied in Fig.~\ref{fig:9}(b). As
  an illustrative example, Zn elasticity with HCP2 ideal plasticity is
  addressed with different $\tau_0$ values for basal, prismatic and
  pyramidal second order slip modes defined in Tab. \ref{tab2}.
  Following the approach in \citep{Lebensohn07}, a single contrast
  parameter is introduced, descriptive of the single crystal plastic
  anisotropy, given by the ratio of the initial critical resolved
  shear stress between pyramidal (assumed here to be equal to
  prismatic) and basal slip modes, $\tau_0^{pyr}/\tau_0^{bas}$.
  Although $\tau_0^{pyr}/\tau_0^{bas}>50$ in some single hcp crystals
  (\textit{e.g.}, $48-96$ for Mg \citep{Hutchinson10}), the ratio is
  usually greatly reduced in polycrystalline arrangement
  (\textit{e.g.}, $2-5$ for Mg \citep{Hutchinson10}), as explained in
  \citep{Hutchinson10}. In this view, a close vicinity of the uniform
  case, $0.5\le \tau_0^{pyr}/\tau_0^{bas}\le 2.0$, is investigated in
  Fig.~\ref{fig:9}(b). While a uniform-like response is reproduced for
  relatively small (up to $\sim$20$\%$) variations in $\tau_0$, the
  results for moderate $\tau_0^{pyr}/\tau_0^{bas}=0.5$ and 2.0 already
  demonstrate substantial deviations of the calculated intergranular
  normal stresses from the proposed HCP2 universal behavior. In order
  to extend the validity of Eq.~(\ref{eqfinal}), to apply also for the
  general realistic hcp polycrystal, the non-uniformity of $\tau_0$
  would need to be considered also in Taylor factor calculation. This
  is left for future investigations.

  Following the results presented in Fig.~\ref{fig:9}, it is
  reasonable to assume that universal log-like behavior of
  $s(\sigma_{nn}/\Sigma)$ with $\epsilon/\epsilon_{el}$, proposed and
  demonstrated in Sec. \ref{section3} for Taylor hardening, applies
  also for other more sophisticated hardening laws as long as they
  provide a uniform (or close to uniform) strength evolution among the
  slip systems.}

}

%---------------------------------------
\section{Conclusions}
%---------------------------------------

\begin{figure}[t]
\centering
\includegraphics[height = 8cm]{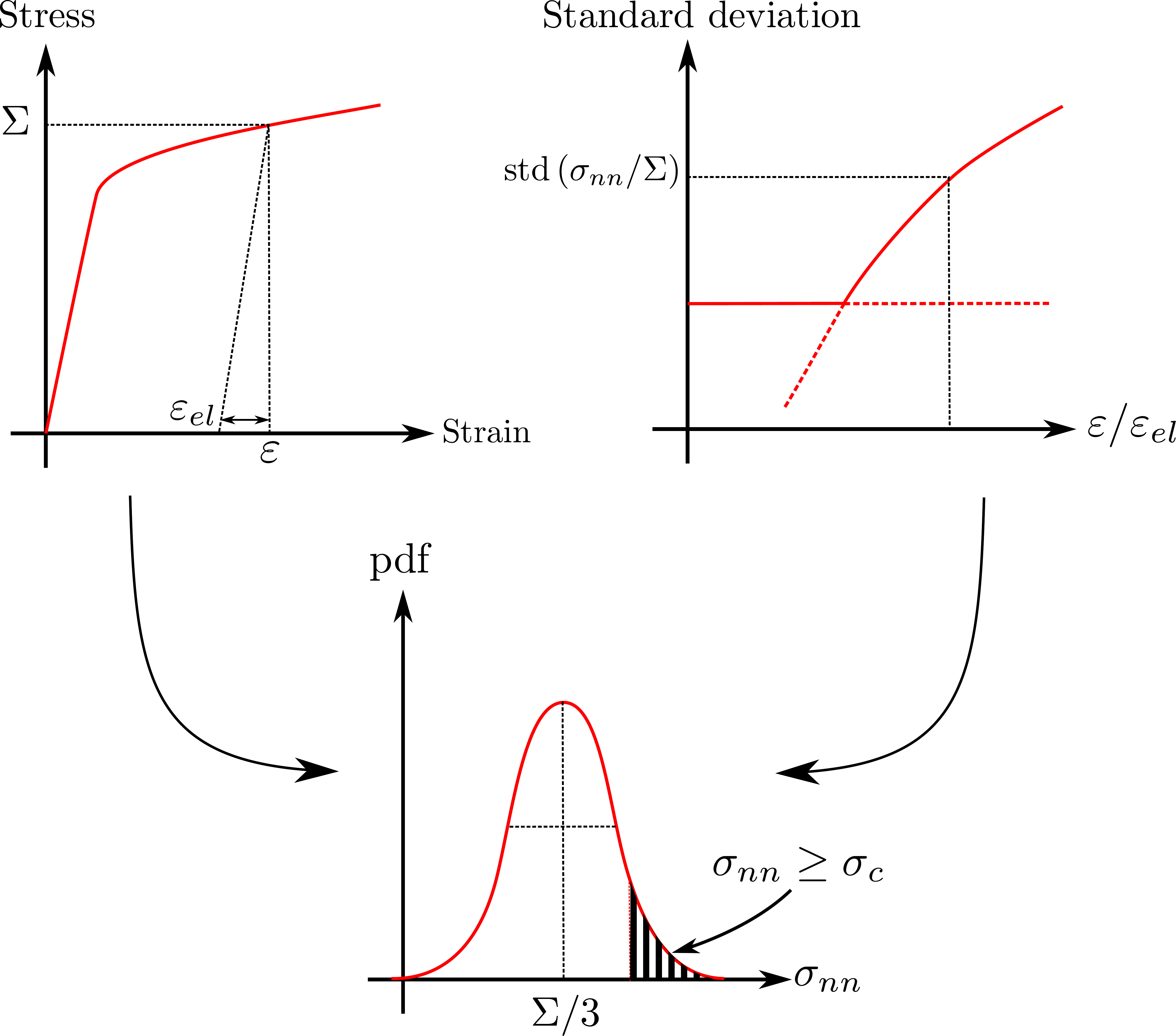}
\caption{
Methodology for estimating the probability of finding intergranular
normal stress higher than a critical value based on the knowledge of
the material properties and Eq.~(\ref{eqfinal}).}
\label{fig:conclusion}
\end{figure}

{Intergranular normal stresses are likely to be one of the key {
    ingredients required for quantitative modeling of} intergranular
  stress corrosion cracking {(IGSCC)} of polycrystalline materials.
  Crystal plasticity finite element simulations have been performed in
  order to assess the distributions of intergranular normal stresses
  as a function of elastic and plastic properties at the grain scale
  {for two different macroscopic} loading conditions. Untextured
  Voronoi aggregates have been considered with initially equiaxed
  grain shapes and random crystallographic orientations, under
  relatively low applied strain. Moreover, {\red small variations in
    the initial critical resolved shear stress have been assumed among
    different slip systems}, {with} self and latent hardening being of
  the same order of magnitude. A wide variety of elastic/plastic
  properties have been {studied}. Although not covering all the
  situations where IGSCC phenomenon is observed experimentally, {the
    investigated} conditions are relevant to a wide class of problems
  where IGSCC initiation is involved, for example, to austenitic
  stainless steels or nickel-based alloys.

Depending on material properties, complex intergranular normal stress
distributions have been obtained and characterised through their first
two statistical moments, \textit{i.e.}, mean and standard
deviation. For all simulations presented in this study, the mean
normal stress remains approximately constant at $\Sigma/3$
(\textit{resp.} $2\Sigma/3$) for uniaxial (\textit{resp.} equibiaxial)
loading conditions, where $\Sigma$ characterizes the macroscopic
stress {amplitude}. Standard deviation {normalized by macroscopic
stress} has been shown to depend only on elastic properties for
applied strain {below} the macroscopic yield strain, and only on
plastic properties {in fully plastic regime when plotted as a
function of applied strain rescaled by elastic strain}. The evolution
of the standard deviation during the transition between the {elastic}
regime and fully plastic regime depends on both elastic and plastic
properties, leading to an increase or decrease depending on the
relative {measure of} elastic \textit{vs.} plastic {anisotropy}.

In the elastic regime, a simple correlation has been found and
explained through simple arguments between the standard deviation of
intergranular normal stress and a universal anisotropy index recently
proposed. In the plastic regime, the standard deviation increases with
strain due to grain-grain interactions {resulting from plastic
anisotropy. The increasing behavior of standard deviation with strain
has been found to be decoupled into (i) a strain-independent part that
correlates linearly with the standard deviation of Taylor factor
characterizing the slip systems and (ii) a universal strain-dependent
function
.}
A simple phenomenological formula {(Eq.~(\ref{eqfinal}))} has been
proposed that captures well all the numerical data obtained in this
study. As shown in Fig.~\ref{fig:conclusion}, {this formula leads} to
an easy tool to estimate the intergranular normal stresses in a
polycrystalline aggregate. Knowing the single crystal elastic
properties and standard deviation of Taylor factor (for most cases
tabulated in reference textbook) as well as the loading applied to the
material ({macroscopic} stress and strain), standard deviation of
normalized intergranular normal stress can be readily obtained,
leading to an estimate of the probability density function of the
intergranular stress. The probability of finding a stress higher than
a critical value (related to grain boundary strength) can finally be
assessed. Such approach can be used to quickly {and reliably}
estimate the propensity of an aggregate to develop high intergranular
{normal} stresses without having to perform finite element
simulations and/or to build an IGSCC fracture model once knowing the
grain boundary strength.  }

\begin{figure}[t]
\centering
\includegraphics[height = 6cm]{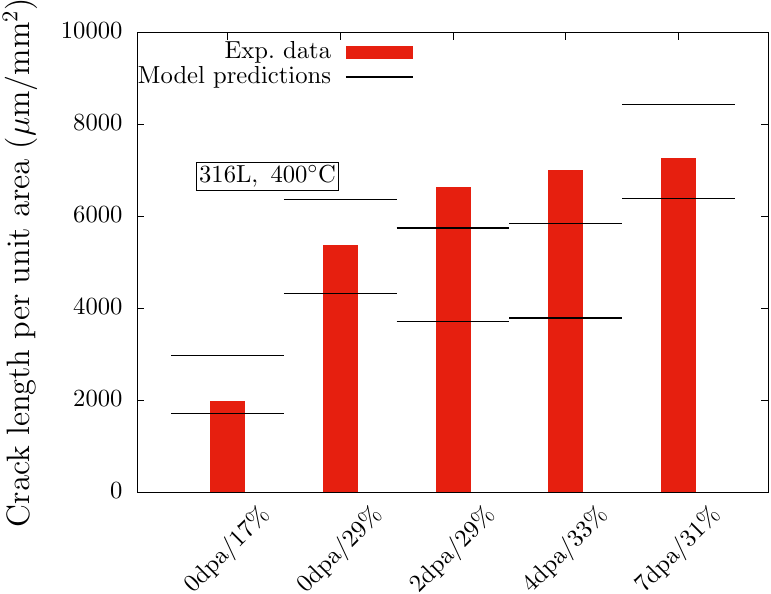}
\caption{\red Crack length per unit area for different irradiation
  levels/applied strain \citep{Zhou}. Numerical predictions have been
  computed assuming an applied true stress of [$455 - 515$] MPa and
  [$555 - 630$] MPa at the unirradiated state (denoted 0 dpa)
  (corresponding to the stress range for conventional strain of 17\%
  and 29\% on the tensile curves given in \citep{Zhou}). For the
  irradiated materials (2/4/7 dpa), the applied true stress is set
  equal to the yield stress computed as the sum of the yield stress of
  the unirradiated material at test temperature ($\sim$180 MPa) and
  the irradiation hardening reported in \citep{Zhou} (as low
  strain-hardening is expected for such material).  This leads to
  applied true stress of [$530 - 610$] MPa and [$630 - 710$] MPa for
  the 2/4 dpa and 7 dpa irradiation levels. The grain boundary length
  per unit area is approximated by $G_f \approx 2/\phi$, where $\phi$
  is the grain size. The critical stress $\sigma_c$ was calibrated to
  a value of 550 MPa.}
\label{fig:17b}
\end{figure}

% revision: Jeremy
{\red As a working example of this approach, IGSCC results presented
  in \citep{Zhou} of (un-)irradiated austenitic stainless steel tested
  in supercritical water are used. This study reports crack lengths
  per unit area under uniaxial loading conditions after slow strain
  rate tests, as a function of strain and irradiation levels (that
  correspond at first approximation to different yield stresses of the
  material) (Fig.~\ref{fig:17b}). Crack lengths reported are close to
  the grain size, so that these data correspond mainly to IGSCC
  initiation regime. Under the assumption of normal distribution, the
  probability of finding intergranular normal stress higher than a
  critical stress $\sigma_c$ can be expressed as
  $P_f(\sigma_{nn}/\Sigma\ge\sigma_c/\Sigma) = 1/2\ {\rm erfc}[(\sigma_c/\Sigma - 1/3)/\sqrt{2}std]$, 
  where ${\rm erfc}$ is the complementary error function, $\Sigma$ the
  macroscopic stress, and $std$ the standard deviation as given by
  Eq.~(\ref{eqfinal}). This probability is assumed to be the fraction
  of grain boundaries subjected to normal stress higher than
  $\sigma_c$, which leads to an estimation of the crack length per
  unit area $G_f P_f$, where $G_f$ is the grain boundary length per
  unit area. Figure~\ref{fig:17b} shows that a constant $\sigma_c$ is
  able to capture the trend of the data presented in \citep{Zhou} for
  both dependence on strain and irradiation (hardening) levels.
  Although in this example critical grain boundary stress might also
  depend on irradiation as a result of microchemistry modification
  (which might explain the discrepancy observed in Fig.
  \ref{fig:17b}) as well as the rather high strain level, such comparison appears promising and call for
  further assessment of this methodology.

  In addition, the standard deviations obtained in this study could be
  easily used to improve the IGSCC model developed recently by Couvant
  \textit{et al.}  \citep{Couvant1,Couvant2}. This model, that
  describes the incubation, initiation and propagation stages of
  IGSCC, has been shown to give promising results once calibrated on
  Nickel based alloy 600. The initiation phase modeling uses standard
  deviation of intragranular axial stress (under uniaxial tensile
  conditions) computed through CPFEM simulations, and assuming that
  the upper tail is somehow related to grain boundaries perpendicular
  to the loading directions where high stresses are expected to occur.
  A critical stress is then calibrated based on experimental data. The
  main drawback of this approach is that the physical meaning of this
  critical stress may not be so obvious -- as how this critical stress
  is related to grain boundary fracture energy ? -- that would be
  clearer when using the standard deviations of intergranular normal
  stress obtained in this study.  }

{Although Eq.~(\ref{eqfinal}) captures well most of the numerical data
  presented in this study, the theoretical justification of the
  evolution of the standard deviation of intergranular normal stress
  with strain remains an open question. Moreover, the effects of {\red
    a general non-uniform critical resolved shear stress
    distribution}, unloading and texture (either from initial
  crystallographic orientations/grain shapes or induced by
  deformation/hardening), relevant for specific applications, are
  still to be {further} assessed, which is left for future work.}

\section*{Acknowledgments}

The authors acknowledge the financial support from Slovenian Research
Agency and French Atomic Energy Commission through the bilateral
project ``Towards quantitative predictions of stress corrosion
cracking initiation stress threshold for PWR's internals'', {
No. BI-FR/CEA/15-17-007}, between CEA and JSI in years 2015-2017.

%---------------------------------------
\appendix

%---------------------------------------
\section{Influence of loading conditions on intergranular stress distributions}
\label{app1}
%---------------------------------------

In this study, tensile loading is simulated by applying an incremental
tensile displacement along the $X$ axis to all the nodes on the front
surface of the Voronoi model, while keeping the nodes on the back
surface constrained to have zero axial displacement (for equibiaxial
loading also the $Y$ axis is loaded in the same manner). Such loading
conditions keep the opposite surfaces flat, thus providing additional
(unwanted) stresses close to these surfaces.

\begin{figure}[H]
\centering
\includegraphics[height = 5.5cm]{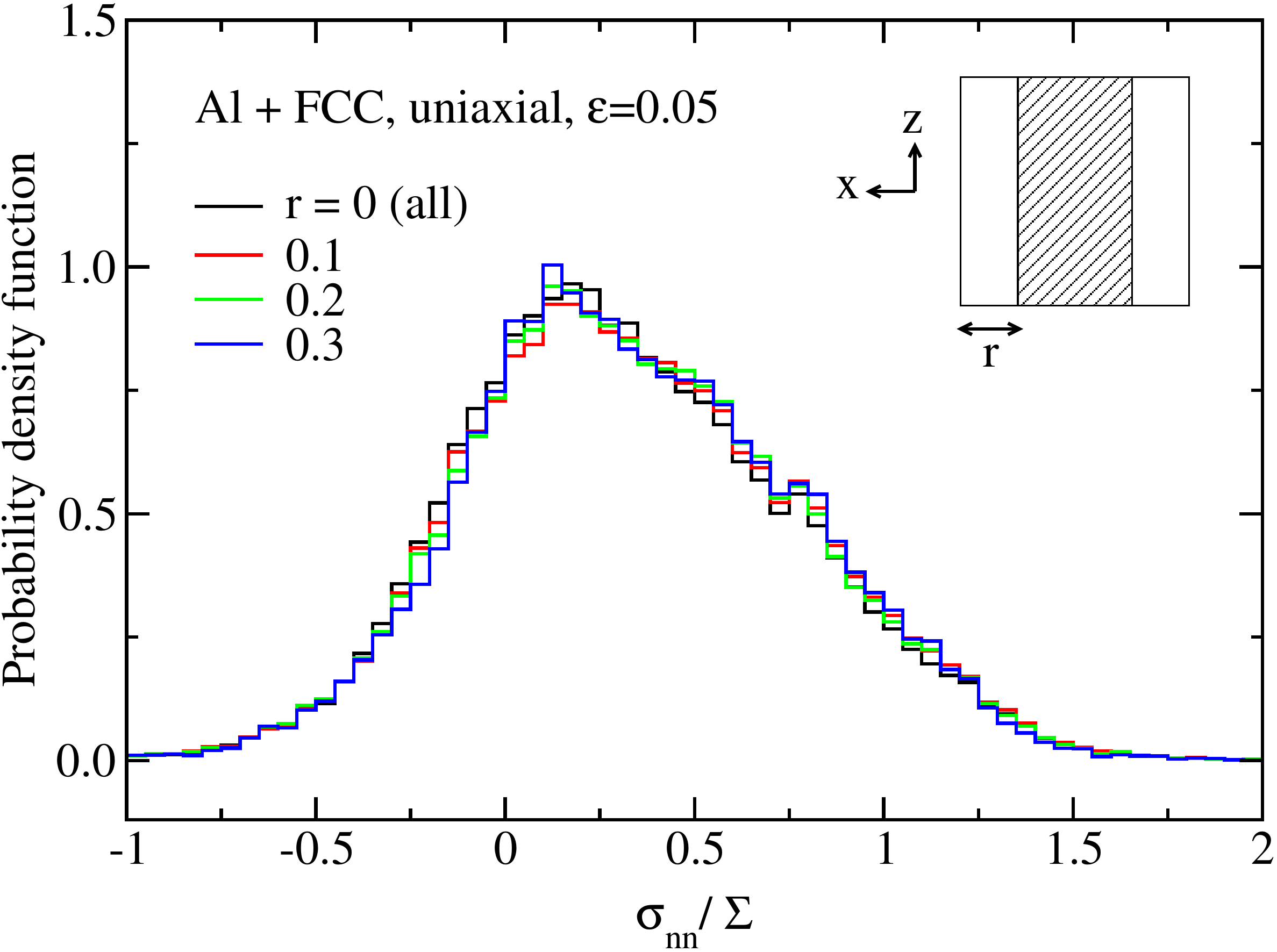}
\caption{
Probability density functions of normalized intergranular normal
stress calculated on grain boundaries with a minimum distance $r$
from the loading surfaces ($r=0$ denotes entire aggregate). Voronoi
finite element simulations were performed for $\epsilon=0.05$ assuming
crystal elasticity and ideal plasticity for Al and FCC under uniaxial
loading conditions.}
\label{fig:A1}
\end{figure}

Figure \ref{fig:A1} shows the influence of such loading conditions on
the intergranular normal stress distributions when calculated away (by
distance $r$) from the loading surfaces. Practically no effects can be
observed on the distributions for the increasing $r$. It is therefore
confirmed that proposed loading conditions have negligible effects on
the results presented in this study.

%---------------------------------------
\section{Comparison of different approaches to calculate intergranular stress}
\label{app2}
%---------------------------------------

\begin{figure}[H]
\centering
\includegraphics[height = 5.5cm]{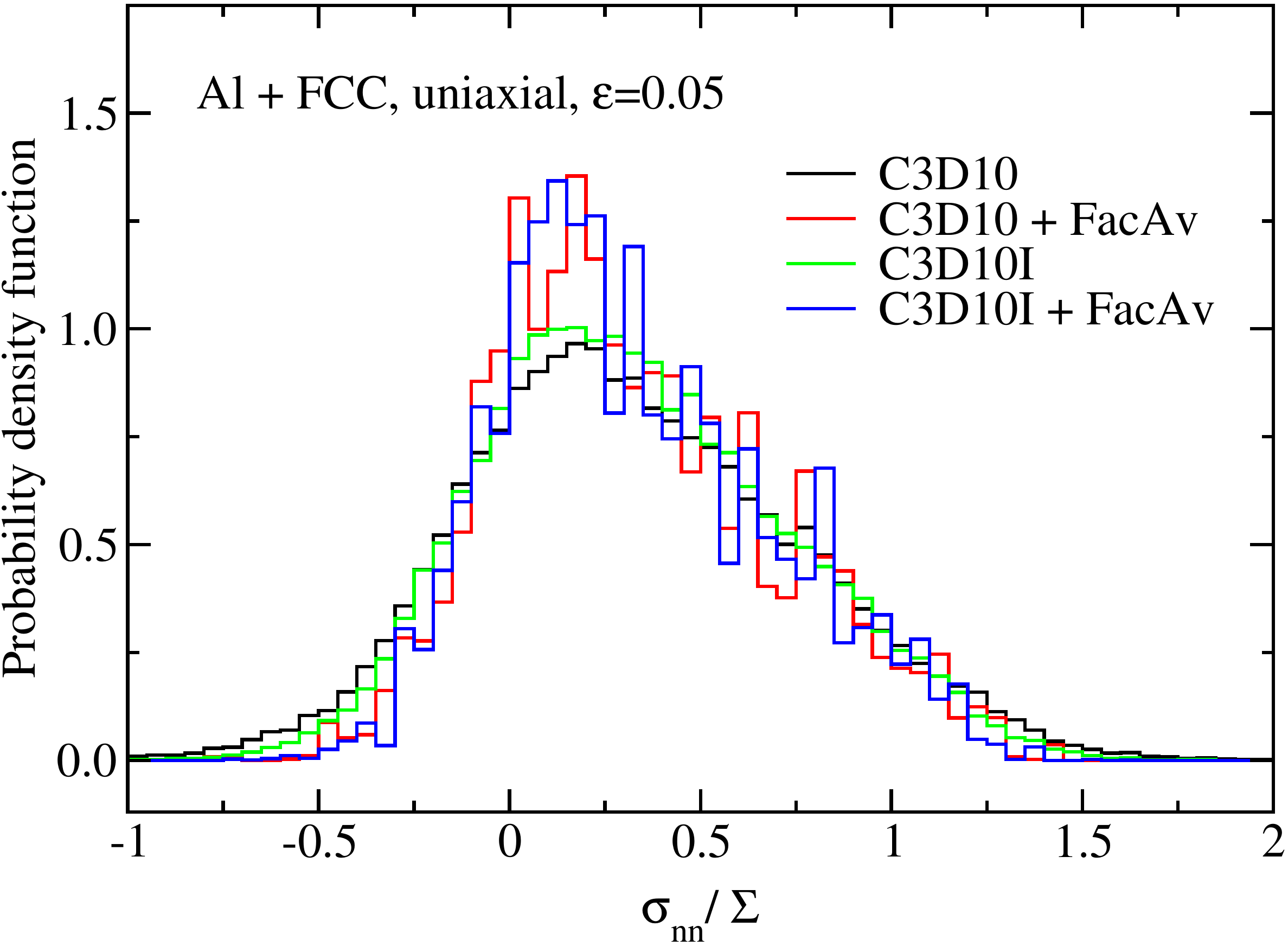}
\caption{
Probability density functions of normalized intergranular normal
stress calculated with Voronoi finite element simulations at
$\epsilon=0.05$ assuming crystal elasticity and ideal plasticity for
Al and FCC under uniaxial loading conditions. The curves correspond to
different methods used to obtain intergranular normal stress. Same
aggregate model is used in all cases. Legend: C3D10: at closest Gauss
points using standard tetrahedral element C3D10; C3D10I: at Gauss
points exactly at the intergranular position using special tetrahedral
element C3D10I; FacAv: averaging on grain facet using weights from
triangular element areas. }
\label{fig:A2}
\end{figure}

Figure \ref{fig:A2} shows a comparison between different strategies in
obtaining intergranular normal stresses. Relatively small differences
are observed between calculated \textit{pdf}s when using C3D10 or
C3D10I tetrahedra. When using a special tetrahedral element C3D10I
with 10 Gauss points positioned at the nodes (and 1 at the element
center), the projected stresses may be obtained exactly at grain
boundaries. However, as grain boundary node is usually shared by
several tetrahedra (associated tetrahedra can either share a common
grain boundary facet, edge or node), additional (weighted) averaging
of the stresses may be needed in order to obtain a unique stress value
per grain boundary node. This brings some ambiguity to intergranular
stress calculation. For this reason and since the differences between
using C3D10 and C3D10I tetrahedra are small, C3D10 tetrahedra have
been used through the paper to calculate the \textit{pdf}s.

Figure \ref{fig:A2} also shows the effect of using an additional
grain-facet averaging, used \textit{e.g.} in \citep{gonzalez2014}, where
only one stress value is retained per grain facet. In this way,
extreme local intergranular stresses average out to produce (slightly)
narrower \textit{pdf} but with retained global shape. 
Also, as the number of grain facets is much smaller than the number of
element pairs on the grain boundaries, larger aggregate models would
be needed to obtain smoother distributions. For this reason no
additional averaging over grain facets has been used throughout the
paper to calculate the \textit{pdf}s.

%---------------------------------------
\section{Intergranular normal stress distribution in polycrystal with isotropic grains}
\label{app2b}
%---------------------------------------

In a polycrystal with isotropic grains local stress $\bm{\sigma}$ is
uniform and equal to the applied stress $\bm{\Sigma}$. The
intergranular normal stress may therefore be calculated as
$\sigma_{nn}=n.\bm{\Sigma}.n$, where $n=(\sin\theta \sin\phi,
\sin\theta \cos\phi, \cos\theta)$ is grain boundary normal written
with spherical coordinates. Assuming diagonal stress state,
$\bm{\Sigma}=\Sigma_{xx} e_X \otimes e_X + \Sigma_{yy} e_Y \otimes e_Y
+ \Sigma_{zz} e_Z \otimes e_Z$, the intergranular normal stress
becomes $\sigma_{nn}=\Sigma_{zz}\cos^2\theta +
\sin^2\theta(\Sigma_{xx}\cos^2\phi+\Sigma_{yy}\sin^2\phi)$,
which simplifies to
\be
  \sigma_{nn}=\Sigma \cos^2\theta
  \hspace{1cm}\hbox{and}\hspace{1cm}
  \sigma_{nn}=\Sigma \sin^2\theta
\ee
for, respectively, tensile ($\Sigma=\Sigma_{zz}$,
$\Sigma_{xx}=\Sigma_{yy}=0$) and equibiaxial loading
($\Sigma=\Sigma_{xx}=\Sigma_{yy}$, $\Sigma_{zz}=0$).

To calculate the distribution function of $\sigma_{nn}/\Sigma$,
$f_2(\sigma_{nn}/\Sigma)$, it is assumed that $\cos\theta$ is
distributed uniformly, $f_1(\cos\theta)=1$, on a range
$0\le\cos\theta\le 1$ (the negative values of $\cos\theta$ can be
omitted without loss of generality since $n$ and $-n$ denote the same
grain boundary). This corresponds to a polycrystal with random grain
boundary topology or random grain shapes (with no preferred
direction).

Following the theorem of change of variables, $f_1(x_1) |d x_1| =
f_2(x_2) |d x_2|$, where $x_1=\cos\theta$, $x_2=\sigma_{nn}/\Sigma$
and $f_1(x_1)=1$, the distribution function becomes
$f_2(\sigma_{nn}/\Sigma)=|d(\cos\theta)/d(\sigma_{nn}/\Sigma)|$.
Writting $\cos\theta=\sqrt{\sigma_{nn}/\Sigma}$ for tensile and
$\cos\theta=\sqrt{1-\sigma_{nn}/\Sigma}$ for equibiaxial loading,
respectively,
\be
  f_2(\sigma_{nn}/\Sigma)=\frac{1}{2\sqrt{\sigma_{nn}/\Sigma}}
  \hspace{1cm}\hbox{and}\hspace{1cm}
  f_2(\sigma_{nn}/\Sigma)=\frac{1}{2\sqrt{1-\sigma_{nn}/\Sigma}}
\ee
on a range $0\le\sigma_{nn}/\Sigma\le 1$.

%---------------------------------------
\section{Universal elastic anisotropy index calculation}
\label{app3}
%---------------------------------------

The universal elastic anisotropy index is calculated as
\citep{ranganathan}
\be
  A^u=5\frac{G^V}{G^R}+\frac{K^V}{K^R}-6 \label{eq_au}
\ee
where $G$ and $K$ denote shear and bulk moduli, and superscripts $V$
and $R$ represent the Voigt and Reuss estimates, respectively. In the
following, the estimates for the elastic moduli are given for a
quasi-isotropic polycrystal aggregate composed of randomly oriented
grains whose size is small relative to the size of the polycrystal.

The general forms for $K^V$ and $G^V$ can be expressed as
\citep{tromans}
\ba
  &&K^V = \frac{C_{11}+C_{22}+C_{33}+2(C_{12}+C_{13}+C_{23})}{9}\\
  &&G^V = \frac{C_{11}+C_{22}+C_{33}+3(C_{44}+C_{55}+C_{66})-(C_{12}+C_{13}+C_{23})}{15}\nonumber
\ea
where $C_{ij}$ are stiffness matrix components in Voigt notation.

In a similar way, the general forms for $K^R$ and $G^R$ can be written
as \citep{tromans}
\ba
  &&K^R = \frac{1}{S_{11}+S_{22}+S_{33}+2(S_{12}+S_{13}+S_{23})}\\
  &&G^R = \frac{15}{4(S_{11}+S_{22}+S_{33})+3(S_{44}+S_{55}+S_{66})-4(S_{12}+S_{13}+S_{23})}\nonumber
\ea
where $S_{ij}$ denote compliance matrix components. $S_{ij}$ are
related to $C_{ij}$ via matrix inversion.

Elastic moduli can be further simplified when accounting for the
symmetries of the crystal lattice. For cubic symmetry,
$C_{11}=C_{22}=C_{33}$, $C_{44}=C_{55}=C_{66}$ and
$C_{12}=C_{13}=C_{23}$, the moduli can be expressed (solely with
$C_{ij}$) as
\ba
  &&K^V=\frac{C_{11}+2C_{12}}{3}\label{eq_KGcub}\\
  &&G^V=\frac{C_{11}+3C_{44}-C_{12}}{5}\nonumber\\
  &&K^R=K^V\nonumber\\
  &&G^R=\frac{5C_{44}(C_{11}-C_{12})}{3C_{11}+4C_{44}-3C_{12}},\nonumber
\ea
while for hexagonal symmetry, $C_{11}=C_{22}$, $C_{44}=C_{55}$,
$C_{66}=(C_{11}-C_{12})/2$ and $C_{13}=C_{23}$, the moduli follow as
\ba
  &&K^V=\frac{2C_{11}+C_{33}+2C_{12}+4C_{13}}{9}\label{eq_KGhex}\\
  &&G^V=\frac{7C_{11}+2C_{33}+12C_{44}-5C_{12}-4C_{13}}{30}\nonumber\\
  &&K^R=\frac{C_{33}(C_{11}+C_{12})-2C_{13}^2}{C_{11}+2C_{33}+C_{12}-4C_{13}}\nonumber\\
  &&G^R=\frac{15C_{44}(C_{11}-C_{12})(C_{33}(C_{11}+C_{12})-2C_{13}^2)}
              {6(C_{11}-C_{12})(C_{33}(C_{11}+C_{12})-2C_{13}^2) + 2C_{44}(2C_{11}^2 - 2C_{12}^2 + 
               5C_{12}C_{33} + C_{11}(4C_{13}+7C_{33}) - 4C_{13}(C_{12}+3C_{13}))}.\nonumber
\ea
Note that in isotropic limit (\textit{i.e.}, cubic symmetry with
$C_{44}=(C_{11}-C_{12})/2$) Voigt and Reuss estimates become the same,
$K^V=K^R=(C_{11}+2C_{12})/3$ and $G^V=G^R=(C_{11}-C_{12})/2$.

\begin{table}[H]
\begin{tabular}{c|c|c|c|c|c|c|c|c|c}
Material & Crystal system & $C_{11}$ & $C_{12}$ & $C_{13}$ & $C_{33}$ & $C_{44}$  & $A^u$ & $\tilde{A}^u$ & $|\tilde{A}^u-A^u|/A^u$ \\
\hline
\hline
Al       & FCC  & 107.3    & 60.9    &        &        & 28.3  & 0.048  & 0.051 & 0.077 \\
Ni       & FCC  & 246.5    & 147.3   &        &        & 127.4 & 1.149  & 1.171 & 0.019 \\
Au       & FCC  & 192.9    & 163.8   &        &        & 41.5  & 1.443  & 1.636 & 0.133 \\
Cu       & FCC  & 168.4    & 121.4   &        &        & 75.4  & 1.824  & 1.932 & 0.059 \\
Pb       & FCC  & 49.5     & 42.3    &        &        & 14.9  & 2.857  & 3.182 & 0.114 \\
\hline
Nb       & BCC  & 240.2    & 125.6   &        &        & 28.2  & 0.629  & 0.698 & 0.110 \\
Fe $\al$ & BCC  & 231.4    & 134.7   &        &        & 116.4 & 0.987  & 1.006 & 0.019 \\
Li       & BCC  & 13.5     & 11.44   &        &        & 8.78  & 7.970  & 8.272 & 0.038 \\
Na       & BCC  & 6.15     & 4.96    &        &        & 5.92  & 9.660  & 9.380 & 0.029 \\
\hline
Mg       & HCP  &  59.7    & 26.2    & 21.7   &  61.7  &  16.4 & 0.036  & 0.037 & 0.031 \\
Be       & HCP  & 292.3    &  26.7   &  14.0  & 336.4  & 162.5 & 0.048  & 0.042 & 0.121 \\
Zr       & HCP  & 143.4    & 72.8    & 65.3   & 164.8  &  32.0 & 0.108  & 0.115 & 0.059 \\
Co       & HCP  & 307.0    & 165.0   & 103.0  & 358.1  &  78.3 & 0.259  & 0.271 & 0.043 \\
Cd       & HCP  & 115.8    & 39.8    & 40.6   & 51.4   &  20.4 & 1.170  & 1.171 & 0.001 \\
Zn       & HCP  & 161.0    & 34.2    & 50.1   & 61.0   &  38.3 & 1.774  & 1.772 & 0.001 \\
\hline
\end{tabular}
\caption{
Elastic constants $C_{ij}$ (in units of GPa) and universal elastic
anisotropy index $A^u$ and its approximation $\tilde{A}^u$ of cubic
(FCC, BCC) and hexagonal (HCP) single crystals used in this
study. Elastic constants are taken from \citep{bower}. Note that
$C_{33}=C_{11}$ and $C_{13}=C_{12}$ for cubic crystal symmetry.}
\label{tab_au}
\end{table}

Equation (\ref{eq_au}) is used together with Eqs. (\ref{eq_KGcub}) and
(\ref{eq_KGhex}) to compute the universal elastic anisotropy index
$A^u$. Table \ref{tab_au} lists the elastic constants $C_{ij}$ and the
calculated $A^u$ of all the metals used in this study. In addition,
the approximated value of $A^u$, denoted by $\tilde{A}^u$, is
calculated as (see Eq.~(\ref{auapprox}))
\be
  A^u \approx \tilde{A}^u=6\left(\frac{Y^V}{Y^R} -1 \right)
      = 6\left(\frac{K^V G^V (3K^R+G^R)}{K^R G^R (3K^V+G^V)} -1 \right)
\ee
and compared to $A^u$ in Tab. \ref{tab_au}. A good agreement is
observed between $A^u$ and $\tilde{A}^u$ for all the materials
studied. The corresponding error is found to be less than 15\% with
respect to the true $A^u$.

%---------------------------------------
\section{Calculation of Taylor factor}
\label{app4}
%---------------------------------------

{Following the uniform strain assumption proposed by Taylor to
estimate the stress-strain curve of a polycrystalline aggregate,
Taylor factor is used to characterize the relation between local slips
to the average strain
\be
  M_i=\frac{d\Gamma_i}{d E_{p,eq}}=\frac{\sum_\al d\gamma_i^\al}{d E_p},
\label{deftaylor}
  \ee
where $d\Gamma_i$ represents the increment of total accumulated slip
in grain $i$ and $d E_{p,eq}$ is the increment of macroscopic
equivalent (von Mises) plastic strain. For uniaxial loading, however,
$d E_{p,eq}$ reduces to the increment of macroscopic uniaxial plastic
strain $d E_p$. {The estimate for the equivalent (von Mises)
grain stress $\sigma_{i,eq}$ can be obtained from the equality
requirement of plastic work increment on local and grain scales,
\be
  \sigma_{i,eq}\ d E_{p,eq} = \tau_0\ d\Gamma_i,\quad \sigma_{i,eq} = M_i \tau_0,
\ee
%
%\be
%\Sigma_i\,E_p = \tau_0 d\Gamma_i  \Rightarrow \Sigma_i = M_i \tau_0
%\ee
%
where} identical critical resolved shear stress is assumed for all
slip systems. Averaging the above relation leads to the classical
estimate of yield stress in a polycrystalline aggregate {under
tension}, $\Sigma_y \approx \langle M\rangle\tau_0$, where $\langle
M\rangle$ is the average Taylor factor.

For a given crystallographic orientation and equivalent plastic
strain, Taylor factor can be computed using Eq.~(\ref{deftaylor}) by
applying minimum work principle (or minimum microscopic shear
principle) to get $d\Gamma_i$. This relies on linear programming, and
dedicated tools are available {in literature} (see,
\textit{e.g.}, \citep{mtex}). In order to obtain the distributions of
Taylor factor in a polycrystalline aggregate for a specific {
macroscopic} stress tensor, the associated equivalent plastic strain
should be guessed to apply this methodology. {In this study, an
equivalent procedure has been used using Abaqus UMAT subroutine to
calculate $d\Gamma_i$. The procedure is detailed below.}}

The macroscopic tensile loading is applied in $X$ direction. In the
model, the grains are assumed to be randomly oriented and described
within crystal elasticity and ideal plasticity ($H=0$), accounting for
different slip systems (FCC, BCC or HCPn). To obtain local accumulated
slip ($\Gamma_i$) and local stresses ($\bm{\sigma}^i$), the applied
uniform total strain $\bm{E} = \bm{\epsilon}^i$ is identified
self-consistently as to recover the macroscopic uniaxial stress state
of the model, $\bm{\Sigma} = \sum_i\bm{\sigma}^i/N \approx \Sigma e_X
\otimes e_X$. When the number of grains $N$ is large enough
(\textit{e.g.}, $N\gtrsim 1000$) and grain orientations are randomly
distributed (providing zero texture), the macroscopic response of %Lin
the aggregate becomes quasi-isotropic. In this limit, the macroscopic
uniaxial stress state is realized (approximately) when the applied
total strain is of the form $\bm{E}=E e_X \otimes e_X - \nu E (e_Y
\otimes e_Y + e_Z \otimes e_Z)$. To achieve self-consistency, hence, a
macroscopic Poisson number $\nu(E)$ needs to be determined as a
function of total strain $E$. The optimal $\nu(E)$ is selected when
$\bm{\Sigma}$ becomes closest to the uniaxial stress solution $\Sigma
e_X \otimes e_X$. In this respect, a code for the identification of
$\nu(E)$ was written, based on the minimization of $\sum_{i,j}
\Sigma_{ij}^2-\Sigma_{11}^2$, and coupled to UMAT subroutine for the
calculation of local stresses $\bm{\sigma}^i$. The UMAT subroutine
followed the implementation of Huang \citep{huang} where slight
modifications were needed to accommodate the shear flow law of
Eq. (\ref{eq_1}). Once $\nu(E)$ is determined, $\Gamma_i$ follows
immediately from the UMAT.
{It can be noted that the same Taylor factor is obtained for
equibiaxial loading conditions due to the equivalence of applied
macroscopic strains.}

\begin{figure}[H]
\centering
\subfigure[]{\includegraphics[height = 5.5cm]{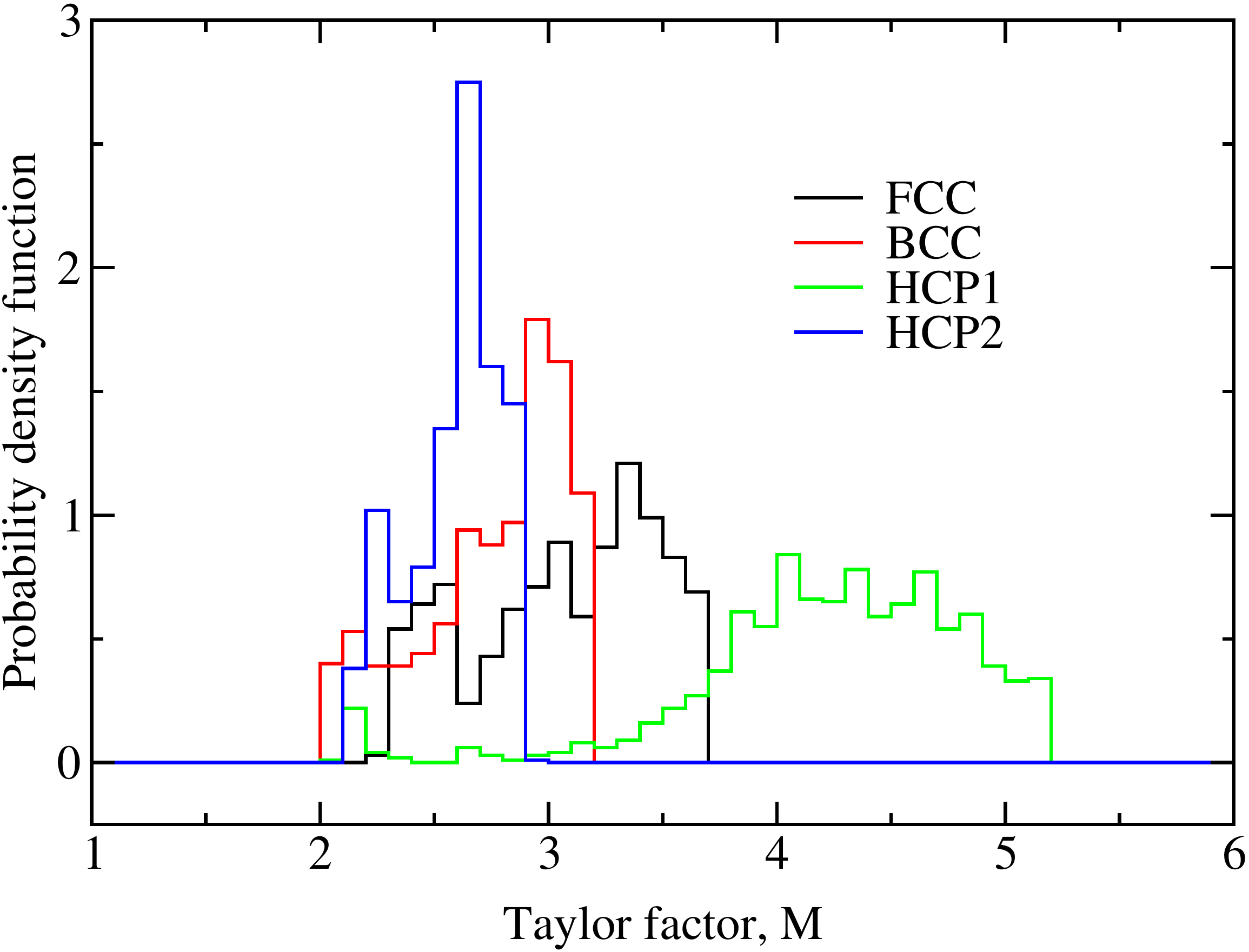}}
\caption{
Probability density function of Taylor factor calculated in
a 1000-grain
%
%Lin 
%
aggregate model with uniform strain assuming crystal elasticity and
ideal plasticity with various slip systems under uniaxial {(or
equibiaxial)} loading conditions. Results are shown for total strain
of 0.05. }
\label{fig:D1}
\end{figure}

Figure \ref{fig:D1} shows the distributions of Taylor factor
calculated in
%
%Lin 
%
a uniform-strain polycrystal model with $N=1000$ grains and different
slip systems: FCC, BCC, HCP1 and HCP2. Rather unusual, multi-modal
distributions are observed which seem to be intrinsic to (random) %Lin
aggregate model and not a statistical error (note that each
distribution is composed of 1000 data points). However, the average
Taylor factor, $\langle M\rangle=\sum_i M_i/N$, seems to be well
correlated with the distribution width measured, \textit{e.g.}, by its
standard deviation, $s(M)=\sqrt{\langle M^2\rangle - \langle
M\rangle^2}$. Smaller $\langle M\rangle$ implies also smaller
$s(M)$. It is therefore reasonable to assume that distributions can be
well characterized by first two statistical moments, see
Tab. \ref{tabA1}.

\begin{table}[H]
\begin{tabular}{c|c|c}
Crystal system &  $\langle M\rangle$  &  $s(M)$ \\
\hline
\hline
FCC          &  3.07  &  0.391 \\
\hline
BCC          &  2.76  &  0.320 \\
\hline
HCP1         &  4.23  &  0.603 \\
HCP2         &  2.58  &  0.197 \\
\hline
\end{tabular}
\caption{ 
Average Taylor factor, $\langle M\rangle$, and standard deviation of
Taylor factor, $s(M)$, calculated in a 1000-grain 
%
%Lin 
%
aggregate model with uniform strain assuming crystal elasticity and
ideal plasticity with various slip systems under uniaxial {(or
equibiaxial)} loading conditions. Results are shown for total strain
of 0.05. }
\label{tabA1}
\end{table}

The average Taylor factors for FCC and BCC may be compared with the
values published in literature. In Tab.~1 of \citep{rosenberg},
$\langle M\rangle=3.067$ and 2.754 are found to be in excellent
agreement with the values shown in Tab. \ref{tabA1}.

%---------------------------------------

\newpage

\bibliographystyle{plainnat}
\bibliography{spebib2}

\end{document}